\documentclass[amssymb,aps,amsmath,floatfix,pre, twocolumn,superscriptaddress,nolongbibliography]{revtex4-2}
\usepackage[latin1]{inputenc}

\usepackage{mathptmx}
\usepackage[T1]{fontenc}

\usepackage{graphicx}
\usepackage{amsmath,amssymb,amsfonts, MnSymbol}
\usepackage{mathrsfs}
\usepackage{physics}
\usepackage{bm}
\usepackage{hyperref}
\usepackage{bbold}
\usepackage{xcolor}
\usepackage{tikz}
\usetikzlibrary{shapes.geometric,arrows.meta,positioning,fit,backgrounds}
\usepackage{enumitem}

\newcommand{\algn}[1]{\begin{align} #1 \end{align}}
\newcommand{\sbeqs}[1]{\begin{subequations} #1 \end{subequations}}
\renewcommand{\pmat}[1]{\begin{pmatrix} #1 \end{pmatrix}}
\newcommand{\css}[1]{\begin{cases} #1 \end{cases}}

\newcommand{\ve}[1]{\boldsymbol{#1}}
\newcommand{\nn}{\nonumber}
\newcommand{\ee}{\ensuremath{\text{e}}}
\newcommand{\ed}{\ensuremath{\text{d}}}
\newcommand{\DD}[1]{\ensuremath{\frac{\text{d}}{\text{d} #1}}}

\DeclareMathOperator{\erfc}{erfc}
\newcommand{\mc}[1]{\ensuremath{\ms{#1}}}
\newcommand{\ms}[1]{\ensuremath{\mathscr{#1}}}
\newcommand{\mbb}[1]{\ensuremath{\mathbb{#1}}}
\newcommand{\kb}{\ensuremath{k_\text{B}}}
\newcommand{\eqnlab}[1]{\label{eq:#1}}
\newcommand{\seclab}[1]{\label{sec:#1}}
\newcommand{\figlab}[1]{\label{fig:#1}}
\newcommand{\eqnref}[1]{\eqref{eq:#1}}
\newcommand{\Eqnref}[1]{Eq.~\eqref{eq:#1}}
\newcommand{\Eqsref}[1]{Eqs.~\eqref{eq:#1}}
\newcommand{\secref}[1]{\ref{sec:#1}}
\newcommand{\Secref}[1]{Sec.~\ref{sec:#1}}
\newcommand{\Secsref}[1]{Secs.~\ref{sec:#1}}
\newcommand{\figref}[1]{\ref{fig:#1}}
\newcommand{\Figref}[1]{Fig.~\ref{fig:#1}}
\newcommand{\Figsref}[1]{Figs.~\ref{fig:#1}}

\begin{document}
\title{Relaxation speed in quenched-random force fields}
\author{Jan Meibohm}
\affiliation{Technische Universit\"at Berlin, Institut f\"ur Physik und Astronomie, Fachgruppe Theoretische Physik, Hardenbergstra\ss{}e 36, 10623 Berlin, Germany}
\author{Sabine H. L. Klapp}
\affiliation{Technische Universit\"at Berlin, Institut f\"ur Physik und Astronomie, Fachgruppe Theoretische Physik, Hardenbergstra\ss{}e 36, 10623 Berlin, Germany}
\begin{abstract}
Disorder is commonly expected to slow down relaxation, as in glasses and disordered magnets. Here we show that this need not be the case in general, by analysing the asymptotic relaxation rate of a Brownian particle in a harmonic background potential and a quenched-random Gaussian force field with conservative and non-conservative components. Combining weak-disorder perturbation theory, random-matrix theory, and numerical solutions of the Fokker-Planck equation, we show that the outcome depends on the geometry of the background and on the properties of the random forces. We identify two distinct mechanisms that shape the relaxation speed: an effective stiffening of the background by the disorder, as previously reported in one dimension, and interactions (repulsive and attractive) between the levels of the Fokker-Planck operator. Through level attraction, a common feature of non-Hermitian operators, non-conservative driving forces pull the slowest mode towards faster ones, which facilitates faster relaxation. In anisotropic backgrounds, both effects occur at second order in the disorder strength, enabling either slower or faster relaxation. In isotropic backgrounds, level interactions dominate and the disorder affects the degenerate slowest mode already at first order. The shift is then given by the left-most eigenvalue of a Gaussian random matrix and relaxation is generically slowed down. In backgrounds whose anisotropy is comparable to the disorder strength, level attraction is most effective by merging the slowest mode with its neighbour into an oscillating complex-conjugate pair. In this case, relaxation becomes faster on average once the random forces are sufficiently non-conservative. Our results draw a nuanced picture of how disorder affects the asymptotic relaxation speed in complex systems and uncover general mechanisms that are likely to apply beyond the current setting.
\end{abstract}
\maketitle

\section{Introduction}\seclab{intro}
Relaxation, one of the most fundamental processes in statistical mechanics, describes the dynamics by which a physical system evolves towards its equilibrium or a steady state. Close to equilibrium or nearby a steady state, relaxation is governed by linear-response theory~\cite{Kub57a,Kub66,Cal51,Mar08} and its non-equilibrium extensions~\cite{Bai13}, which connect relaxation to stationary correlation functions.  Far from steady states, no such general frameworks exist, and non-equilibrium relaxation gives rise to a rich variety of phenomena, many of which remain poorly understood. A prominent example is the Mpemba effect~\cite{Mpe69,Lu17,Las17,Kli19,Kum20,Bai19}, the observation that an initially hotter system can reach a cold temperature faster than a system at an intermediate initial temperature. Others include memory effects such as ageing~\cite{Arc22,Cal06} and the Kovacs effect~\cite{Kov63}, coarsening in phase-ordering kinetics~\cite{Bra02}, and the anomalously slow relaxation observed in structural and spin glasses~\cite{Mez87,Bou97,Cug94}.

Recent theoretical progress has come from studying relaxation in simplified model systems with few degrees of freedom.  For example, the so-called Markovian Mpemba effect~\cite{Lu17,Kli19,Kum20,Kum22,Che21} as well as heating--cooling asymmetries~\cite{Lap20,Mei21b,Vu21,Iba24,Die23} are typically studied in simplified models that can be analysed in great detail. The same holds for dynamical phase transitions in the relaxation process~\cite{Mei22a,Mei23b,Tez23b,Vad24}, and for similar relaxation anomalies in quantum systems~\cite{Car21,Cha23,Wu24,Jos24,Ryl24,Pta24}.  A recurring theme is that the structure of the energy landscapes is a crucial ingredient for non-trivial relaxation behaviour: the Markovian Mpemba effect, for instance, is often associated with double-well potentials in which one well acts as a metastable funnel~\cite{Lu17,Kum20}. Simplified models have also guided the search for strategies to speed up relaxation, for example by preheating~\cite{Gal20,Pem21,Pem24}, or by optimised driving protocols that act as shortcuts to equilibrium~\cite{Def14,Mar16,Gue23,Ray23,Tez23a}.

While minimal models have greatly improved our understanding of non-equilibrium relaxation, they cannot fully capture the physics of realistic complex systems, in which the free-energy landscape is typically high-dimensional, possesses many local minima and maxima~\cite{Mez87,Fra91,Yu14}, and may even be dynamical~\cite{Mak26}.  For example, the free-energy landscapes of spin and structural glasses display a hierarchical structure of local minima of exceptional complexity~\cite{Mez87,Bou97,Cha17}. Similar pictures emerge in the dynamics of macromolecules such as proteins, whose conformations explore a high-dimensional free-energy landscape~\cite{Fra91}, or in neural-network learning where the gradient-descent dynamics takes place in a rugged loss landscape imposed by the training data~\cite{Win25}. The complexity of such landscapes has been quantified by the number of stationary points of $d$-dimensional random potentials with harmonic background. For $d\to\infty$, such landscapes undergo a transition~\cite{Fyo04,Fyo13} where either the background or the disorder dominates. In the first case, the global minimum is the only stationary point, whereas for strong disorder the number of stationary points grows exponentially with $d$. This transition coincides with the onset of glassy dynamics and with the breakdown of replica symmetry~\cite{Fyo04,Fyo13}. An analogous transition occurs for non-gradient random forces with conservative and solenoidal components, where the vast majority of the exponentially many equilibria are unstable~\cite{Fyo16,Ben21}.%

A widely held intuition, supported by many observations, is that disorder \emph{generically slows down} relaxation, glasses and disordered magnets being the prime examples~\cite{Mez87,Cug94,Bou97,Zha22}. Quenched disorder has also been observed to slow down relaxation in the random-field Ising model~\cite{Nat88}, in fluids in mesoporous materials~\cite{Kla01,Woo03,Det03}, and in polar active flocks~\cite{Dua21}, see Ref.~\cite{Bir15} for an overview of slow relaxation in classical and quantum systems.  In a recent study~\cite{Mei25a}, we have challenged this intuition by an in-depth analysis of a one-dimensional model for a disordered system, based on a Brownian particle in a quenched-random potential with a harmonic background. The harmonic background confines the particle, so that relaxation rather than diffusion~\cite{Lif62,Zwa88,Dea94,Dee94,Mak26} becomes the relevant process at long times. Such a background can be interpreted as a ``megabasin'' of the free energy of a structural glass~\cite{Zha22}, or as a regulariser of the loss landscape of a neural network~\cite{Win25}. While disorder does slow relaxation in many regions of parameter space, we found the surprising result that it can also \emph{accelerate} it.  In low dimensions, this model is realisable in experiments, for example on a colloidal particle confined in an optical trap superimposed with laser speckles~\cite{Bew16a,Bew16b,Vol14}, synthetic laser fields~\cite{Han12,Zun22,Vos26}, or by introducing a disordered substrate~\cite{Su17}. Related models have also been studied for quantum wave packets and Bose-Einstein condensates in disordered harmonic traps~\cite{Dri10,Bho10,Hsu18,Hsu20,Schu22}.
\begin{figure}
	\includegraphics[width=\linewidth]{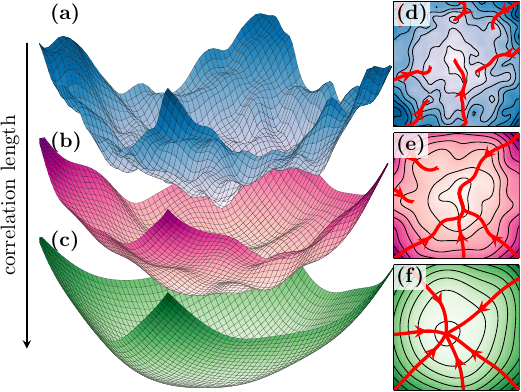}
	\caption{(a)--(c) Harmonic background potential supplemented by a disordered potential part with correlation lengths increasing from top to bottom, as indicated by the arrow. (d)--(f) Contour plots of the potentials shown in (a)--(c) with smallest values of the potentials shown in light colours. Noise-free particle trajectories are shown in red, with arrows indicating the direction of motion.}\figlab{potential}
\end{figure}

The one-dimensional setting~\cite{Mei25a}, however, imposes a strong geometric constraint: a particle must traverse every energy barrier on its way toward equilibrium, with no possibility of going around. Furthermore, in higher dimensions, extrema are most often saddle points~\cite{Fyo04,Fyo16,Ben21}, where the dynamics, albeit slower, do not get trapped. Figure~\figref{potential} shows a sketch of two-dimensional realisations of a random potential with a harmonic background, where the correlation length of the quenched-random part decreases from bottom to top. We observe that small correlation lengths (top) favour the formation of additional local minima where particle trajectories (red) tend to be trapped in the noise-free limit. The additional space in $d\geq2$, however, in principle allows particles to move around these minima.

These considerations seem to suggest that one-dimensional disorder is exceptionally effective in changing the relaxation speed, while the effect is weakened in higher dimensions. By opening additional escape pathways, it is also tempting to believe that higher dimensions overall facilitate faster relaxation.

To test these hypotheses, we here consider the more general higher-dimensional case $d\geq 2$, where the geometric constraints of one dimension are lifted. Our work shows that the intuitive expectations developed above are in general too naive, at least when focussing on \emph{long-time} relaxation, which is controlled by the slowest relaxation modes. Throughout, we quantify relaxation by its asymptotic rate, the smallest non-vanishing real part of the spectrum of the Fokker-Planck operator. We do not address relaxation at finite times, which depends also on the initial state and on the faster modes. It turns out that in many, though not all, circumstances, higher dimensions in fact generate \emph{more} ways for the system to be slower than in the unperturbed case. We furthermore identify two key factors that are decisive in determining whether disorder accelerates relaxation or not.

The first key factor is the geometry of the harmonic background potential, i.e., whether it is isotropic or anisotropic. In an isotropic background without disorder, the slowest relaxation mode is $d$-fold degenerate, and the disorder lifts this degeneracy in every disorder realisation. We show below that for weak disorder, the corresponding disorder-induced shift is given by the eigenvalues of a random matrix. The slowest relaxing mode then corresponds to selecting the eigenvalue with smallest real part of a set of random eigenvalues. Because this selection process biases the real part of the mode towards negative values, quenched-random force fields in isotropic backgrounds generically make the slowest modes slower than in the unperturbed background, as described above.

An anisotropy background, by contrast, lifts this degeneracy. The shift in the relaxation speed is then no longer obtained through such an extremal selection, and it may take either sign, so that relaxation can be genuinely sped up by disorder, analogous to the one-dimensional case~\cite{Mei25a}. We furthermore identify an in-between regime of weak anisotropy, where the anisotropy of the background forces is of the same order as the quenched-disordered forces. This regime combines aspects of both the isotropic and anisotropic backgrounds and enables either slower or faster relaxation.

The second key factor is the nature of the quenched-random forcing. Considering dimensions higher than one opens up the possibility of non-conservative (solenoidal or curl) forces, which enrich the phenomenology. Non-conservative forces drive a system away from equilibrium and lead to non-equilibrium steady states with non-vanishing currents in the long-time limit. They occur in systems with external drive such as colloidal particles in rotating laser traps~\cite{Bli07b} but also in transport phenomena in random media~\cite{Isi92} or advection-diffusion through porous materials~\cite{Cha97}. We find that non-conservative forces generally increase the speed of relaxation, as they help particles to escape potential minima. This is reminiscent of the acceleration of relaxation by non-reversible drifts that break detailed balance~\cite{Hwa93,Ich13}, which is exploited in Monte Carlo sampling~\cite{Suw10,Cog21}.
\begin{figure*}
\centering
\includegraphics{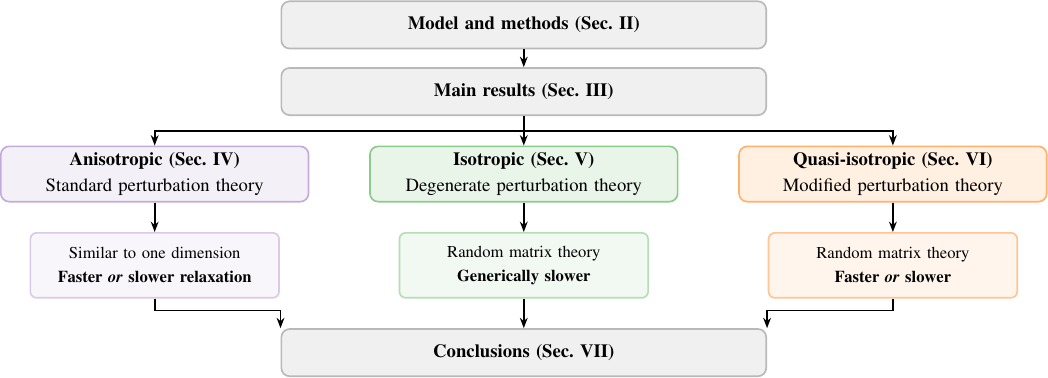}
\caption{Schematic overview of the logical structure of the paper.
  Starting from the model (top), weak-disorder perturbation theory leads to
  three regimes determined by the degeneracy of the unperturbed eigenvalues.}
\figlab{flowdiagram}
\end{figure*}

A central result of this paper is that when combining these key factors, disorder either decreases \emph{or} increases the relaxation speed, depending on the amount of anisotropy, non-conservative forcing, and on the disorder correlation length. Faster relaxation arises from two distinct mechanisms. The first is a stiffening of the background on average, the mechanism of the one-dimensional case~\cite{Mei25a}. The second is level attraction, by which non-conservative forces pull the slowest mode towards faster ones. It has no counterpart in one dimension and is most effective in quasi-isotropic backgrounds, where it merges the slowest mode with its neighbour into an oscillating complex-conjugate pair.%
\subsection{Structure of the paper}
The remainder of the paper is structured as shown schematically in Fig.~\ref{fig:flowdiagram}. In \Secref{model} we introduce the model: a
Brownian particle in a harmonic background potential, subject to a quenched-random Gaussian force field. We introduce the dimensionless control parameters~$\zeta_T$,
$\kappa$, and~$\gamma$, and the Fokker--Planck formalism used to quantify the
asymptotic relaxation speed through the shift~$\Delta\lambda_1$ of the smallest non-vanishing eigenvalue~$\lambda_1$ of the Fokker--Planck operator. In particular, \Secref{perturbation} develops the weak-disorder perturbation theory based on coherent states of the $d$-dimensional quantum harmonic oscillator that underpins all analytical results. After having laid out the model and methods in detail, we provide in \Secref{main} a summary of the main results.

A central finding of our analysis is that the relaxation behaviour crucially depends on the (an)isotropy of the background forces, which determines the degeneracy of the smallest non-vanishing eigenvalue $\lambda_1^0$ in the unperturbed background. We distinguish three cases each of which is discussed in a dedicated section:
an anisotropic background (\Secref{aniso}, purple in \Figref{flowdiagram}), where $\lambda_1^0$ is
non-degenerate, an isotropic background (\Secref{iso}, green in \Figref{flowdiagram}), where $\lambda_1^0$ is $d$-fold degenerate, and a quasi-isotropic background
(\Secref{qiso}, orange in \Figref{flowdiagram}), where the background anisotropy is of the same order as the disorder strength. Roughly speaking, higher-dimensional anisotropic systems show similar, albeit more intricate, behaviour to the one-dimensional case~\cite{Mei25a}. For the anisotropic case, we derive explicit
expressions for the mean eigenvalue shift and its variance, obtain universal limits at small and large correlation lengths, construct state diagrams delineating regions of faster and slower relaxation, and compare all results with numerical simulations.
The isotropic and quasi-isotropic cases exhibit new physical mechanisms for relaxation slow-downs and speedups through level attraction and repulsion, which require a different kind of tool-box provided by random matrix theory~\cite{Meh04}. For the isotropic case we connect the perturbation
matrix with the Gaussian elliptic random matrix ensembles and discuss the large-$d$ limit. The quasi-isotropic case provides an intermediate regime between the isotropic and anisotropic ones where the different behaviours mix. Section~\secref{conclusions} summarises the results and discusses experimental
implications. The Appendices contain detailed calculations that were omitted in the main text.
\section{Model and methods}\seclab{model}
We consider a single Brownian particle in a $d$-dimensional harmonic potential distorted by quenched disorder, which has Gaussian statistics and a finite correlation length. The disorder enters in the form of a Gaussian random force field $\ve F$ with both conservative and non-conservative components.
\subsection{Langevin equation}
The Langevin equation for a Brownian particle in a harmonic background potential subject to a random force field $\ve F$ in $d$ dimensions reads
\algn{\eqnlab{langevin}
	\dot{\ve x} = -\mu_\text{p} \mbb{A}\ve x + \frac{\mu_\text{p}\zeta}{\ell_c} \ve F\left(\frac{\ve x}{\ell_c}\right) + \sqrt{\mu_\text{p}/\beta}\ve \xi(t)\,,
}
with mobility $\mu_\text{p}$, force strength $\zeta$, and correlation length $\ell_c$. The term $\ve \xi(t)$ denotes $d$-dimensional Gaussian white noise with correlation function
\algn{
	\langle \xi_i(t)\xi_j(t')\rangle = 2\delta_{ij}\delta(t-t')\,.
}
The appearance of $\mu_\text{p}$ and $\beta=1/(\kb T)$ multiplying $\ve \xi(t)$, with $T$ being the temperature of the bath and $\kb$ Boltzmann's constant, accounts for the Einstein relation.

The symmetric stiffness matrix $\mbb{A}$ defines the shape of the harmonic potential background. Both $\ve \xi(t)$ and the forces $\ve F$ are assumed to be spatially isotropic, so we can bring $\mbb{A}$ into diagonal form, $\mbb{A} = \text{diag}(A_1,A_2,\ldots)$, with $A_1\leq A_2\leq\ldots\leq A_d$.
\subsection{Quenched-random forces}
We assume $\ve F$ to be quenched-random with Gaussian statistics, zero mean and an isotropic and homogeneous correlation function. These properties stipulate that the quenched-random forces are statistically invariant under spatial rotations and translations. We construct $\ve F$ from the gradient of a random scalar potential $V(\ve x)$ and from the divergence of an antisymmetric rank-two tensor $\Phi_{ij}(\ve x) = -\Phi_{ji}(\ve x)$ as
\algn{\eqnlab{force}
	F_i(\ve x) = -\sqrt{1-\gamma}\, \nabla_i V(\ve x) - \sqrt{\gamma} \sum_{j=1}^d\nabla_j \Phi_{ji}(\ve x)\,.
}
Here the first term corresponds to the conservative potential part of the force, while the second one describes a non-conservative solenoidal part. In one, two, and three dimensions, $\Phi_{ij}$ can be brought into the more familiar forms
\sbeqs{\eqnlab{potentials}
\algn{
	\Phi_{ij}(\ve x) &= 0\,, & (d=1)\,,\\
	\Phi_{ij}(\ve x) &= \epsilon_{ij}\phi(\ve x)\,, & (d=2)\,,\\
	\Phi_{ij}(\ve x) &= \sum_{k=1}^d\epsilon_{ijk}A_k(\ve x)\,, & (d=3)\,.
}
}
 First, in $d=1$ we have $\Phi_{ij}=0$, because all forces can be derived from a potential in one dimension. In $d=2$, we can write $\Phi_{ij}$ in terms of the stream function $\phi(\ve x)$ and the Levi-Civita symbol $\epsilon_{ij}$, where $\epsilon_{12}=-\epsilon_{21}=1$ and zero otherwise. Finally, in three dimensions $\Phi_{ij}$ can be expressed as the curl of the vector potential $\ve A$. The solenoidal part of the force $\ve F$ then reduces to $\sqrt{\gamma}\, \nabla\times\ve A$, as is familiar in electromagnetism.

 The relative strength of the potential and solenoidal contributions to $\ve F$ is controlled by the parameter $\gamma\in[0,1]$. For $\gamma=0$ the forces are purely potential and thus conservative. The case $\gamma=1$, by contrast, corresponds to purely non-conservative forces. Values for $\gamma$ between these limits represent mixtures of these limiting cases. Potential forces are common in relaxation problems towards an equilibrium state. Examples with disordered harmonic potentials range from relaxation of trapped colloidal particles in disordered environments~\cite{Bew16a,Su17} to models for neural-network learning~\cite{Win25}. For noiseless dynamics with purely potential forces $\ve F(\ve x) = - \nabla V(\ve x)$, the potential $V(\ve x)$ serves as a Lyapunov function, i.e.,
 \algn{
	 \DD{t} V[\ve x(t)] = -\nabla V[\ve x(t)]\cdot\nabla V[\ve x(t)]<0\,,
}
which implies that the dynamics corresponds to gradient descent towards the minimum of $V(\ve x)$, see right panels in \Figref{potential}.

Non-conservative forces drive non-equilibrium fluxes in the system. For each realisation, the forces $\ve F$ in the model --- including the non-conservative ones --- are time-independent, so for any finite value $\gamma>0$ the system relaxes toward a non-equilibrium steady state in the long-time limit.

Similar to the discussion of potential forces (\Figref{potential}) above, we find it illustrative to consider the motion of a particle in a noiseless $d$-dimensional purely solenoidal field. For this purpose, it is convenient to express $\Phi_{ij}$ locally as
\algn{
	\Phi_{ij}(\ve x) = \sum_{k_1,\ldots,k_{d-2}=1}^d\epsilon_{ij k_1\ldots k_{d-2}}\phi_1(\ve x)\prod_{n=1}^{d-2}\nabla_{k_n}\phi_{n+1}(\ve x)\,,
}
with the ``stream functions''~\cite{Mor98} or ``Hamiltonians''~\cite{Nam73} $\phi_n(\ve x)$. Along particle trajectories, all $\phi_n(\ve x)$ are conserved
\algn{
	\DD{t}\phi_n[\ve x(t)] = \sum_{j=1}^d\nabla_{j}\phi_n(\ve x) F_j(\ve x)=0\,,
}
which shows that particle motion occurs on the intersections of the level surfaces of the $d-1$ stream functions $\phi_n(\ve x)$. This mechanism becomes particularly simple in $d=2$, where a single stream function $\phi(\ve x)$ suffices. In this case, a purely solenoidal $\ve F$ moves the particles on a level line of $ \phi(\ve x)$. In the presence of thermal noise, the continuous motion of particles driven by non-conservative forces leads to non-equilibrium probability fluxes even at steady state. Consequently, driven systems never equilibrate.

Figure~\figref{contours} shows contours of the stream function $\phi(\ve x)$ in $d=2$ together with the trajectories of particles transported in the corresponding non-conservative force field. Particles move on the level lines of $\phi(\ve x)$, and are forced on closed orbits.
\begin{figure}
	\includegraphics[width=\linewidth]{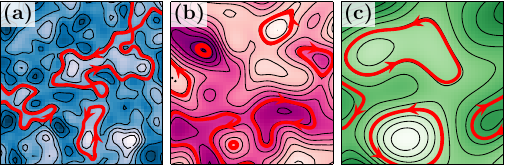}
	\caption{Effects of quenched-random non-conservative forces in two dimensions. Colours indicate the magnitude of the stream function $\phi$ with darker colours corresponding to larger values. The red lines show particle trajectories in the corresponding force field with arrows indicating the direction of particle motion. Panels (a)--(c) correspond to increasing correlation lengths from left to right. }\figlab{contours}
\end{figure}

The correlation functions of $V(\ve x)$ and $\Phi_{ij}(\ve x)$ upon averaging over disorder realisations are assumed to take the forms
\sbeqs{\eqnlab{corrs}
\algn{
	\langle V(\ve x) V(\ve y)\rangle_F &= C_F(|\ve x - \ve y|)\,,\\
	\langle \Phi_{ij}(\ve x) \Phi_{kl}(\ve y)\rangle_F &= (\delta_{ik}\delta_{jl}-\delta_{il}\delta_{jk})C_F(|\ve x - \ve y|)\,.
}
}
The a priori arbitrary function $C_F(x)$ has the symmetry and normalisation properties
\sbeqs{
\algn{
	C_F(0)  &= 1\,,\\
	C_F(-x) &= C_F(x)\,,\\
	\int_0^\infty\!\!\!\ed x \ C_F(x) &= 1\,.
}
}
In addition, the requirement that $\langle [\int\!\!\ed \ve x f(\ve x)V(\ve x)]^2\rangle_F>0$ (analogous conditions hold for $\Phi_{ij}$), implies that $C_F$ must be positive definite~\cite{Yag86}, i.e.,
\algn{\eqnlab{posdef}
	\int\!\!\!\ed^d x\!\! \int\!\!\!\ed^d x'\, v(\ve x) C_F(|\ve x - \ve x'|) v(\ve x') &> 0\,,
}
for an arbitrary function $v(\ve x)$. This is ensured by requiring that the Fourier transform $\hat C_F(|\ve k|)$ of $C_F(|\ve x|)$ is positive for all $\ve k$~\cite{Boc55}. When convenient, we will parametrise the isotropic correlation function $C_F(|\ve x|)$ as
\algn{
	C_F(|\ve x|) = \rho(r^2)\,,
}
where $r^2 = \sum_{n=1}^d x_n^2$. %
Positive-definiteness~\eqnref{posdef} becomes increasingly restrictive for increasing number of dimensions $d$. In fact, as $d\to\infty$ all correlation functions can be written in the form~\cite{Sch38}
\algn{\eqnlab{schoenberg}
	\rho(r^2) = \int_0^\infty\!\!\!\ed \eta\ \ee^{-\eta^2r^2} \Omega(\eta)\,,
}
where $\Omega(\eta)\geq 0$ and both $\Omega(\eta)$ and $\Omega(\eta)\eta^{-2}$ are integrable over $\eta$. Consequently, in high dimensions $d\to\infty$, an isotropic correlation function $\rho(r^2)$ is a superposition of Gaussians with variance $2/\eta^2$. Here, we will mainly focus on the case $\rho(r^2) = \ee^{-\pi r^2/4}$, where $\Omega(\eta) = \delta(\eta-\sqrt{\pi}/2)$.

Combining \Eqsref{force} with \eqnref{corrs}, we obtain for the force correlation functions
\begin{multline}\eqnlab{Fcorr}
	\langle F_i(\ve x) F_j(\ve y)\rangle_{F} = \left[(1-\gamma)\grad^{(\ve x)}_i \grad^{(\ve y)}_j \right.\\
	\left.+ \gamma\left(\delta_{ij} \grad^{(\ve x)}\cdot \grad^{(\ve y)} - \grad^{(\ve y)}_i \grad^{(\ve x)}_j\right)\right]C_F\left(|\ve x - \ve y|\right)\,.
\end{multline}
In addition to $\gamma$, the correlation length $\ell_c$ enters $\ve F$ through its argument in the Langevin equation~\eqnref{langevin}.

For  $\ve y=0$, we write the correlation function~\eqnref{Fcorr} in terms of $\rho$ as
\begin{multline}\eqnlab{Fcorr2}
	\langle F_i(\ve x) F_j(0)\rangle_{F} = -2\left\{\delta_{ij}[1+\gamma(d-2)]\rho'(r^2)\right.\\ +\left. 2\left[\gamma r^2\delta_{ij} + (1-2\gamma)x_ix_j\right]\rho''(r^2)\right\}\,.
\end{multline}

We now introduce dimensionless coordinates that simplify the exposition that follows.
\subsection{Dimensionless coordinates}\seclab{dimensionless}
We introduce the dimensionless time $\tilde t = t/\tau_\text{R}$, where $\tau_\text{R} = (\mu_\text{p} A_1)^{-1}$ denotes the largest relaxation time of the unperturbed system. We furthermore dedimensionalise space as $\ve{\tilde x} = \ve x/\ell_T$ with the thermal length scale $\ell_T = 1/\sqrt{A_1 \beta}$. With these coordinates and dropping the tildes, we find the dimensionless dynamics
\algn{\eqnlab{laneqn}
	\dot{\ve x} = -\grad V_0(\ve x) + \frac{\zeta_T}{\kappa} \ve F\left(\frac{\ve x}{\kappa}\right) + \ve \xi(t)\,,
}
where $V_0$ denotes the dimensionless background potential
\algn{\eqnlab{V0eqn}
	V_0(\ve x) = \frac12\ve{x}^{\sf T}\!\mbb{a}\,\ve x\,, \qquad \mbb{a} = \text{diag}(a_1,a_2,\ldots,a_d)\,,
}
where $1=a_1\leq a_2\leq\ldots\leq a_d$. Furthermore, $\zeta_T = \beta\zeta$ denotes the dimensionless coupling, and $\kappa = \ell_c/\ell_T$ denotes the dimensionless force correlation length.

In addition to these two parameters, the model depends on the dimensionless parameter $\gamma$ introduced earlier. Hence, in dimensionless formulation, the model depends on the three parameters $\zeta_T$, $\kappa$, and $\gamma$, as well as on the $d-1$ dimensionless background stiffnesses $a_2\leq a_3\leq\ldots\leq a_d$.

The two parameters $\zeta_T$ and $\gamma$ set the strengths of the potential and the solenoidal parts of $\ve F$ separately: $\gamma$ assigns the weight $\zeta_T^2(1-\gamma)$ to the potential component of the forces and $\zeta_T^2\gamma$ to each of the $d-1$ solenoidal components. For $\gamma=1/2$ all $d$ directions have equal weight, the correlation function~\eqnref{Fcorr} becomes proportional to $\delta_{ij}$, and the components of $\ve F$ are statistically independent. As a consequence, the variance of each force component, obtained from \Eqnref{Fcorr2} at $\ve x=0$, is proportional to $\zeta_T^2[1+\gamma(d-2)]$ and grows with $d$ for $\gamma>0$.

When considering the behaviour in high dimensions, it is therefore useful to consider also a parametrisation that keeps the force variance fixed. To this end, we define
\algn{\eqnlab{hatgamma}
	\hat\zeta_T^2 = \left[1+\gamma(d-2)\right]\zeta_T^2\,,\qquad \hat\gamma = \frac{(d-1)\gamma}{1+\gamma(d-2)}\,,
}
where $\hat\zeta_T$ measures the strength of each force component, and $\hat\gamma\in[0,1]$ denotes the fraction of the total force variance carried by the solenoidal part. The weights of the two kinds of directions then read $\zeta_T^2(1-\gamma)=\hat\zeta_T^2(1-\hat\gamma)$ and $\zeta_T^2\gamma=\hat\zeta_T^2\hat\gamma/(d-1)$, so that all results can be translated between the two parametrisations. They coincide for $d=2$, but lead to different limits as $d\to\infty$. At fixed $\gamma>0$, we have $\hat\gamma\to1$, i.e., the solenoidal part dominates the forces in high dimensions. Keeping $\hat\gamma$ fixed instead amounts to scaling the solenoidal part of $\ve F$ in \Eqnref{force} with $d^{-1/2}$ relative to the potential part. This is the convention used in Ref.~\cite{Fyo16}, where the number of equilibria in random dynamical systems with forces of the form~\eqnref{force} was studied. Throughout this paper we use $\zeta_T$ and $\gamma$. Whenever the dependence on $d$ matters, we state the results also in terms of $\hat\zeta_T$ and $\hat\gamma$.
\subsection{Fokker-Planck equation}\seclab{fpe}
The relaxation behaviour of a probability density of particles is conveniently studied through the spectrum of the Fokker-Planck operator $\ms{L}$. We find it convenient to work with the adjoint $\ms{L}^\dagger$, which satisfies the eigenvalue equation
\algn{\eqnlab{Leig}
	\ms{L}^\dagger l_n(\ve x) = -\lambda_n l_n(\ve x)\,,
}
where $\lambda_n$, $n=0,1,\ldots$ are the eigenvalues of the operator $\ms{L}^\dagger$ given by
\algn{\eqnlab{FPop}
	\ms{L}^\dagger = \left[-\mbb{a}\ve x + \frac{\zeta_T}{\kappa} \ve F\left(\frac{\ve x}{\kappa}\right)\right]\cdot \grad +\grad\cdot\grad = \ms{L}_0^\dagger + \zeta_T \ms{L}_{F}^\dagger\,.
}
Both the operator $\ms{L}^\dagger$ and its eigenvalues $\lambda_n$ are random due to their dependence on the random force $\ve F$. Consequently, each realisation of $\ve F$ will in general yield different eigenvalues.

To analyse the connection between the eigenvalues and long-time relaxation, we consider an observable $f$ of the state $\ve x$. The noise average $\langle f(\ve x)\rangle$ of $f$ evolves in time $t$ with $\ms{L}^\dagger$ according to
\algn{
	\partial_t\langle f[\ve x(t)]\rangle = \langle \ms{L}^\dagger f(\ve x)\rangle\,.
}
In the long-time limit $t\gg\tau_\text{R}$, the average $\langle f[\ve x(t)]\rangle$ is dominated by its time-independent stationary value $\langle f(\ve x)\rangle_{ss}$ and exponential corrections
\algn{\eqnlab{fasympt}
	\langle f(\ve x)\rangle \sim \langle f(\ve x)\rangle_{ss} + C_1 \ee^{-\lambda_1 t/\tau_\text{R}} + C_2 \ee^{-\lambda_2 t/\tau_\text{R}} + \ldots\,,
}
where $\lambda_n$ are the eigenvalues of $\ms{L}$. For purely conservative forces ($\gamma=0$~\footnote{Note that in $d=1$, the case studied in Ref.~\cite{Mei25a}, the force field $\ve F$ is always conservative.}), the spectrum of $\ms{L}^\dagger$ (and $\ms{L}$) is real~\cite{Ris89}. In general, however, non-conservative forces ($\gamma>0$) allow the eigenvalues $\lambda_n$ to occur in complex-conjugate pairs, denoted by
\algn{\eqnlab{realimag}
	\lambda_n = \mu_n + i\omega_n\,.
}
For convenience, we choose a numbering of the spectrum such that $0=\mu_0\leq \mu_1\leq\ldots$, as implicitly done in \Eqnref{fasympt}.

If $\lambda_1$ occurs as a complex-conjugate pair, this enables long-lived oscillatory behaviour of \Eqnref{fasympt} in the long-time limit:
\algn{\eqnlab{oscill}
	\langle f(\ve x)\rangle - \langle f(\ve x)\rangle_{ss} \sim C_{12} \ee^{-\mu_1 t}\cos(\omega_1 t+\varphi_{12})+\ldots\,,
}
with constants $C_{12}$ and $\varphi_{12}$. Consequently, finite $\omega_{1}$ enables oscillations on top of exponential relaxation induced by $\mu_1>0$. For a concrete example, consider the case $f[\ve x(t)] = (\ve x(t)-\langle\ve x\rangle_{ss})\cdot (\ve x(0)-\langle\ve x\rangle_{ss})$, for which the left-hand side of \Eqnref{oscill} becomes the correlation function $C_x(t) = \langle\ve x(t)\cdot \ve x(0)\rangle -\langle\ve x\rangle_{ss}^2$. Equation~\eqnref{oscill} then corresponds to the long-time decay of $C_{x}(t)$,
\algn{
	C_x(t) \sim C_{12} \ee^{-\mu_1 t}\cos(\omega_1 t+\varphi_{12})+\ldots\,,
}
which is also governed by the eigenvalue $\lambda_1$ with the smallest non-vanishing real part. Importantly, the long-time decay of $C_x(t)$ can be measured from long trajectories in the stationary state.

In most of this work, we focus on the relaxation speed measured by $\mu_1$ and disregard the frequency $\omega_1$, as explained in more detail in \Secref{relspeed}. However, this does not mean that complex-conjugate pairs are irrelevant. On the contrary, we find in \Secref{qiso} that the formation of complex-conjugate pairs is the driving mechanism for enabling faster relaxation in quasi-isotropic backgrounds.

To determine the relevant eigenvalues of $\ms{L}^\dagger$ it is convenient to transform the operator $\ms{L}^\dagger$ in such a way that the unperturbed $\ms{L}_0^\dagger$ becomes Hermitian. This is done by applying the transformation
\algn{
	\ms{H}^\dagger_0 	&= -\ee^{-V_0(\ve x)/2}\ms{L}_0^\dagger\ee^{V_0(\ve x)/2}\,, \\
			&= -\Delta +\frac14[\grad V_0(\ve x)]^2 -  \frac{\tr\mbb{a}}2  =   \ve c^\dagger\! \mbb{a}\, \ve c\,, \eqnlab{Ham}
}
where $\ve c = (c_1,\ldots,c_d)^{\sf T}$ and $\ve c^\dagger = (c_1^\dagger,\ldots,c_d^\dagger)$ denote vectors of the lowering and raising (ladder) operators
\sbeqs{\eqnlab{ladders}
\algn{
 	c_m =& \frac1{\sqrt{2}}\left(\sqrt{\frac{a_m}2}x_m + \sqrt{\frac{2}{a_m}}\grad_{\!\! m}\right)\,,\\
 	c^\dagger_m =& \frac1{\sqrt{2}}\left(\sqrt{\frac{a_m}2}x_m - \sqrt{\frac{2}{a_m}}\grad_{\!\! m}\right)\,,
}
}
of the quantum harmonic oscillator. The ladder operators are normalised so that $[c_n,c^\dagger_m] = \delta_{nm}$.

In this picture, the (quantum) harmonic oscillator describes the dynamics of the particle in the unperturbed harmonic background. In particular, in the special case when the dimensionless stiffness matrix corresponds to the identity matrix, $\mbb{a} = \mbb{1}$, see \Eqnref{V0eqn}, the background is isotropic. Otherwise it is anisotropic. We can write $\ms{H}^\dagger_0$ in terms of its eigenbasis $\left\{|\ve m\rangle\right\}$ with eigenstates $|\ve m\rangle = |m_1,\ldots,m_d\rangle$, $m_i=0,1,\ldots$, as
\algn{\eqnlab{H0dagm}
	\ms{H}^\dagger_0 = \sum_{\{\ve m\}}\ve a\cdot \ve m\, |\ve m\rangle\langle \ve m|\,,
}
where the sum runs over all combinations of $\ve m$. Similarly to $\ms{H}_0^\dagger$, we transform the $\ve F$-dependent part $\ms{L}^\dagger_{F}$ of $\ms{L}^\dagger$ according to
\algn{
		\ms{H}^\dagger_{F} &= -\ee^{-V_0(\ve x)/2}\ms{L}_{F}^\dagger\ee^{V_0(\ve x)/2} = -\kappa^{-1}[\sqrt{\mbb{a}}\ve F\left(\ve x/\kappa\right)]\cdot\ve c\,,
}
and express $\ve F$ as
\algn{\eqnlab{fourierforce}
	\ve F\left(\ve x/\kappa\right) = \int_{\mbb{R}^d}\!\!\ed^d k\,\ee^{i\kappa^{-1}\ve k\cdot \ve x}\, \ve{\hat{F}}(\ve k)\,,
}
where $\ve{\hat{F}}$ denotes the Fourier transform of $\ve F$. Disorder correlations of the Fourier-transformed forces $\ve{\hat{F}}$ are obtained from \Eqsref{Fcorr} as
\begin{equation}\eqnlab{hfcorr}
	\langle \hat F_n(\ve k) \hat F_m(\ve k')\rangle_{F} = \hat C_F(\ve k)\delta(\ve k + \ve k') \ms{G}_{nm}(\ve k)\,,
\end{equation}
where
\algn{
		\ms{G}_{nm}(\ve k)= (1-\gamma)k_{n}k_{m} + \gamma( \delta_{nm} \ve k^2-k_nk_m)\,.
}

The expression~\eqnref{fourierforce} in terms of $\ve{\hat{F}}$ is convenient, because it allows us to express the $\ve x$ dependence in terms of the ladder operators $\ve c$ and $\ve c^\dagger$. This way, the eigenvalue equation~\eqnref{Leig} takes a representation-independent form and it enables us to use simpler operator algebra to compute the eigenvalues.

The position $x_m$ in \Eqnref{fourierforce} can be expressed in terms of ladder operators as
\algn{
	x_m = a_m^{-\frac12}\left(c_m + c_m^\dagger\right)\,,
}
so we write
\algn{\eqnlab{Dalpha}
	\ee^{i\kappa^{-1}\ve k\cdot \ve x}  = \ee^{i\kappa^{-1}\sqrt{\mbb{a}}^{-1}\ve k\cdot (\ve c + \ve c^\dagger)} = \ee^{\ve c^\dagger\ve z  -\ve z^\dagger \ve c} = D(\ve z)\,,
}
where $D$ denotes the so-called displacement operator~\cite{Gla63,Ger23} associated with the purely imaginary vector $\ve z = i \ve y$, where
\algn{\eqnlab{xydef}
	\ve y = \kappa^{-1}\mbb{a}^{-\frac12}\ve k\,.
}
It follows from its definition~\eqnref{Dalpha} that $D(\ve z)$ is unitary, $D(\ve z)^\dagger D(\ve z) = 1$ and that $D(\ve z)^\dagger = D(-\ve z)$. Upon acting on the ground state $|0\rangle$ of the harmonic oscillator, $D(\ve z)$ creates the coherent state $|\ve z)$~\cite{Gla63,Ger23}, i.e.,
\algn{\eqnlab{coh}
	D(\ve z) |0\rangle = |\ve z) = %
	\ee^{-\frac{|\ve z|^2}2}\sum_{\ve m=0}^\infty \frac{\ve z^{\ve m}}{\sqrt{\ve m!}}|\ve m\rangle\,,
}
where we have defined
\algn{
	\ve z^{\ve m} = \prod_{n=1}^d z_n^{m_n}\,,
}
and $\ve m! = \prod_{n=1}^d m_n!$. The coherent states $|\ve z)$ are eigenstates of the lowering operators
\algn{
	c_n|\ve z) = z_n|\ve z), \qquad 	(\ve z|c^\dagger_n = z^*_n(\ve z|\,.
}
And with respect to the raising operators $\ve c^\dagger$, we furthermore have
\algn{
	c^\dagger_n|\ve z) = \left(\frac{\partial}{\partial z_n}+\frac{ z_n^*}2\right)|\ve z)\,,\quad 	(\ve z| c_n = \left(\frac{\partial}{\partial z_n^*}+\frac{ z_n}2\right)(\ve z|\,.
}
Upon using \Eqnref{xydef} we can equivalently write the last equation in terms of $\ve y$ as
\algn{
	c^\dagger_n|i\ve y) = -i\left(\frac{\partial}{\partial y_n}+y_n\right)|i\ve y)\,,\quad 	(i\ve y| c_n = i\left(\frac{\partial}{\partial y_n}+y_n\right)(i\ve y|\,.}

We finally have the additional properties
\algn{
	D^\dagger(\ve z)c_n D(\ve z) = c_n +  z_n\,, \quad D(\ve z)c_n D^\dagger(\ve z) = c_n -  z_n\,,
}
from which the commutation relations between $D(\ve z)$ and $\ve c$ and $\ve c^\dagger$ follow. Using $D(\ve z)$, we express $\ms{H}^\dagger_{F}$ as
\algn{\eqnlab{HdagF}
	\ms{H}^\dagger_{F} &= -\kappa^{-1}\!\!\int_{\mbb{R}^d}\!\! \ed ^d\ve k\, [\sqrt{\mbb{a}}\hat{\ve{F}}(\ve k)]\cdot \left[D(\ve z)\ve c\right]\,,
}
without explicit dependence on $\ve x$. In terms of the complete operator
\algn{
	\ms{H}^\dagger &=  \ms{H}^\dagger_0 + \zeta_T \ms{H}^\dagger_{F}\\
	 &= \ve c^\dagger\! \mbb{a}\, \ve c -\frac{\zeta_T}{\kappa}\!\!\int_{\mbb{R}^d}\!\! \ed ^d\ve k\, [\sqrt{\mbb{a}}\hat{\ve{F}}(\ve k)]\cdot \left[D(\ve z)\ve c\right]\,,\eqnlab{Hdag}
}
the eigenvalue equation~\eqnref{Leig} takes the representation-independent form
\algn{\eqnlab{Heig}
	\ms{H}^\dagger |\psi_n\rangle &= \lambda_n|\psi_n\rangle\,,
}
where $|\psi_n\rangle$ corresponds to the right eigenket of $\ms{H}^\dagger$. In position representation, this eigenket is related to the left eigenfunction $l(\ve x)$ in \Eqnref{Leig} by $\langle \ve x |\psi_n\rangle = \ee^{-V_0(\ve x)/2}l_n(\ve x)$.

The transformation into the eigenvalue equation~\eqnref{Heig} with $\ms{H}^\dagger_0$ is convenient, because $\ms{H}_0^\dagger=\ms{H}_0$ is the (Hermitian) Hamiltonian of the quantum harmonic oscillator, whose eigenstates and spectrum are well known explicitly. In particular, the formulation~\eqnref{Hdag} in terms of $\ve c$ and $\ve c^\dagger$ facilitates perturbation theory in the weak-disorder limit $\zeta_T\ll1$.
\subsection{Asymptotic relaxation rate}\seclab{relspeed}
Before treating the eigenvalue problem~\eqnref{Heig} in the weak-disorder limit, we discuss how to characterise the relaxation speed in the long-time limit. The logarithmic distance from the stationary state is dominated by the eigenvalue with the smallest real part:
\algn{
	-\frac1t\log\left|\langle f(\ve x)\rangle-\langle f(\ve x)\rangle_{ss}\right|\sim \mu_1\,,
}
where we neglect logarithmic terms that arise from oscillatory contributions to \Eqnref{oscill} as they play a role only in time windows that are exponentially small in $t\gg 1$. Considering the logarithmic distance relative to the same quantity in the unperturbed system, we obtain
\algn{
	-\frac1t\log\frac{\left|\langle f[\ve x(t)]\rangle-\langle f(\ve x)\rangle_{ss}\right|}{\left|\langle f(\ve x)\rangle^{F=0}-\langle f(\ve x)\rangle^{F=0}_{ss}\right|}\sim \Delta\mu_1\,,
}
which is a measure of the real part of $\Delta\lambda_1=\lambda_1-\lambda_1^0 = \Delta\mu_1 + i\omega_1$. This quantity characterises whether the relaxation speed is larger $\Delta\mu_1>0$ or smaller $\Delta\mu_1<0$ than in the system without disordered forces ($\ve F=0$). For the specific choice $f[\ve x(t)] = (\ve x(t)-\langle\ve x\rangle_{ss})\cdot (\ve x(0)-\langle\ve x\rangle_{ss})$ introduced above, we have
\algn{
	-\frac1t\log\left|\frac{C_x(t)}{C_x^{F=0}(t)}\right|\sim \Delta\mu_1\,,
}
where $C_x^{F=0}(t)$ denotes the correlation function in the background potential. Throughout this paper, faster and slower relaxation refer to this asymptotic rate, that is, to $\Delta\mu_1>0$ and $\Delta\mu_1<0$, respectively. It governs the decay once all modes with larger real parts have died out. At earlier times, the decay depends also on the initial state and on the faster modes, and need not follow the sign of $\Delta\mu_1$. In isotropic and quasi-isotropic backgrounds, the slowest levels lie within $O(\zeta_T)$ of each other, so that the asymptotic decay emerges on time scales of order $\zeta_T^{-1}$.

As the quenched-random forces $\ve F$ vary across realisations, $\Delta \mu_1$ is a random variable. To make realisation-independent statements about the impact of $\ve F$ on the relaxation speed, we need to characterise the statistical properties of $\Delta\mu_1$. Our previous work~\cite{Mei25a} focused on the mean and variance of $\Delta \mu_1$ over disorder realisations in one spatial dimension.

To refine this analysis, we introduce the asymmetry $\Delta P$ between the \textit{probabilities} of predominantly faster or predominantly slower relaxation, given by
\algn{\eqnlab{DP}
	\Delta P = P(\Delta\mu_1>0)-\frac12 = \langle \theta (\Delta\mu_1)\rangle_F-\frac12\,,
}
where $\theta(x)$ denotes the Heaviside function. In the definition~\eqnref{DP} of $\Delta P$, we have subtracted $1/2$ from the probability $P(\Delta\mu_1>0)$ in order to obtain $\Delta P=0$ when positive and negative $\Delta\mu_1$ are equally likely. Consequently, positive $\Delta P>0$ corresponds to the case when it is more likely to observe faster relaxation in the long-time limit, and $\Delta P<0$ otherwise. The probability shift $\Delta P$ can be measured in experiments, e.g., by observing the frequencies of faster and slower relaxation over many realisations of the relaxation process.

Importantly, in the weak-disorder limit $\zeta_T\ll 1$, we may express $\Delta P$ in terms of the mean and variance of $\Delta\mu_1$ at short and long correlation lengths and in high dimensions, and in terms of these and one additional covariance in general, as we will show in \Secref{aniso}.
\subsection{Weak-disorder perturbation theory}\seclab{perturbation}
 \begin{figure*}
	\includegraphics{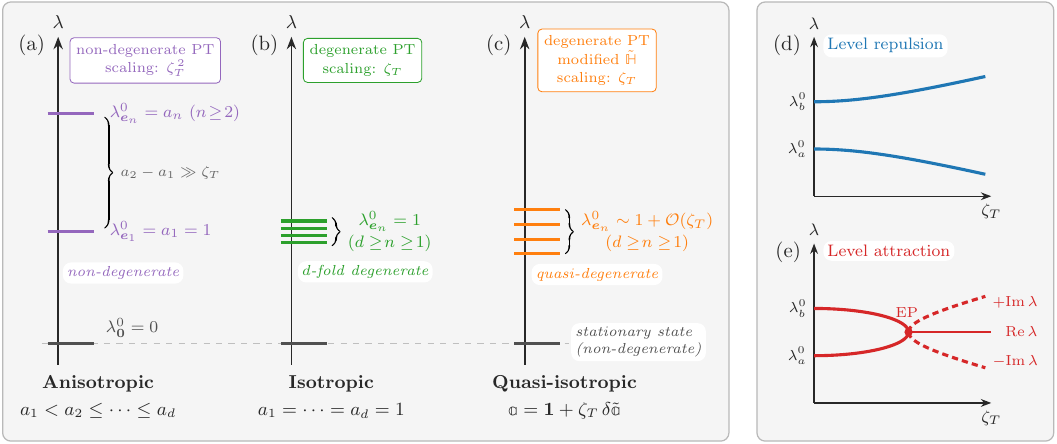}
	\caption{Schematic of the unperturbed spectrum of $\smash{\ms{L}_0^\dagger}$ in the three cases distinguished in the text. In all cases, the stationary state $\smash{\lambda^0_{\bm 0}=0}$ (dashed line) is non-degenerate and separated from the rest of the spectrum. (a) Anisotropic background, where the smallest non-zero eigenvalue $\smash{\lambda^0_{\ve{e}_1}}$ is non-degenerate and separated from the next level $\smash{\lambda^0_{\ve{e}_2}}$ by a gap $a_2-a_1\gg\zeta_T$. Non-degenerate perturbation theory applies and the mean eigenvalue shift is quadratic in $\zeta_T$. (b) Isotropic background, where the $d$ lowest non-vanishing levels $\smash{\lambda^0_{\bm{e}_n}}$ coincide exactly. Degenerate perturbation theory is required and the mean eigenvalue shift becomes linear in $\zeta_T$. (c) Quasi-isotropic background, where the $d$ lowest non-vanishing levels are split by an amount $\sim\zeta_T$. Degenerate perturbation theory applies but with an additional deterministic contribution from the stiffness differences. (d),(e) Level repulsion and attraction: Two levels $\smash{\lambda^0_a<\lambda^0_b}$ coupled by the disorder, whose distance $s$ obeys $\smash{s^2=\delta^2+4\zeta_T^2H_{ab}H_{ba}}$, see \Secref{perturbation}. (d) For symmetric coupling, $H_{ab}H_{ba}>0$, the levels repel. (e) For antisymmetric coupling, $H_{ab}H_{ba}<0$, their real parts attract until the levels merge at an exceptional point (EP). Beyond it they form a complex-conjugate pair with a common real part and imaginary parts $\pm\mathrm{Im}\,\lambda$ (dashed).}\figlab{spectrum}
\end{figure*}
In the limit of weak disorder, $\zeta_T \ll 1$, the term
$\zeta_T \ms{H}^\dagger_F$ in \eqnref{Hdag} can be treated as a
perturbation of the $d$-dimensional quantum harmonic
oscillator~\cite{Dav76}. The unperturbed operator $\ms{H}^\dagger_0$ has eigenvalues
$\lambda^0_{\bm{m}} = \bm{a} \cdot \bm{m}$, where
$\bm{m} = (m_1, \ldots, m_d)$ with $m_n = 0, 1, \ldots$ is the
occupation vector.  The ground state $\bm{m} = \bm{0}$ carries
eigenvalue $\lambda^0_{\bm{0}} = 0$ and corresponds, in the
Fokker--Planck picture, to the stationary distribution for a Brownian particle in the harmonic background $V_0(\ve x)$ with stiffness matrix $\mbb{a}$.  Since all forces in the unperturbed problem derive from
$V_0$, this stationary distribution is the Boltzmann equilibrium
distribution
\begin{align}
  P^0_\mathrm{eq}(\bm{x})
    = \frac{e^{-V_0(\bm{x})}}{Z^0_\mathrm{eq}}\,,
  \qquad
  Z^0_\mathrm{eq} = \sqrt{(2\pi)^d/\det(\mathbb{a})}\,.
\end{align}
Because the Fokker-Planck operator $\mc{L}$ always has a unique stationary state~\cite{Ris89}, we have $\lambda^0_{\bm{0}} = \lambda_{\bm{0}} = 0$, which remains unaffected by
any perturbation.

Disorder does, however, shift the non-vanishing eigenvalues that characterise the relaxation dynamics, governed by states with
$\bm{m} \neq \bm{0}$.  The slowest relaxing modes correspond to the
smallest non-vanishing eigenvalue, which in the unperturbed system is
$\lambda^0_1 = a_1 = 1$, associated with excitations along the
direction of least stiffness, $\bm{m} = \bm{e}_1$.  The key
question is whether $\lambda^0_1$ is degenerate, which
depends entirely on the anisotropy of the background stiffnesses
$\mbb{a} = \text{diag}(a_1, \ldots, a_d)$. For this, we distinguish the three cases mentioned in \Secref{intro} and illustrated schematically in \Figref{spectrum}(a)--(c):

\begin{itemize}
  \item \emph{Anisotropic background}
    ($a_1 < a_2 \leq \cdots \leq a_d$, with $a_2 - a_1 \gg \zeta_T$):
    The eigenvalue $\lambda^0_1 = 1$ is non-degenerate, and the gap
    to the next level is large compared to the disorder strength $\zeta_T$,
    as shown in \Figref{spectrum}(a).

  \item \emph{Isotropic background}
    ($a_1 = a_2 = \cdots = a_d = 1$):
    The eigenvalue $\lambda^0_1 = 1$ is $d$-fold degenerate, since
    $\lambda^0_{\bm{e}_n} = 1$ for all $n = 1, \ldots, d$, as shown in
    \Figref{spectrum}(b).

  \item \emph{Quasi-isotropic background}
    ($\mbb{a} = \mbb{1} + \zeta_T \delta\tilde{\mbb{a}}$):
    the stiffness differences are of the same order as the disorder
    strength, so the eigenvalue $\lambda^0_1 = 1$ is effectively
    $d$-fold degenerate, and we include $\zeta_T \delta\tilde{\mbb{a}}$ into the perturbation; see \Figref{spectrum}(c).
\end{itemize}

The degeneracy structure above determines which form of perturbation
theory applies. We now briefly discuss these cases.

\paragraph{Anisotropic case.}
Standard non-degenerate perturbation theory applies to $\lambda_1$.
The eigenvalue shift $\Delta\lambda_1 = \lambda_1 - \lambda^0_1$ to
second order is~\cite{Dav76}
\begin{align}\eqnlab{ev1}
  \Delta\lambda_1 \sim
    \zeta_T \langle \bm{e}_1 | \ms{H}^\dagger_F | \bm{e}_1 \rangle
    - \zeta_T^2 \sum_{\bm{m} \neq \bm{e}_1}
      \frac{\langle \bm{e}_1 | \ms{H}^\dagger_F | \bm{m} \rangle
            \langle \bm{m} | \ms{H}^\dagger_F | \bm{e}_1 \rangle}
           {\bm{a} \cdot \bm{m} - a_1}\,.
\end{align}
In \Secref{aniso} we analyse this case in detail, showing that the first-order term vanishes on average, so the
leading contribution to $\langle \Delta\lambda_1 \rangle_F$ is
quadratic in $\zeta_T$.

\paragraph{Isotropic case.}
The $d$-fold degeneracy of $\lambda^0_1$ requires degenerate
perturbation theory.  To first order in $\zeta_T$, the eigenvalue shifts $\Delta \lambda_n$ of the operator $\ms{H}^\dagger$ are given by $\zeta_T$ times the eigenvalues of
the $d \times d$ perturbation matrix~\cite{Dav76}
\begin{align}
  H_{nm} = \langle \bm{e}_n | \ms{H}^\dagger_F | \bm{e}_m \rangle\,.
\end{align}
In \Secref{iso}, we characterise $\mbb{H}$ as a Gaussian random matrix
with non-standard correlation of the entries, where $\Delta\lambda_1$ is given by $\zeta_T$ times the eigenvalue of $\mbb{H}$ with the
smallest real part. One of the consequences of this is that the mean eigenvalue shift $\langle\Delta\lambda_1\rangle_F$ is linear in $\zeta_T$ for isotropic backgrounds.

\paragraph{Quasi-isotropic case.}
The same degenerate scheme applies, but the perturbation matrix
acquires an additional deterministic diagonal contribution from the
small anisotropy.  Writing $\ms{H}^\dagger_0$ as the isotropic
part plus an $O(\zeta_T)$ correction, the latter enters the
perturbation at the same order as $\ms{H}^\dagger_F$, giving
the modified perturbation matrix
\begin{align}
  \delta\tilde{\mbb{a}} + \mbb{H}\,.
\end{align}
The diagonal term partially lifts the degeneracy and, as we show in \Secref{qiso},
opens additional parameter regimes in which relaxation can be faster
than in the unperturbed system, with a mean shift $\langle\Delta\lambda_1\rangle_F$ that is linear in $\zeta_T$.

In all three cases, the shift of the stationary eigenvalue
$\lambda_0 = 0$ vanishes to all orders, as required for consistency with probability
conservation.

\paragraph{\texorpdfstring{Level repulsion and attraction.}{Level repulsion and attraction.}}
What distinguishes the three cases physically is how strongly the disorder couples $\lambda_1$ to the neighbouring levels. The basic mechanism can be explained in terms of two closeby levels $\lambda^0_a<\lambda^0_b$~\cite{Hei12}, that the disorder couples through the off-diagonal elements $H_{ab}$ and $H_{ba}$, defined precisely in \Secref{iso}. Their distance $s=\lambda_b-\lambda_a$ obeys $s^2=\delta^2+4\zeta_T^2H_{ab}H_{ba}$, where $\delta^2>0$ is always positive, containing the unperturbed distance and diagonal couplings. Consequently, the sign of $H_{ab}H_{ba}$ decides between two opposite effects, sketched in \Figsref{spectrum}(d) and~(e). For symmetric coupling, $H_{ab}H_{ba}>0$, the levels repel. For antisymmetric coupling, $H_{ab}H_{ba}<0$, the real parts of the levels attract, and once $4\zeta_T^2|H_{ab}H_{ba}|$ exceeds $\delta^2$, the two levels merge at an exceptional point~\cite{Kat66,Hei12} and continue as a complex-conjugate pair with a common real part. This phenomenon, also called  ``level attraction'' is a property of non-Hermitian operators~\cite{Ash20,Yu24}. We show in \Secref{iso} that for the random forces considered here, $\langle H_{ab}H_{ba}\rangle_F\propto1-2\gamma$, so that predominantly conservative forces make the levels repel and predominantly non-conservative forces make them attract. In the isotropic and quasi-isotropic cases, $\delta\lesssim\zeta_T$, so that repulsion and attraction play a significant role in shaping the relaxation speed. In the anisotropic case of \Secref{aniso}, by contrast, the levels are too distant to merge, i.e., $\delta\gg\zeta_T$, and they interact only at second order. Their attraction then gives a small speedup in addition to the stiffening of the background known from one dimension, see \Secref{large}.
\subsection{Numerical simulations}\seclab{numsim}
All analytical results established below are perturbative in the disorder strength $\zeta_T$. Beyond the perturbative regime $\zeta_T\ll 1$, we need to resort to numerical methods. To this end, we solve the eigenvalue problem for the full operator $\ms{H}^\dagger$ in \Eqnref{Hdag} numerically, for individual realisations of the quenched-random force field $\ve F$. In order to do this, we discretise the operator on two- and three-dimensional periodic grids of spacing $a$ and box size $L$. The number of grid points along each direction follows as $N=(L/a)$ with $N_\text{tot} = N^d$ as the total number of points. We use a pseudospectral method to represent the operator $\ms{H}^\dagger$ as a matrix whose eigenvalues we calculate numerically using an iterative Krylov method. Correlated random disorder with Gaussian statistics is generated using random Fourier sums involving fast-Fourier transforms. The numerical method and the parameters used for each figure are described in more detail in Appendix~\secref{numerics}.
\section{Main results}\seclab{main}
\begin{figure*}
	\centering
	\includegraphics{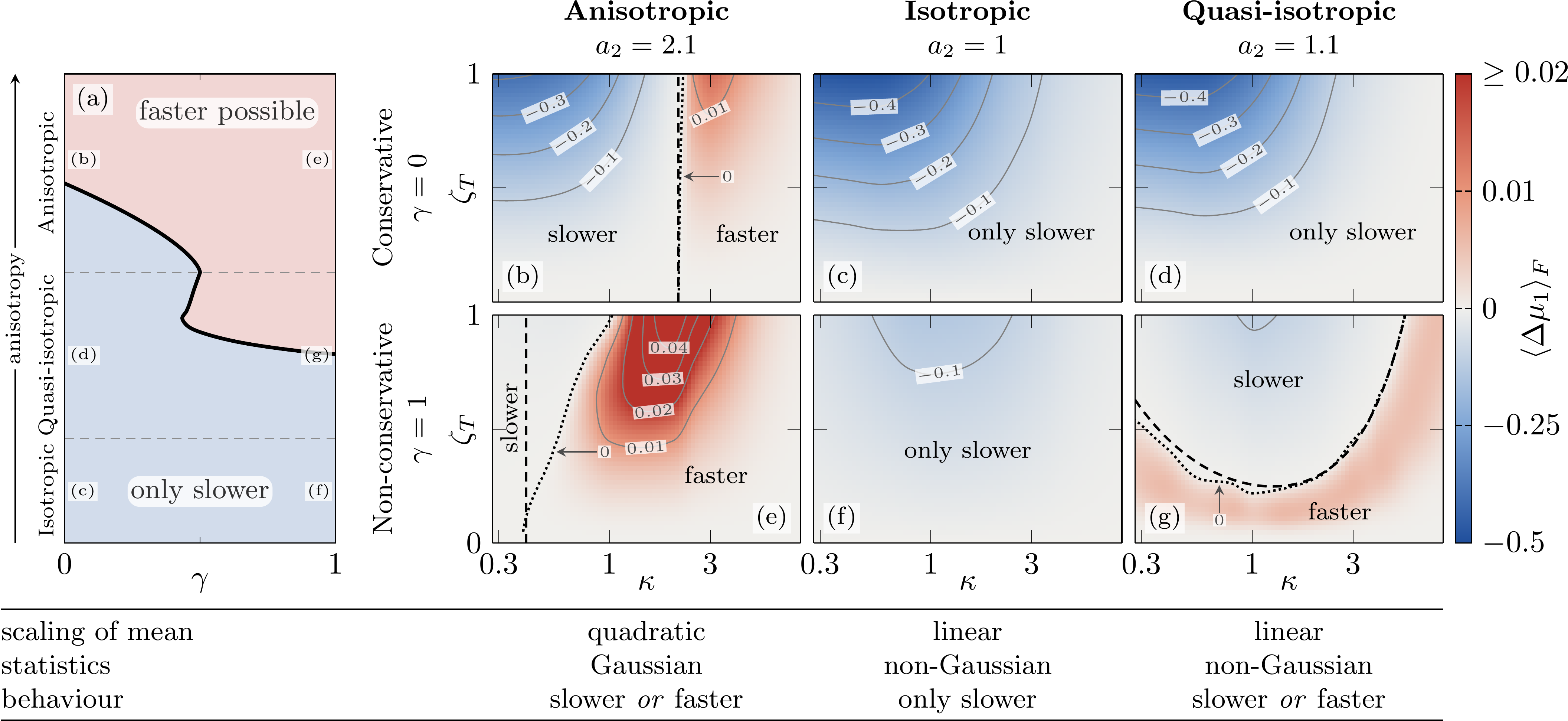}
	\caption{Summary of the main results. (a) Schematic of the regions in which disorder can lead to faster relaxation (red) or only to slower relaxation (blue), as a function of the fraction $\gamma$ of non-conservative forces and of the anisotropy $a_2-1$ of the background compared with the disorder, $\zeta_T\Sigma(\kappa)d$, for $d=2$. The letters mark the parameters of panels (b)--(g). (b)--(g) Mean shift $\langle\Delta\mu_1\rangle_F$ of the asymptotic relaxation rate in $d=2$ as a function of the correlation length $\kappa$ and the disorder strength $\zeta_T$, for conservative forces, $\gamma=0$ [(b)--(d)], and non-conservative forces, $\gamma=1$ [(e)--(g)], in an anisotropic background, $a_2=2.1$ [(b),(e)], an isotropic one, $a_2=1$ [(c),(f)], and a quasi-isotropic one, $a_2=1.1$ [(d),(g)]. The colours show direct numerical solutions of \Eqnref{Leig}, computed with the method of \Secref{numsim} for $10^3$ to $9\times10^4$ pairs of realisations $\pm\ve F$ per parameter set, see Appendix~\secref{numerics}. Blue indicates slower relaxation, $\langle\Delta\mu_1\rangle_F<0$, and red faster relaxation, $\langle\Delta\mu_1\rangle_F>0$. Since the speedups are smaller than the slowdowns, the colour scale is $25$ times finer for $\langle\Delta\mu_1\rangle_F>0$ and saturates at $0.02$. The thin grey lines are level lines, spaced by $0.1$ for $\langle\Delta\mu_1\rangle_F<0$ and by $0.01$ for $\langle\Delta\mu_1\rangle_F>0$. The dotted lines, marked $0$, show the simulated boundary $\langle\Delta\mu_1\rangle_F=0$ for $\zeta_T\geq0.05$, the smallest simulated disorder strength. The dashed lines show the weak-disorder predictions of this boundary, the zero $\kappa^*$ of the second-order mean~\eqnref{mu1} in (b) and (e), and $\delta\hat a=1.05$, the zero of the random-matrix mean of \Figref{qisoscal}(a) for $\gamma=1$, in (g). In the other panels, the theory predicts no change of sign. The table at the bottom lists the scaling of the mean eigenvalue shift with the disorder strength, the statistics of the shift, and the generic effect of the disorder on the mean asymptotic relaxation rate in each background.}
	\figlab{statemap}
\end{figure*}
Before we proceed, we summarise the main results of this paper. In \Secsref{aniso}--\secref{qiso} we provide the details of our analysis; see \Figref{flowdiagram} for the structure of the paper.

As discussed previously, the results strongly depend on the properties of the harmonic background. Figure~\figref{statemap}(a) shows schematically where disorder can speed relaxation up, as a function of the fraction $\gamma$ of non-conservative forces and of the anisotropy of the background compared with the disorder. Figures~\figref{statemap}(b)--(g) give an overview of the results in $d=2$, based on numerical solutions of the eigenvalue problem (\Secref{numsim}). They show the mean shift $\langle\Delta\mu_1\rangle_F$ of the asymptotic relaxation rate as a function of the correlation length $\kappa$ and of the disorder strength $\zeta_T$, up to $\zeta_T=1$, for conservative ($\gamma=0$) and non-conservative forces ($\gamma=1$). The boundaries between slower and faster relaxation were derived for weak disorder, but they describe the simulations up to $\zeta_T=1$, except in \Figref{statemap}(e). The table at the bottom of \Figref{statemap} lists the characteristics of the three background types. We summarise the result of each type of background in turn.

\subsection{Anisotropic backgrounds}
	\begin{enumerate}
		\item For weak disorder, $\Delta\lambda_1$ is real for each realisation. The levels shown in \Figref{spectrum}(a) can only become complex by forming complex-conjugate pairs, which requires interactions between the levels and $\gamma>0$. Since $a_2-a_1\gg \zeta_T$ here, the levels interact only at second order and are too distant to merge.
		\item We show that the statistics of $\Delta\lambda_1$ is Gaussian for typical values of order $\zeta_T$, so the mean and variance characterise the asymptotic relaxation rate completely, up to a covariance that enters the asymmetry $\Delta P$ between faster and slower relaxation at intermediate correlation lengths. We then derive perturbative expressions~\eqnref{mu1} and \eqnref{sigmain} for the mean and variance of the eigenvalue shift $\Delta\lambda_1$. The disorder-averaged shift is quadratic in the dimensionless disorder strength $\zeta_T$ while typical fluctuations are linear in $\zeta_T$, suggesting that disorder-induced fluctuations dominate. These results hold in arbitrary dimension $d$.

		\item The limiting behaviours of \Eqsref{mu1} and \eqnref{sigmain} of small and large correlation lengths $\kappa$, given in \Eqsref{mushort} and~\eqnref{sigshort}, and \Eqsref{mulong} and~\eqnref{siglong}, respectively,
		reveal that disordered forces give rise to distinct regions in parameter space, in which relaxation is either slower \emph{or} faster
		than in the unperturbed background.  Using these limiting behaviours, we construct a
		diagram [shown schematically in the upper part of \Figref{statemap}(a)] delineating the regions where only slower
		relaxation is possible from those where faster relaxation can also occur.
		Crucially, this diagram is independent of the functional form of the correlation
		function. In \Figsref{statemap}(b) and~(e), for $a_2=2.1$, relaxation is faster on average beyond the correlation length $\kappa^*$ at which the second-order mean~\eqnref{mu1} changes sign. Non-conservative forces lower this threshold from $\kappa^*\approx2.1$ for $\gamma=0$ to $\kappa^*\approx0.4$ for $\gamma=1$. For $\gamma=1$, the region of faster relaxation recedes from $\kappa^*\approx0.4$ at weak disorder to $\kappa\approx1$ at $\zeta_T=1$.
		\item The effect of non-conservative forces depends on the dimension and on the background. For $d<4$, increasing $\gamma$ either increases or decreases the relaxation speed, depending on the background. For $d\geq d_\text{th}=4$, increasing $\gamma$ always increases it. This threshold dimension refers to a fixed weight of each solenoidal polarisation direction. When the total force variance is kept fixed instead, the effect depends on the background in every dimension.
	\end{enumerate}
\subsection{Isotropic backgrounds}
	\begin{enumerate}
		\item In isotropic backgrounds, the unperturbed eigenvalue $\lambda_1^0$ is $d$-fold degenerate. The disorder lifts this degeneracy, and for non-conservative forces ($\gamma>0$) the split eigenvalues can interact and form complex-conjugate pairs. The eigenvalue(s) $\Delta\lambda_1$ with the smallest real part $\Delta\mu_1$ determine the asymptotic relaxation rate.
		\item Using degenerate perturbation theory, we find that the shifts $\Delta\lambda_1$ are determined by the
		eigenvalues of a random perturbation matrix $\mbb{H}$, given in \Eqnref{Hnmdef}. Although the entries of $\mbb{H}$ are Gaussian, the statistics of the eigenvalues $\Delta\lambda$ are strongly non-Gaussian.  The smallest of these eigenvalues has typically a negative real part $\Delta\mu_1<0$, implying that relaxation is \emph{generically slowed down} in disordered isotropic backgrounds [lower part of \Figref{statemap}(a)]. In \Figsref{statemap}(c) and~(f), relaxation is slower on average for all $\kappa$ and all $\zeta_T\leq1$. The mean shift now scales \emph{linearly} with the disorder strength, and so does its standard deviation, in contrast to the quadratic scaling of the mean in the anisotropic case.

  		\item The perturbation matrix $\mbb{H}$ belongs to a Gaussian random matrix ensemble with
		correlated entries.  Upon a suitable rescaling, the rescaled eigenvalue shifts
		depend only on the dimension $d$ and on the non-conservative character of the forces.  For conservative forces, the matrices are drawn from the Wigner
		ensemble~\cite{Meh04} of symmetric matrices, supplemented by an additional random trace. For equal proportions of conservative and non-conservative forces, the perturbation matrix is drawn from the Ginibre ensemble.

		\item In high dimensions $d\gg 1$ the perturbation matrix turns into either essentially a large Wigner matrix~\cite{Wig55} (for strictly conservative forces) or a large Ginibre matrix~\cite{Meh04} (whenever the forces have a finite non-conservative contribution). In this limit, fluctuations of $\Delta\mu_1$ are suppressed and it approaches a \emph{deterministic value}, set by the edge of the limiting spectrum. As a consequence, for forces with non-conservative contribution, the disorder-induced slowdown saturates at a finite value proportional to $\sqrt{\gamma}$, whereas for strictly conservative forces it vanishes as $d^{-1/2}$. The saturation reflects the growth of the solenoidal force variance with $d$ at fixed $\gamma$. At fixed total force variance, the slowdown vanishes as $d^{-1/2}$ for any mixture of conservative and non-conservative forces.
	\end{enumerate}

\subsection{Quasi-isotropic backgrounds}
	\begin{enumerate}
		\item Although the unperturbed levels are non-degenerate, they are close enough to interact in the presence of disordered forces. For $\gamma>0$, they can form complex-conjugate pairs.
		\item The eigenvalue shifts $\Delta\lambda$ are determined by a random perturbation matrix $\mbb{H}$, which, however, acquires a deterministic diagonal component. Consequently, the mean and standard deviation of the eigenvalue shifts scale linearly in $\zeta_T$ and the statistics is strongly non-Gaussian.
		\item Although the linear-in-$\zeta_T$ scaling of the isotropic case is retained, the
		additional structure introduced by the anisotropy generates new scenarios in
		which relaxation can be \emph{faster}. The rescaled mean shift collapses onto a single curve as a function of the rescaled stiffness offsets~\eqnref{dahat}, which becomes positive, i.e., relaxation becomes faster on average, only for $\gamma\gtrsim0.44$ in $d=2$ and $d=3$ [middle part of \Figref{statemap}(a)]. In \Figsref{statemap}(d) and~(g), the background with $a_2=1.1$ is quasi-isotropic at larger $\zeta_T$, where $\delta\hat a$ is of order unity, and becomes anisotropic as $\zeta_T$ decreases. For $\gamma=0$, relaxation is slower throughout. For $\gamma=1$, faster relaxation at small $\zeta_T$ gives way to slower relaxation at $\delta\hat a\approx1$, where the rescaled mean changes sign. In high dimensions, the speedup reduces to a correction of order $1/d$.
	\end{enumerate}
\subsection{\texorpdfstring{Two mechanisms of faster relaxation}{Two mechanisms of faster relaxation}}
Across the three backgrounds, faster relaxation arises from two distinct mechanisms. The first is a stiffening of the background on average, the mechanism of the one-dimensional case~\cite{Mei25a}. It occurs at large correlation lengths in anisotropic backgrounds, where the disorder acts as an additional linear force around the displaced minimum of the background. This mechanism does not require non-conservative forces. The second mechanism is level attraction, which has no counterpart in one dimension. Non-conservative forces make the coupling between the slowest mode and its neighbours antisymmetric on average, so that it pulls the slowest level towards the faster ones, see \Secref{perturbation}. In anisotropic backgrounds this attraction is a small correction of second order, the counterpart of the acceleration of Gaussian diffusions by non-reversible drifts~\cite{Hwa93}. In quasi-isotropic backgrounds it acts at first order and merges the slowest mode with its neighbour into an oscillating complex-conjugate pair, whose decay rate lies between the two unperturbed rates. It is therefore most effective at intermediate anisotropy, where the levels are close enough to merge in many realisations, yet far enough apart so that the relaxation rate is raised substantially.
\section{Anisotropic background}\seclab{aniso}
\begin{figure}
	\includegraphics[width=\linewidth]{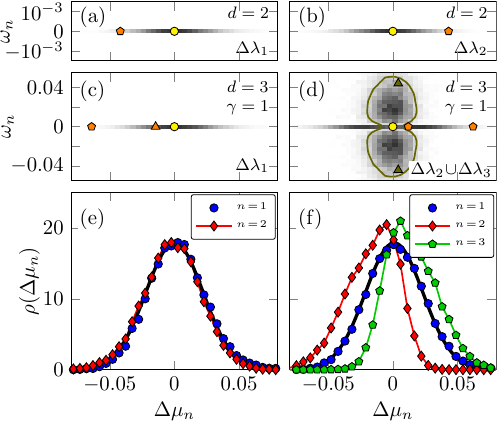}
	\caption{Numerically obtained shifts $\Delta\lambda_n = \lambda_n-\lambda^0_n$ of the $d$ lowest non-vanishing eigenvalues of the Fokker-Planck operator~\eqnref{FPop} in an \textit{anisotropic} background, computed with the method of \Secref{numsim}, using purely solenoidal forces, $\gamma=1$, correlation length $\kappa=1$, and disorder strength $\zeta_T=10^{-1}$. The stiffnesses are $\mbb{a}=\text{diag}(1,5/4)$ for $d=2$ and $\mbb{a}=\text{diag}(1,5/4,5/4)$ for $d=3$. The simulation box is of size $L=20$ with $N=100$ ($d=2$) and $N=40$ ($d=3$) grid points per dimension, and $2.5\times10^4$ disorder realisations. The yellow bullet marks the unperturbed eigenvalue at $\Delta\lambda=0$, onto which each density collapses as $\zeta_T\to0$. Coloured markers show the shifts in single realisations. Their shape indicates the type of the realisation, which is set by its number of real eigenvalues: pentagons for a realisation in which all shifts are real, triangles for one in which $\lambda_2$ and $\lambda_3$ form a complex-conjugate pair. Real shifts are drawn in orange, the members of a pair in olive. (a) Density of $\Delta\lambda_1$ in the complex plane for $d=2$. The density of shifts is shown as a grey scale normalised to its maximum. Since the shifts are exactly real, the density is confined to the line $\mathrm{Im}(\Delta\lambda)=0$. (b) Same as (a), but for $\Delta\lambda_2$. (c) Same as (a), but for $d=3$ and the non-degenerate lowest level, $\Delta\lambda_1$. (d) Same as (c), but for the two-fold degenerate pair $\Delta\lambda_2$ and $\Delta\lambda_3$, whose samples are combined into a single density. In contrast to (a)--(c), the vertical axis is now on the same scale as the horizontal one, and a fraction of the shifts build complex-conjugate pairs and leave the real axis. The olive line is the contour of these pairs at a quarter of their maximum. (e) Densities $\rho(\Delta\mu_n)$ of the real parts for $d=2$. The thick black line shows the result of the perturbation theory developed in \Secref{anisoPT}. (f) Same as (e), but for $d=3$.}
	\figlab{numsimaniso}
\end{figure}
Before developing a theory for the eigenvalue shifts $\Delta\lambda_n$ in an anisotropic background, we illustrate the phenomenology with direct numerical solutions of~\Eqnref{Leig} for a large number ($\sim10^4$) of disorder realisations. When the background is anisotropic, the $d$ lowest non-vanishing levels of the unperturbed operator $\ms{L}_0^\dagger$ are separated even in the absence of disorder~[\Figref{spectrum}(a)]. Specifically, the anisotropy of the background splits $\lambda_1^0$ from the rest of the spectrum already at $\zeta_T=0$, with unperturbed levels at $\lambda^0_n = a_n$ with $1=a_1<a_2\leq\ldots\leq a_d$. The disorder therefore shifts each non-degenerate level individually.

Figure~\figref{numsimaniso} shows the resulting distributions of the shifts $\Delta\lambda_n = \lambda_n-\lambda^0_n$, for $\mbb{a}=\text{diag}(1,5/4)$ in $d=2$ [\Figsref{numsimaniso}(a), (b) and~(e)] and $\mbb{a}=\text{diag}(1,5/4,5/4)$ in $d=3$ [\Figsref{numsimaniso}(c), (d) and~(f)]. Panels (a)--(c) show that whenever the levels are non-degenerate, their shift is \textit{real}, i.e., for all measured realisations, the imaginary part vanishes identically, and the density forms a line on $\text{Im}(\Delta\lambda)=0$. The reason is that because the discretised operator $\ms{L}^\dagger$ is real, a real eigenvalue can acquire an imaginary part only for $\gamma>0$ and only by direct interaction with another real eigenvalue. For an anisotropic background, $\lambda_1$ is held apart from $\lambda_2$ by the background, $a_2-a_1 = 1/4$, while the disorder-induced shifts are of order $10^{-2}$ [see~\Figref{numsimaniso}(e), (f)], so no collision occurs at this disorder strength. This corresponds to the conditions $|a_2-a_1|\gg\zeta_T$ under which non-degenerate perturbation theory may be applied.

The exception is a \textit{residual} degeneracy among the stiffer directions $a_2=a_3$ for $d=3$, which the anisotropic case does not exclude. In \Figref{numsimaniso}(d), the two upper levels of the three-dimensional background coincide, $a_2=a_3=5/4$. For this subspace, degenerate perturbation theory applies and the disorder splits the degenerate pair $\lambda_2^0=\lambda_3^0$. For a finite fraction of realisations, the two shifts leave the real axis, forming a complex-conjugate pair. The markers in \Figsref{numsimaniso}(c) and~(d) show one realisation of each type. In the one marked by pentagons all three shifts are real, whereas in the one marked by triangles $\lambda_2$ and $\lambda_3$ form a complex-conjugate pair. In both, $\lambda_1$ is real. The combined density of $\Delta\lambda_2$ and $\Delta\lambda_3$ in \Figref{numsimaniso}(d) accordingly consists of a line of real shifts together with a two-dimensional cloud, whereas the non-degenerate level in \Figref{numsimaniso}(c) shows only the line. The smallest non-vanishing eigenvalue $\lambda_1$, which governs the asymptotic relaxation speed, belongs to the non-degenerate direction $a_1=1$ and is therefore unaffected by this. We discuss in detail how degeneracies affect the spectrum when we discuss the isotropic background in \Secref{iso} below.

Projecting the shifts onto the real axis gives \Figsref{numsimaniso}(e) and~(f). In $d=2$ the densities $\rho$ of the shifts $\Delta\mu_1$ and $\Delta\mu_2$ overlap almost completely, because each is centred close to its own unperturbed level. In particular, their means are more than an order of magnitude smaller than their common width $\sim10^{-2}$. This asymmetry between mean and standard deviation is a central feature of the anisotropic case, as we show in the perturbative expansion below. The theory shows that within perturbation theory, the fluctuations of $\Delta\lambda_1$ are of first order in $\zeta_T$, whereas its mean is of second order. In $d=3$ the same holds for the non-degenerate branch $n=1$, while the degenerate pair $n=2,3$ is split symmetrically about its common mean by an amount \emph{comparable} to its width, reflecting the fact that degenerate perturbation theory splits a level already at first order in $\zeta_T$ [detailed discussion in \Secref{iso}].

\subsection{Perturbation theory}\seclab{anisoPT}
For anisotropic backgrounds, $\lambda^0_1$ is non-degenerate, because $1=a_1<a_2$ such that $|a_2-a_1|\gg\zeta_T$. In this case, we apply non-degenerate perturbation theory, as in \Eqnref{ev1}, to $\lambda_1$.

First, however, we aim to obtain a simplified expression for $\Delta P$ in \Eqnref{DP} in the limit $\zeta_T\ll1$. To this end we write \Eqnref{ev1} schematically as $\Delta\lambda_1 \sim \zeta_T X_1 + \zeta_T^2 X_2$, where $X_1$ and $X_2$ are random variables that depend on the disorder realisation. Due to the Gaussian disorder statistics, $X_1$ is Gaussian with zero mean, $\langle X_1\rangle_F=0$, and variance $\sigma^2_{X_1}$. In Appendix~\secref{gaussian}, we expand the distribution function of $\Delta\lambda_1$ to first order in $\zeta_T$, which gives
\algn{\eqnlab{DP2}
	\Delta P\sim\frac{\zeta_T\langle X_2|X_1=0\rangle_F}{\sqrt{2\pi}\sigma_{X_1}} =\frac{\zeta_T}{\sqrt{2\pi}\sigma_{X_1}}\left(\mu_{\Delta\lambda_1}+\frac{S}{\sigma_{\Delta\lambda_1}^2}\right)\,,
}
where $\langle X_2|X_1=0\rangle_F$ denotes the mean of $X_2$ over the realisations with $X_1=0$, and $S$ is a correction that arises because $X_1$ and $X_2$ are uncorrelated but not independent. If $X_1$ and $X_2$ were independent, $\langle X_2|X_1=0\rangle_F$ would equal the unconditional mean $\langle X_2\rangle_F$, and $S=0$. It turns out that $S$ is indeed negligible ($S/\sigma^2_{\Delta\lambda_1}\ll \mu_{\Delta\lambda_1}$) in the limits $\kappa\to\infty$ and, for $\gamma<1$, $\kappa\to0$, as well as for $d\gg1$, which we analyse below, but it changes $\Delta P$ noticeably at intermediate $\kappa$. In these limits, the probability shift $\Delta P$ can thus be expressed in terms of the lowest moments, the mean and variance, of $\Delta\lambda_1$, and \Eqnref{DP2} reduces to
\algn{\eqnlab{probshift}
	\Delta P \sim \frac{1}{\sqrt{2\pi}}\frac{\langle \Delta\lambda_1\rangle_F}{\sqrt{\langle\Delta\lambda_1^2\rangle_F-\langle \Delta\lambda_1\rangle_F^2}}\,.
}
This gives additional motivation for us to focus on these quantities in particular. Appendix~\secref{gaussian} also shows that the distribution of $\Delta\lambda_1$ is near-Gaussian for $\Delta\lambda\sim\zeta_T$, with a first correction that involves $S$. This is confirmed by the numerical simulations in \Figref{numsimaniso}. We discuss this aspect in more detail below.

We now compute the mean and variance perturbatively. To proceed, it is convenient to express $(\ve a\cdot\ve m -a_1)^{-1}$ in \Eqnref{ev1} as
\algn{\eqnlab{amint}
	\frac1{\ve a \cdot \ve m -a_1} = \int_0^1\!\!\ed s \, s^{\ve a \cdot \ve m -2}\,,
}
so that
\begin{multline}\eqnlab{ev2}
	\Delta\lambda_1 \sim \zeta_T \langle \ve e_1|\ms{H}_{F}^\dagger|\ve e_1\rangle\\ - \zeta_T^2 \int_0^1\!\!\ed s \, s^{-2}\Big(\sum_{\ve m}s^{\ve a \cdot \ve m}\langle\ve e_1|\ms{H}_{F}^\dagger|\ve m\rangle\langle \ve m|\ms{H}_{F}^\dagger|\ve e_1\rangle\\
	-s^{a_1}\langle\ve e_1|\ms{H}^\dagger_{F}|\ve e_1\rangle^2\Big)\,.
\end{multline}
Note that we have removed the constraint in the sum by subtracting the corresponding $\ve m = \ve e_1$ term in the last line.

From \Eqnref{ev2} we find that we need to evaluate two types of matrix elements, $\langle\ve e_1|\ms{H}_{F}^\dagger|\ve m\rangle$ and $\langle \ve m|\ms{H}_{F}^\dagger|\ve e_1\rangle$. Once these elements are computed, we will perform the sum over $\ve m$ and finally the integral over $s$ to obtain an expression for $\lambda_1$. The form~\eqnref{HdagF} of $\ms{H}^\dagger_{F}$ in turn suggests that we must evaluate expressions of the form $\langle\ve e_1|D(\ve z)\ve c|\ve m\rangle$ and $\langle\ve m|D(\ve z)\ve c|\ve e_1\rangle$. Using the properties of $\ve c$ and $D(\ve z)$ summarised in \Secref{fpe} we find
\algn{\eqnlab{element1}
	\langle \ve m|D(\ve z)\ve c|\ve e_1\rangle =& \ve e_1 \langle \ve m|i\ve y)\,,
}
where $\ve y$ is given in \Eqnref{xydef}, and
\begin{equation}\eqnlab{element2}
	\langle \ve e_1|D(\ve z')\ve c|\ve m\rangle = -(\ve e_1\cdot \ve{\nabla}_{\ve y'})(\ve{\nabla}_{\ve{y'}}+\ve y') \langle \ve m|i\ve y')\,.
\end{equation}
The elements $\langle m|i\ve y^({'}{}^))$ (with or without the prime) are now evaluated using \Eqnref{coh}, to give
\algn{
	 \langle \ve m|i\ve{y})=\ee^{-\frac{\ve y^2}2}\frac{(i\ve y)^{\ve m}}{\sqrt{\ve m!}}\,.
}
The expressions in \Eqsref{element1} and \eqnref{element2} depend on $\ve m$ only through $\langle \ve m|i\ve y^({'}{}^))$, which implies that the sum over $\ve m$ in \Eqnref{ev2} can be evaluated as
\algn{
	\sum_{\ve m}s^{\ve a\cdot\ve m}\langle \ve m|i\ve y)\langle \ve m|i\ve y') &= \ee^{-\frac{\ve y^2+{\ve y'}^2}2}\sum_{\ve m}\frac{(-{s^{\ve a}})^{\ve{m}}\ve y^{\ve m}{\ve y'}^{\ve m}}{\ve m!}\nn\\
	&= \ee^{-\frac12\left(\ve y^2+{\ve y'}^2+2\ve y^{(s)}\cdot\ve y'\right)}\,,
}
where we defined the vector $\ve y^{(s)}$ with elements
\algn{
	y^{(s)}_n = s^{a_n} y_n\,.
}
For $\ve m =  \ve e_1$ and $\ve z=\ve z'$, the expressions in \Eqsref{element1} and \eqnref{element2} agree to give
\algn{
	 \langle \ve e_1|D(i\ve y)\ve c|\ve e_1\rangle = i\ve e_1(\ve e_1\cdot \ve y)\ee^{-\frac{\ve y^2}2}\,.
}
Combining these results, we arrive at the following expression for $\Delta\lambda_1 = \lambda_1-1$
\begin{multline}\eqnlab{Dlam_N}
	\Delta\lambda_1 	\sim -\frac{\zeta_T}{\kappa}\int_{\mbb{R}^d}\!\!\ed^dk\, \hat F_1(\ve k)(iy_1)\ee^{-\frac{\ve y^2}2}\\
	-\frac{\zeta^2_T}{\kappa^2}\int_0^1\!\!\ed s\, s^{-2}\!\!\int_{\mbb{R}^d}\!\!\ed^dk\int_{\mbb{R}^d}\!\!\ed^dk'\, \ee^{-\frac{\ve y^2}2} \hat F_1(\ve k)\\
	\times\ve{\hat{F}}(\ve k')\cdot\partial_{y'_1}\left[\sqrt{\mbb{a}}\ve y^{(s)}\ee^{-\ve y^{(s)}\cdot\ve y'}-\ve e_1 y^{(s)}_1	\right]\ee^{-\frac12{\ve y'}^2}\,.
\end{multline}
This formula holds for arbitrary time-independent force fields $\ve F(\ve x)$, independently of whether $\ve F(\ve x)$ is quenched-random, or not. It can easily be generalised to the higher eigenvalue shifts $\Delta\lambda_n$ with $n=2,\ldots,d$, which are not our focus here. Instead, we now explore the consequences of \Eqnref{Dlam_N} for the mean and variance of $\Delta\lambda_1$, using quenched-random force fields $\ve F(\ve x)$.

We start by computing the rescaled disorder average $\mu_{\Delta\lambda_1} = \zeta_T^{-2}\langle \Delta\lambda_1\rangle_F$. Using the correlation function of the Fourier transforms $\ve{\hat{F}}$ in \Eqnref{hfcorr} and performing the integral over $\ve k'$, we obtain
\begin{multline}
	\mu_{\Delta\lambda_1} \sim-\kappa^{-4}\sum_{n=1}^d\int_0^1\!\!\ed s\, s^{a_n-2}\!\!\int_{\mbb{R}^d}\!\!\ed^dk\ \ee^{-\ve y^2}\hat C_F(\ve k)\\
	\times\ms{G}_{1n}(\ve k)k_1k_n\left[(1-s)\ee^{\ve y^{(s)}\cdot\ve y}-\delta_{1n}\right]\,.\eqnlab{mu1}
\end{multline}
This expression enables computation of the scaled average eigenvalue shift for arbitrary correlation functions $\hat C_F(\ve k)$. Moreover, it yields closed-form asymptotic expressions in the two important limits $\kappa\to0$ and $\kappa\to\infty$ of the dimensionless correlation length $\kappa$.

The rescaled variance, $\sigma_{\Delta\lambda_1}^2 = \zeta_T^{-2}(\langle \Delta\lambda_1^2\rangle_F - \langle\Delta\lambda_1\rangle_F^2)$ is obtained to lowest order in $\zeta_T$ by taking a disorder average over the square of \Eqnref{Dlam_N}, which gives
\sbeqs{\eqnlab{sigmain}
\algn{
	\sigma_{\Delta\lambda_1}^2 &\sim \kappa^{-4}\int_{\mbb{R}^d}\!\!\ed^dk\ \ee^{-\ve y^2} \hat C_F(\ve k) k_1^2\left[k_1^2 + \gamma\left(\ve k^2-2k_1^2\right)\right]\,,	\eqnlab{sigmain1}\\
	&= \kappa^{d}\int_{\mbb{R}^d}\!\!\ed^dk\ \ee^{-\kappa^2\ve y^2} \hat C_F(\kappa \ve k) k_1^2\left[k_1^2 + \gamma\left(\ve k^2-2k_1^2\right)\right]\,. \eqnlab{sigmain2}
}
}
As discussed above, the mean and the variance together fix the shift $\Delta P$ in the probability of observing faster relaxation in the limits $\kappa\to0$ and $\kappa\to\infty$, and for $d\gg1$. Through \Eqnref{probshift}, $\Delta P$ is then determined entirely by $\mu_{\Delta\lambda_1}$ and $\sigma^2_{\Delta\lambda_1}$ and so carries no independent information in the anisotropic weak-disorder limit. At intermediate $\kappa$, it depends in addition on the covariance $S$ of Appendix~\secref{gaussian}. Its asymptotic limits are nevertheless worth discussing, and we do so alongside those of $\mu_{\Delta\lambda_1}$ and $\sigma^2_{\Delta\lambda_1}$ below. Before taking limits, however, we compare \Eqsref{mu1} and~\eqnref{sigmain1} with numerical simulations of the eigenvalue problem, based on the spectral method summarised in Appendix~\secref{numerics}.

\begin{figure}
	\includegraphics[width = \linewidth]{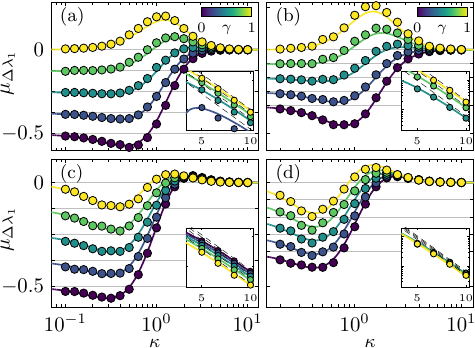}
	\caption{Scaled mean eigenvalue shift $\mu_{\Delta\lambda_1}$ from theory (\Eqnref{mu1}, lines) and simulations of the eigenvalue equation (markers) for up to $10^3$ realisations or $2\times10^3$ pairs of fields $\pm\ve F$ as functions of $\kappa$ for different $\gamma$ [colour bars in panels (a) and (b)] and for anisotropic backgrounds in $d=2$ (left) and $d=3$ (right). The simulations use fits over eleven $\zeta_T$ values between $10^{-4}$ and $10^{-2}$ or, at small and large $\kappa$, the pairs at two values of $\zeta_T\leq0.02$, see Appendix~\secref{numerics} and Table~\ref{tab:numerics}. (a) $a_2 = 5/4$. (b) $a_2=a_3=5/4$. (c) $a_2 = 10$. (d) $a_2=a_3=10$. The grey lines mark the small-$\kappa$ limits $-(1-\gamma)/d$ of \Eqnref{mushort}. The insets show $|\mu_{\Delta\lambda_1}|$ at large $\kappa$ on double-logarithmic axes, the abscissa again being $\kappa$; dashed lines are the asymptotics of \Eqnref{mulong}.}\figlab{mean}
\end{figure}
Figure~\figref{mean} shows $\mu_{\Delta\lambda_1}$ as a function of $\kappa$ in $d=2$ and $d=3$, for $\gamma\in[0,1]$ and for a weakly anisotropic background, $a_2=5/4$ [panels (a), (b)], and a strongly anisotropic one, $a_2=10$ [panels (c), (d)], with $a_3=a_2$ in $d=3$. We observe that $\mu_{\Delta\lambda_1}$ has a non-monotonic dependence on $\kappa$ for all $\gamma$. Furthermore, for intermediate values of $\gamma$ in \Figref{mean}(a),(b) and for all values in \Figref{mean}(c),(d), $\mu_{\Delta\lambda_1}$ changes sign at a finite $\kappa$ value for $\kappa$ of order unity.

\begin{figure}
	\includegraphics[width=\linewidth]{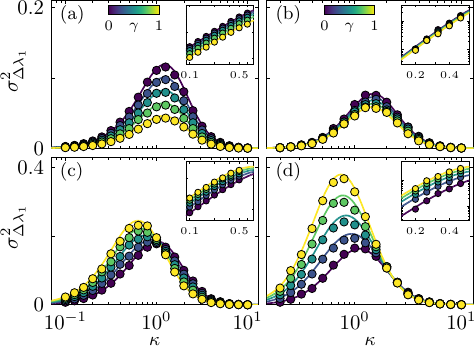}
	\caption{Scaled variance $\sigma^2_{\Delta\lambda_1}$ from theory (\Eqnref{sigmain1}, lines) and simulations of the eigenvalue equation (markers) for the same data as in \Figref{mean}, as functions of $\kappa$ for different $\gamma$ [colour bars in panels (a) and (b)] and for anisotropic backgrounds in $d=2$ (left) and $d=3$ (right). (a) $a_2 = 5/4$. (b) $a_2=a_3=5/4$. (c) $a_2 = 10$. (d) $a_2=a_3=10$. The insets show the small-$\kappa$ behaviour on double-logarithmic axes, the abscissa again being $\kappa$.}\figlab{var}
\end{figure}
Figure~\figref{var} shows $\sigma^2_{\Delta\lambda_1}$ as a function of $\kappa$ in $d=2$ and $d=3$, for $\gamma\in[0,1]$ and for the same numerical data as in \Figref{mean}. We observe that $\sigma^2_{\Delta\lambda_1}$ becomes small for small and large $\kappa$ and reaches a maximum at intermediate values.

Overall, the agreement between theory and numerics is good across the whole range of correlation lengths for both $\mu_{\Delta\lambda_1}$ and $\sigma^2_{\Delta\lambda_1}$, in $d=2$ and in $d=3$. At small and large $\kappa$, this agreement requires discretisations adapted to $\kappa$; see Appendix~\secref{numerics} including Table~\ref{tab:numerics}. The individual features of the curves in \Figsref{mean} and \figref{var} are discussed in more detail below, when we compare them with closed-form predictions in asymptotic limits.

Together with the result that the eigenvalue shift $\Delta\lambda_1=\Delta\mu_1$ is Gaussian distributed for weak disorder [see Appendix~\secref{gaussian}], the mean and variance in \Eqsref{mu1} and \eqnref{sigmain} provide the distribution $\rho(\Delta\mu_1)$ for $\zeta_T\ll 1$. The first correction to this Gaussian, \Eqnref{pdfDlam}, adds a skewness proportional to the covariance $S$ of \Eqnref{Sdef}. In \Figref{numsimaniso}(e) and (f), the so-obtained Gaussian distribution, shown as the black lines agrees excellently with the numerical data even at moderate values of $\zeta_T=0.1$.

We now evaluate \Eqsref{mu1}, \eqnref{sigmain1} and~\eqnref{probshift} in the two limits $\kappa\to0$ and $\kappa\to\infty$ in turn, and then in the limit of high dimensions.
\subsection{Short correlation lengths ($\kappa\to0$)}\seclab{anisoshort}
\subsubsection{Rescaled mean}
Applying a saddle-point approximation of \Eqnref{mu1} (Appendix~\secref{short}), $\mu_{\Delta\lambda_1}$ takes the universal form
\algn{\eqnlab{mushort}
	\mu_{\Delta\lambda_1}  \sim -\frac{1-\gamma}{d}\,,\qquad \kappa\to 0\,,
}
generalising the one-dimensional result $\mu_{\Delta\lambda_1}\sim -1$ (valid for $d=1$, $\gamma=0$) of Ref.~\cite{Mei25a}. The limit is independent of the correlation function and the background stiffnesses and is always negative for $0\leq\gamma<1$. We understand this from the fact that potential forces create many local minima at small $\kappa$, see \Figref{potential}, slowing relaxation. Non-conservative forces, by contrast, do not contribute at this order in $\zeta_T$, which is seen by identifying the factor $1-\gamma$ in \Eqnref{mushort} with the magnitude of conservative forces in \Eqnref{force}. This universal limit is supported by the numerical simulations in \Figref{mean}, where the $d=2,3$ curves approach the corresponding plateau $\mu_{\Delta\lambda_1}\to-(1-\gamma)/d$ (grey lines) as $\kappa\to0$, in both backgrounds and for every $\gamma$.

Furthermore, we can understand \Eqnref{mushort} as the leading eigenvalue shift for an effectively softer harmonic background with stiffness matrix
\algn{\eqnlab{Aeff0}
	\mbb{a}^\text{eff}_{\kappa\to0} = \ee^{-(1-\gamma)\frac{\zeta_T^2}{d}}\mbb{a}\,,
}
which follows from a simple replica calculation, summarised in Appendix~\secref{replica}. The effective stiffness $\mbb{a}^\text{eff}_{\kappa\to0}$ in \Eqnref{Aeff0} implies for $\mu_{\Delta\lambda_1}$
\algn{
	\mu_{\Delta\lambda_1} \sim \frac{(\mbb{a}^\text{eff}_{\kappa\to0})_{11}-1}{\zeta_T^2}\,,
}
which reduces to \Eqnref{mushort} in the limit $\kappa\to0$.

The expression for $\mbb{a}^\text{eff}_{\kappa\to0}$ is universal and independent of the details of the random force. This can be rationalised by noting that for $\kappa\to0$, individual relaxing particles experience a large number of local maxima and minima due to the disordered potential $V(\ve x)$. Consequently, each particle experiences a self-averaged version of the disorder potential that removes its specific details. This self-averaging behaviour in the $\kappa\to0$ limit is also reflected in the variance considered next. In the related but different context of diffusion in periodic potentials~\cite{Lif62} or random potentials without background~\cite{Zwa88,Dea94,Dee94}, expressions analogous to \Eqnref{Aeff0} have been derived for the effective diffusion constant, which was shown to be reduced by the same factor $\exp(-\zeta_T^2/d)$ (for $\gamma=0$) as $\mbb{a}$ in \Eqnref{Aeff0} in the long-time limit.
\subsubsection{Rescaled variance}
We now consider the rescaled variance $\smash{\sigma_{\Delta\lambda_1}^2}$ in the limit $\kappa\to0$, using \Eqnref{sigmain2}. In this case, we can write $\hat C_F(\kappa \ve k)\sim \hat C_F(0)$ and evaluate the Gaussian integral in \Eqnref{sigmain2}, to give
\algn{\eqnlab{sigshort}
	\sigma_{\Delta\lambda_1}^2 \sim \kappa^{d}\frac{\pi^{d/2}}{4} \sqrt{\det\mbb{a}}\,\hat C_F(0)\left[3+\gamma\left(\sum_{n=2}^d a_n-3\right)\right]\,,
}
where
\algn{
	\hat C_F(0) 	= \frac{2}{(4\pi)^{d/2}}\frac{\bar \rho_d}{\Gamma\left(\frac{d}2\right)}\,,
}
and $\bar \rho_d = \int_{0}^\infty\!\!\!\ed r\ r^{d-1}\rho(r^2)$ is a constant that depends on the correlation function. We thus find
\algn{\eqnlab{sDlam}
	\sigma_{\Delta\lambda_1}^2 \sim \left[3+\gamma\left(\sum_{n=2}^d a_n-3\right)\right]\frac{\bar \rho_d \sqrt{\det\mbb{a}}}{2\Gamma\left(\frac{d}2\right)}\left(\frac{\kappa}2\right)^{d}\,,
}
for $\kappa\ll1$, showing that the rescaled variance $\sigma_{\Delta\lambda_1}^2$ decays as $\kappa^{d}$ for $\kappa\to0$. This aligns with the expectation that for $\kappa\ll1$, self-averaging behaviour suppresses fluctuations of the relaxation speed. The insets of \Figref{var} show this decay in $d=2$ (left) and $d=3$ (right).

Equation~\eqnref{sDlam} also shows that the $\gamma$-dependent contribution to $\sigma^2_{\Delta\lambda_1}$ is positive when
\algn{\eqnlab{sDlamcond1}
	\sum_{n=2}^d a_n>3\,.
}
Hence, increasing the non-conservative contribution to the forces increases the fluctuations of $\Delta\lambda_1$ when~\eqnref{sDlamcond1} is fulfilled. Using $a_{2\leq m \leq d}\geq 1$, we find the sufficient condition
\algn{\eqnlab{sDlamcond2}
	d> d_\text{th}=4\,.
}
Hence, for $d>d_\text{th}=4$ non-conservative forces increase the fluctuations around the mean $\langle \Delta\lambda_1\rangle_F$ in every background. We find the same threshold dimension $d_\text{th}$ for the mean and for the probability shift at large $\kappa$ below.

The value $d_\text{th}=4$ is a consequence of how non-conservative contributions are weighted, see the discussion in \Secref{dimensionless}. In \Eqnref{sDlam}, the single potential direction contributes $3(1-\gamma)$ to the bracket, while each of the $d-1$ solenoidal directions contributes $\gamma a_n\geq\gamma$. Increasing $\gamma$ moves weight from the potential direction to the solenoidal ones, which raises the bracket in any background once $d-1>3$. The threshold thus depends on how the forces are parametrised. At fixed force variance, the variance $\zeta_T^2\sigma^2_{\Delta\lambda_1}$ in \Eqnref{sDlam} is proportional to $\hat\zeta_T^2[3(1-\hat\gamma)+\hat\gamma\langle a\rangle]$, with the parameters~\eqnref{hatgamma} and $\langle a\rangle=\sum_{n=2}^da_n/(d-1)$. Non-conservative forces then increase the fluctuations if and only if $\langle a\rangle>3$, in \emph{every} dimension.

\subsubsection{Probability shift}
As shown in \Eqnref{probshift}, where the covariance $S$ is negligible, the probability shift $\Delta P$ can be expressed in terms of the mean and variance as $\Delta P \sim \zeta_T \mu_{\Delta\lambda_1}/\sigma_{\Delta\lambda_1}$. Using the $\kappa\to0$ limits developed above, we can thus find a corresponding limit for $\Delta P$. At short correlation lengths, $\Delta P$ takes the form
\algn{\eqnlab{DPsmall}
	\Delta P \sim{}&   \zeta_T\sqrt{\frac{\Gamma\left(\frac{d}2\right)}{\pi \bar \rho_d}}\frac{\gamma-1}{d}\frac{(\det\mbb{a})^{-1/4}}{ \sqrt{3+\gamma\left(\sum_{n=2}^d a_n-3\right)}}\left(\frac{2}{\kappa}\right)^{d/2}\,,
}
whose derivative with respect to $\gamma$ is positive for every $\gamma$ and
every $d$. Thus, non-conservative forces always increase the
probability of observing faster relaxation at short correlation lengths,
mirroring their effect on $\mu_{\Delta\lambda_1}$. Because \Eqnref{probshift}
is valid only for $\Delta P\ll1$, the $\kappa^{-d/2}$ divergence also means
that the perturbative treatment requires $\zeta_T\ll\kappa$.
\subsection{Large correlation lengths ($\kappa\to\infty$)}\seclab{anisolong}
\subsubsection{Rescaled mean}\seclab{large}
From an expansion of the integral in \Eqnref{Dlam_N} in large $\kappa$ [or small $\ve y$, see \Eqnref{xydef}], we show in Appendix~\secref{long} that $\mu_{\Delta\lambda_1}$ takes the form
\algn{\eqnlab{mulong}
	\mu_{\Delta\lambda_1} \sim\frac{4\rho''(0)}{\kappa^{4}}\left[3+\gamma(d-4)- \sum_{m= 2}^d\frac{1-2\gamma}{a_m(a_m-1)}\right]\,, \quad \kappa\to\infty\,,
}
generalising the one-dimensional result $\mu_{\Delta\lambda_1} \sim 12\rho''(0)\kappa^{-4}$~\cite{Mei25a}. Hence, $\mu_{\Delta\lambda_1}$ decays as $\kappa^{-4}$ for large $\kappa$, which is confirmed by the insets of \Figref{mean}, with the dashed lines giving \Eqnref{mulong}. Such a decay is expected because at large correlation length, the forces become more and more constant across the minimum of the background potential and have therefore less and less of an effect.

Consequently, the result~\eqnref{mulong} can be recovered from a Taylor expansion of $\ve F(\ve x/\kappa)$ for $\kappa\gg1$ (Appendix~\secref{kapexp}), which shows that the force corrections to the background yield an effective linear force field $\ve F(\ve x) \sim \mbb{b}^\text{eff}_{\kappa\to\infty} \ve x$ with matrix
\algn{\eqnlab{linforce}
	(\mbb{b}^\text{eff}_{\kappa\to\infty})_{ij} = -a_i\delta_{ij} + \frac{\zeta_T}{\kappa^2} \partial_jF_i(0)+\frac{\zeta^2_T}{\kappa^4} \sum_{k=1}^d\frac{\partial_k\partial_jF_i(0)F_k(0)}{a_k}\,.
}
The zeroth order corresponds to the unperturbed background, whereas the additional terms are random-force corrections. With this effective linear force, the dynamics is an Ornstein-Uhlenbeck process for each realisation, and $\lambda_1$ is the eigenvalue of $-\mbb{b}^\text{eff}_{\kappa\to\infty}$ with the smallest real part, see Appendix~\secref{linforce}. On average, the first-order term vanishes, and the second-order term stiffens the background, see \Eqnref{Aeffinf} in Appendix~\secref{kapexp}. Unlike $\mbb{a}^\text{eff}_{\kappa\to0}$ in \Eqnref{Aeff0}, however, this mean background does not determine $\mu_{\Delta\lambda_1}$ alone. A perturbative calculation based on linear forces, summarised in Appendix~\secref{linforce}, shows that the fluctuations of the first-order term couple $\lambda_1$ at second order to the levels $\lambda^0_{\ve e_m}=a_m$, $m\geq2$, see \Eqnref{detforcecorr}. Separating the two contributions, \Eqnref{mulong} reads
\algn{\eqnlab{mulongsplit}
	\mu_{\Delta\lambda_1}\sim\frac{4\rho''(0)}{\kappa^4}\left\{\left[3+\gamma(d-4)+\sum_{m=2}^d\frac{1-2\gamma}{a_m}\right]-\sum_{m=2}^d\frac{1-2\gamma}{a_m-1}\right\}\,,
}
where the square bracket is the stiffening of the mean background along its slowest direction, and the last sum is the level coupling. The two forms in \Eqsref{mulong} and \eqnref{mulongsplit} agree because $1/a_m-1/(a_m-1)=-1/[a_m(a_m-1)]$.

The prefactor in \Eqnref{mulong} $4\rho''(0)/\kappa^4$ is positive, so faster relaxation at large $\kappa$ requires
\algn{\eqnlab{prefac}
	3+\gamma(d-4)+ (2\gamma-1)\sum_{m= 2}^d\frac{1}{a_m(a_m-1)}>0\,.
}
This condition depends only on $\gamma$, the background stiffnesses $\mbb{a}$, and $d$, not on the correlation function. For $\gamma>1/2$ the condition~\eqnref{prefac} is always satisfied, meaning that slower-relaxation-only regions can only exist for $\gamma<1/2$.

The two terms in \Eqnref{mulongsplit} contain the two mechanisms of faster relaxation. The first is the stiffening of the background on average along its slowest direction. It is the only contribution in one dimension~\cite{Mei25a}, and it does not require non-conservative forces. Since the square bracket in \Eqnref{mulongsplit} equals $(1-\gamma)(3+\sum_{m=2}^da_m^{-1})+\gamma\sum_{m=2}^d(1-a_m^{-1})$, the stiffening is positive for all $\gamma$ and $a_m>1$. The second term couples $\lambda_1$ at second order to the levels $\lambda^0_{\ve e_m}=a_m$, $m\geq2$, which do not exist in one dimension. As in the two-level picture of \Secref{perturbation}, the symmetric part of this coupling repels the levels and slows relaxation down, whereas its antisymmetric part, which is due to non-conservative forces, attracts them and speeds relaxation up, see \Eqnref{detforcecorr} in Appendix~\secref{linforce}. For potential forces, the sign of this coupling follows from a general argument: $\mbb{b}^\text{eff}_{\kappa\to\infty}$ is then symmetric, and because the smallest eigenvalue of a symmetric matrix is a concave function of its elements, Jensen's inequality gives $\langle\lambda_1\rangle_F\leq(\mbb{a}^\text{eff}_{\kappa\to\infty})_{11}$, the rate of the mean background. Non-conservative forces make $\mbb{b}^\text{eff}_{\kappa\to\infty}$ non-symmetric, which allows its fluctuations to raise $\langle\lambda_1\rangle_F$ above this rate. This level attraction is the counterpart, for random forces, of the acceleration of Gaussian diffusions by non-reversible drifts~\cite{Hwa93}. It outweighs the repulsion for $\gamma>1/2$, and it grows as $a_m\to1$, where the levels approach each other. In quasi-isotropic backgrounds, it becomes a first-order effect, \Secref{qiso}. Because the stiffening is positive, slower relaxation at large $\kappa$ is caused by level repulsion alone, which is why it requires $\gamma<1/2$.

For the simplified case $a_m=a_2$ ($m\geq2$), the boundary between faster and slower relaxation regions, where the left-hand side of \eqnref{prefac} vanishes, reduces to the quadratic condition
\algn{\eqnlab{acond}
	\left[3+\gamma(d-4)\right]a^*_2(a^*_2-1)+ (2\gamma-1)(d-1)=0\,,
}
where $a^*_2=a^*_2(d,\gamma)$ denotes the value of $a_2$ at the boundary.

Figure~\figref{statediag}(a) shows the resulting state diagram in the $(a_2,\gamma)$ plane for $d=2,3,4$ and $d\to\infty$. The yellow-hatched region marks where relaxation is faster at large $\kappa$ across all dimensions. For $\gamma<1/2$, regions with slower relaxation at large $\kappa$ appear. These regions shrink with decreasing dimensionality $d$. For fixed $\gamma$, we find that the smaller $d$, the smaller is the degree of anisotropy needed to observe faster relaxation. These results confirm that lower dimensionality and stronger non-conservative forcing (large $\gamma$) generically favour faster relaxation.
\begin{figure}
	\includegraphics[width=\linewidth]{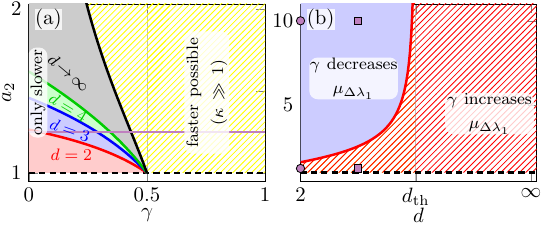}
	\caption{State diagrams as a function of stiffnesses $a_{2\leq m \leq d}=a_2$ and $\gamma$ for different dimensions $d$ in an anisotropic background. (a) State diagram for the occurrence of enhanced relaxation speed. The yellow-hatched region shows where relaxation is faster for $\kappa\gg1$ across all space dimensions $d$. For decreasing $d$, the filled region where relaxation is always slower at large $\kappa$ shrinks. The violet horizontal line marks $a_2=5/4$ of \Figsref{mean}(a) and~(b), and $a_2=10$ of \Figsref{mean}(c) and~(d) lies above the plotted range. (b) State diagram for the impact of non-conservative forces on $\mu_{\Delta\lambda_1}$. The blue-filled region shows parameters for which increasing $\gamma$ leads to a decrease of $\mu_{\Delta\lambda_1}$. The red-hatched region shows where the opposite behaviour occurs. For $d\geq d_\text{th}=4$, the behaviour becomes independent of the details of the background and of the quenched force field. The violet bullets and squares correspond to the $a_2$ and $d$ values in \Figref{mean}.}\figlab{statediag}
\end{figure}
Importantly, this result is independent of the correlation function $C_F$.

Additional information is gained by considering the effect of non-conservative forces on $\mu_{\Delta\lambda_1}$ at small and large $\kappa$. For small $\kappa$, we observe from \Eqnref{mushort} that increasing the fraction of non-conservative forces in \Eqnref{force} always increases $\mu_{\Delta\lambda_1}$. For large $\kappa$, by contrast, \Eqnref{mulong} shows that increasing $\gamma$ increases $\mu_{\Delta\lambda_1}$ if
\algn{\eqnlab{mucond}
	d-4+2\sum_{m= 2}^d\frac{1}{a_m(a_m-1)} > 0\,.
}
Since $a_{m\geq2}>1$, we find that again $d\geq d_\text{th}=4$ [cf. \Eqnref{sDlamcond2}] is a sufficient condition under which non-conservative forces increase $\mu_{\Delta\lambda_1}$ in every background. At fixed force variance, by contrast, increasing $\hat\gamma$ increases $\mu_{\Delta\lambda_1}$ if and only if $\sum_{m=2}^d[a_m(a_m-1)]^{-1}>2(d-1)/d$, a condition that depends on the background in every dimension. For the special case $a_{m\geq2}=a_2$, we can specify the boundary between the regions where non-conservative forces increase or decrease $\mu_{\Delta\lambda_1}$ for $\kappa\gg1$ by solving a quadratic equation for $a_2$. The result is shown in the diagram in \Figref{statediag}(b), where the red curve separates regions in the $(a_2,\gamma)$ parameter space where the two different behaviours occur. For $d\geq d_\text{th}$, the behaviour becomes independent of $a_2$ in accordance with the discussion above.

Figure~\figref{mean} shows both behaviours in $d=2$ and $d=3$. Increasing $\gamma$ raises $\mu_{\Delta\lambda_1}$ at large $\kappa$ for $a_2=5/4$ but lowers it for $a_2=10$, in agreement with the corresponding violet bullets in \Figref{statediag}(b), which indicate their location in parameter space for $d=2$. For $d=3$ the boundary lies at $a_2=(1+\sqrt{17})/2\approx2.56$, so that the two backgrounds fall into the same regions. The two backgrounds differ also at intermediate $\kappa$, where only the weakly anisotropic one develops a pronounced peak whose sign is set by $\gamma$. Purely non-conservative forces ($\gamma=1$) produce a positive excursion, while purely potential forces ($\gamma=0$) remain negative for all $\kappa$. That the $\gamma=0$ curve never turns positive for $a_2=5/4$ is consistent with the violet horizontal line in \Figref{statediag}(a), which crosses the ($d=2$) boundary of \Eqnref{acond} at $\gamma=1/22$ for $a_2=5/4$. In $d=3$ this crossing moves to $\gamma\approx0.29$, so that the curve for $\gamma=1/4$ in \Figref{mean}(b) also remains negative for all $\kappa$.
\subsubsection{Rescaled variance}
We now turn to the rescaled variance $\sigma_{\Delta\lambda_1}^2$ for $\kappa\to\infty$ using \Eqnref{sigmain1}. When the correlation lengths are large, we have $\ee^{-y^2}\sim 1$, and the remaining integral is evaluated as
\algn{
	\sigma_{\Delta\lambda_1}^2 \sim \frac1{\kappa^4}\nabla_1^2\left[\nabla^2_1 + \gamma(\Delta - 2\nabla_1^2)\right] C_F(0)\,.
}
When written in terms of $C_F(|\ve x|) = \rho(r^2)$, we find
\algn{\eqnlab{siglong}
	\sigma_{\Delta\lambda_1}^2 \sim \frac{4\rho''(0)}{\kappa^4}\left[3+(d-4)\gamma\right]\,.
}
This result implies that for large $\kappa$, the rescaled variance $\sigma_{\Delta\lambda_1}^2$ becomes independent of the background, and depends on the correlation function only through $\rho''(0)$, similar to the behaviour of the rescaled mean $\mu_{\Delta\lambda_1}$ in \Eqnref{mulong}.

The impact of non-conservative forces is readily read off from \Eqnref{siglong}. We find that for large $\kappa$, non-conservative forces increase the disorder-induced fluctuations of the relaxation speed when $d>d_\text{th}=4$, independently of the details of the forces and of the background. At fixed force variance, $\zeta_T^2\sigma^2_{\Delta\lambda_1}$ in \Eqnref{siglong} is proportional to $\hat\zeta_T^2(3-2\hat\gamma)$ instead, so that non-conservative forces decrease the fluctuations in every dimension. Together with the condition~\eqnref{sDlamcond1}, we draw a state diagram for the impact of non-conservative forces on $\sigma^2_{\Delta\lambda_1}$ for a background with $a_{2\leq m \leq d}=a_2$. Figure~\figref{statediag_2} shows this diagram, with the red region showing parameters for which increasing $\gamma$ decreases $\sigma^2_{\Delta\lambda_1}$, i.e., non-conservative forces suppress fluctuations of $\Delta\lambda_1$. For $a_{2\leq m \leq d}=a_2$ the boundary (red line) of this region is obtained by solving a quadratic equation for $a_2$ based on~\eqnref{sDlamcond1}. The blue region corresponds to parameters for which non-conservative forces have opposite effects at small and large $\kappa$. When $d>d_\text{th}=4$ non-conservative forces always increase the fluctuations of $\Delta\lambda_1$, independently of the characteristics of the quenched-random forces and of the background.

Figure~\figref{var} locates the two backgrounds of \Figref{mean} in this diagram. For $a_2=5/4$ [\Figref{var}(a),(b)], increasing $\gamma$ lowers $\sigma^2_{\Delta\lambda_1}$ for both small and large $\kappa$, so that non-conservative forces sharpen the relaxation behaviour. This case lies in the red region of \Figref{statediag_2}. For $a_2=10$ [\Figref{var}(c),(d)], the two limits behave differently. Non-conservative forces still sharpen at large $\kappa$ but act dispersively at small $\kappa$, placing this case in the blue region of \Figref{statediag_2}. The change also displaces the maximum, which stays at $\kappa\approx1.2$ in $d=2$ and $\kappa\approx1.5$ in $d=3$ for every $\gamma$ when $a_2=5/4$ but moves to smaller $\kappa$ as $\gamma$ grows when $a_2=10$, while $\sigma^2_{\Delta\lambda_1}$ at $\kappa\approx1$ remains essentially fixed in $d=2$. This produces the crossing point visible in \Figref{var}(c). Since $\sigma^2_{\Delta\lambda_1}$ is linear in $\gamma$, see \Eqnref{sigmain1}, all curves cross at the same $\kappa$. In $d=3$ this crossing moves to $\kappa\approx2.3$, so that the maximum grows with $\gamma$ [\Figref{var}(d)].
\begin{figure}
	\includegraphics[width=\linewidth]{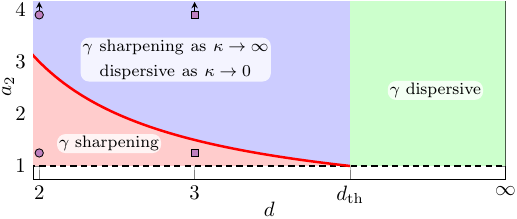}
	\caption{State diagram for the impact of non-conservative forces on $\sigma^2_{\Delta\lambda_1}$ as a function of stiffnesses $a_{2\leq n\leq d}=a_2$ and $d$ for anisotropic backgrounds. The red region shows where $\gamma$ decreases fluctuations of $\Delta\lambda_1$ at small and large $\kappa$. In the blue region, $\gamma$ has opposite effects in both $\kappa$-limits. In the green regions non-conservative forces always increase fluctuations. The purple bullets and squares correspond to the parameter values in \Figref{mean}.}\figlab{statediag_2}
\end{figure}
\subsubsection{Probability shift}
At large correlation lengths, the sign of $\Delta P$ follows that of
$\mu_{\Delta\lambda_1}$, since $\sigma^2_{\Delta\lambda_1}>0$. Faster
relaxation is thus more probable in the same region of parameter space in
which the mean relaxation speed is enhanced, the yellow-hatched region of the
state diagram in \Figref{statediag}(a). The full $\kappa$-dependence of
$\Delta P$, the effect of non-conservative forces at large $\kappa$, and the
comparison with numerical simulations are given in
Appendix~\secref{probapp}.
\subsection{High-dimensional limit $d\gg1$}\seclab{anisohighd}
The bracket in \Eqnref{mulong} grows linearly with the dimension, which suggests that the disorder-induced shift at large correlation lengths becomes ever more effective as $d$ increases. This, however, is an artefact of the order of the limits $\kappa\to\infty$ and $d\to\infty$. In other words, \Eqnref{mulong} is an expansion in $\kappa^{-2}$ whose coefficients themselves grow with $d$, and it is therefore valid only for $\kappa^2\gg d$. Since the shift decays as $\kappa^{-4}$, the regime in which \Eqnref{mulong} holds is also the regime in which the shift is small, and the high-dimensional limit must be taken with the correlation length scaled together with the dimension. We will encounter an analogous behaviour in the isotropic case of \Secref{iso} below.

In order to avoid this issue, we evaluate \Eqnref{mu1} without first expanding in $\kappa$. This can be done in closed form for Gaussian correlations, which is not a strong restriction for $d\to\infty$. This is because by Schoenberg's theorem~\eqnref{schoenberg}, every admissible isotropic correlation function becomes a superposition of Gaussians in this limit. The calculation is given in Appendix~\secref{gauss2}. The structure of the limit is essentially that of \Eqnref{mulong} multiplied by the factor
\algn{\eqnlab{Rmain}
	\hat{\ms{R}} = \prod_{n=1}^d\hat r_n\,,\qquad \hat r_n = \frac1{\sqrt{1+\frac{\pi}{a_n\kappa^2}}}\,.
}
Individually, each factor $\hat r_n$ in $\hat{\ms{R}}$ tends to unity as $\kappa$ grows. Hence $\hat{\ms{R}}\to1$ for large $\kappa$ at finite $d$, consistent with \Eqnref{mu1}. However, the product in $\hat{\ms{R}}$ contains $d$ such factors and is exponentially small in the dimension at any fixed finite $\kappa$.

We write for $\kappa\to\infty$,
\algn{\eqnlab{hakapdef}
	-\log\hat{\ms{R}} \to \frac{\pi}{2\kappa^2}\sum_{n = 1}^d\frac1{a_n} = \frac{\pi}{2\hat\kappa^2}\,,
}
where $\hat \kappa = \ell_\text{c}/\ell_\text{rms}$ denotes the correlation length in units of $\ell_\text{rms}$, the root-mean-square size of thermal fluctuations
\algn{\eqnlab{lrms}
	\ell_\text{rms}^2 = \ell_\text{T}^2\tr (\mbb{a}^{-1}) =  \frac{d\,\ell_\text{T}^2}{\bar a} \,,
}
in the unperturbed background, which grows as $\sqrt{d}$. Here $\bar a = d\,[\tr (\mbb{a}^{-1})]^{-1}$ denotes the harmonic mean of the background stiffnesses. After manipulations detailed in Appendix~\secref{gauss2}, the product $\mu_{\Delta\lambda_1}d$ acquires the scaling form
\algn{\eqnlab{muscalmain}
	\mu_{\Delta\lambda_1}d\sim\Phi_\gamma(\hat\kappa,\mbb{a})\,,
}
as a function of $\hat \kappa$. As we show in Appendix~\secref{gauss2}, the associated scaling function $\Phi_\gamma$ can be written as
\algn{\eqnlab{muscalmain2}
	\Phi_\gamma(\hat\kappa,\mbb{a})= -\left(\frac{\pi \bar a}{2\hat\kappa^2}\right)^2\left[\gamma\ms{J}+(1-2\gamma)\ms{K}\right]\,,
}
with the two one-dimensional integrals
\sbeqs{\eqnlab{Jdefmain}
\algn{
	\ms{J} &= \int_0^1\!\!\ed s\,s^{-1}\left[(1-s)\ee^{-\frac{\pi}{2\hat\kappa^2}\left[1-g(s)\right]}-\ee^{-\frac{\pi}{2\hat\kappa^2}}\right]\,,\\
	\ms{K} &= \frac1d\sum_{n=2}^d\int_0^1\!\!\ed s\, s^{a_n-2}(1-s)\,\ee^{-\frac{\pi}{2\hat\kappa^2}\left[1-g(s)\right]}\,.
}
}
The background enters \Eqsref{Jdefmain} through the dimensionless function
\algn{\eqnlab{gdefmain}
	g(s) = \frac{\ell^2_\text{T}}{\ell^2_\text{rms}}\sum_{n=1}^d\frac{s^{a_n}}{a_n}\,,\qquad g(0)=0\,,\quad g(1)=1\,.
}

Equation~\eqnref{muscalmain} implies that for $d\gg1$ and $\kappa\sim\sqrt{d}$, $\mu_{\Delta\lambda_1}$ decreases as $d^{-1}$ and $d\,\mu_{\Delta\lambda_1}$ depends on $\hat\kappa$ and $\mbb{a}$ alone, so that $d\mu_{\Delta\lambda_1}$ curves taken at different dimensions collapse onto one another when plotted against $\hat\kappa$ at fixed $\mbb{a}$. For $\hat\kappa\to0$ we find $\Phi_\gamma\to-(1-\gamma)$, so $\Phi_\gamma/d$ then recovers the small-$\kappa$ limit in \Eqnref{mushort}, while for $\hat\kappa\to\infty$, $\Phi_\gamma/d$ reduces to the large-$d$ limit of \Eqnref{mulong}.  Hence, $\Phi_\gamma$ as a function of $\hat\kappa$ interpolates between both known limiting cases for large $d$.

We note two important consequences of \Eqnref{muscalmain}. First, $\Phi_\gamma(\hat\kappa,\mbb{a})$ has a maximum at a value $\hat\kappa_{\text{max},\mu}$ of order unity, so that the optimal correlation length and the shift attained there behave as
\algn{\eqnlab{kapoptmain}
	\mu_{\Delta\lambda_1}^\text{max}d \sim \Phi_\gamma(\hat\kappa_{\text{max},\mu},\mbb{a})\,.
}
The correlation length $\kappa_\text{opt}$ that makes the disorder most effective therefore grows as $\sqrt d$, which corresponds to the growth of $\ell_\text{rms}$ in \Eqnref{lrms}. In other words, the disorder is most effective when its correlation length is comparable to the size of the region the particle typically explores. Only in units of $\ell_\text{T}$, which measures the size of fluctuations in a single direction, the correlation length increases. The eigenvalue shift itself decays as $1/d$ which implies that in high-dimensional anisotropic backgrounds, the effect of weak disorder decreases with the dimension.

Second, since $\hat\kappa\to 0$ returns \Eqnref{mushort} and $\hat\kappa\to\infty$ returns \Eqnref{mulong}, whether the disorder speeds relaxation up or slows it down at long correlation lengths is again decided by the condition~\eqnref{prefac}, implying that the state diagram of \Figref{statediag}(a) for $d\to\infty$ applies unchanged.

The variance behaves in a similar way. For a Gaussian correlation function $C_F(|\ve r|) = \ee^{-\pi\ve r^2/4}$, with Fourier transform $\hat C_F(\ve k) = \pi^{-d}\ee^{-\ve k^2/\pi}$, the integral~\eqnref{sigmain1} can also be evaluated in closed form at every $\kappa$ and every $d$, giving
\algn{\eqnlab{siggauss}
	\sigma_{\Delta\lambda_1}^2 = \frac{\pi^2}{4}\frac{\hat{\ms{R}}}{\pi+\kappa^2}\left[\frac{3(1-\gamma)}{\pi+\kappa^2} + \gamma\sum_{j=2}^d\frac{a_j}{\pi+\kappa^2 a_j}\right]\,,
}
with the factor $\hat{\ms{R}}$ in \Eqnref{Rmain}. Equation~\eqnref{siggauss} reduces to \Eqsref{sigshort} and~\eqnref{siglong} in the two limits of the correlation length. Taking $d$ large at fixed $\hat\kappa$, both denominators are dominated by $\kappa^2$ and the sum contributes a factor $d$, which gives
\algn{\eqnlab{sigscalmain}
	\sigma^2_{\Delta\lambda_1} \sim  \Psi^2(\hat\kappa,\bar a)\times\css{\frac{3}{d^2} \,,& \gamma=0	\\ \frac{\gamma}{d}	\,,&  \gamma>0\,,}
}
with the associated scaling function $\Psi>0$ given by
\algn{\eqnlab{psimain}
	\Psi(\hat\kappa,\bar a)=\frac{\pi \bar a}{2\hat\kappa^2}\,\ee^{-\frac{\pi}{4\hat\kappa^2}}\,.
}
This shows that fluctuations are increased for increasing $\gamma$. Importantly, $\sigma^2_{\Delta\lambda_1}$ scales differently with the dimension for conservative or non-conservative forces. In particular, $\gamma=0$ leads to weaker fluctuations that decay as $d^{-2}$ while the fluctuations for $\gamma>0$ are stronger, decaying as $d^{-1}$.

The scaling function $\Psi$ is positive, vanishes at small and large $\hat\kappa$, and has a maximum at
\algn{
	\hat\kappa_{\text{max},\sigma}^2=\frac{\pi}{4}\,,
}
so the fluctuations of the relaxation rate are largest at the same $\hat\kappa$ in every background and at every $\gamma$.

Finally, the probability shift $\Delta P$ in the high-dimensional limit takes the form
\algn{\eqnlab{DPhighd}
	\Delta P \sim \frac{\zeta_T}{\sqrt{2\pi}}\frac{\Phi_\gamma(\hat\kappa,\mbb{a})}{\Psi(\hat\kappa,\bar a)}\times\css{\frac1{\sqrt{3}} \,,& \gamma=0	\\ \frac1{\sqrt{\gamma d}}	\,,&  \gamma>0\,,}
}
inherited from the individual scaling forms for the mean and variance. Here we have neglected the correction $S$ in \Eqnref{DP2}. This is justified for $d\gg1$, because $S/\sigma^2_{\Delta\lambda_1}$ decays as $d^{-2}$ at fixed $\hat\kappa$, faster than $\mu_{\Delta\lambda_1}\propto d^{-1}$ (see Appendix~\secref{gaussian}). For conservative forces, $\Delta P$ takes a finite limit, while it is suppressed as $d^{-1/2}$ for non-conservative forces. The limit for conservative forces is negative, because $\Phi_0(\hat\kappa,\mbb{a})=-[\pi\bar a/(2\hat\kappa^2)]^2\ms{K}<0$ for all $\hat\kappa$ and in every background, the integral $\ms{K}$ in \Eqnref{Jdefmain} being positive. Hence, for conservative forces slower relaxation remains more probable in high dimensions.

The different scalings for $\gamma=0$ and $\gamma>0$ in \Eqsref{sigscalmain} and~\eqnref{DPhighd} originate from the growth of the solenoidal force variance with $d$ at fixed $\gamma$, see \Secref{dimensionless}. Since $\Phi_\gamma=(1-\gamma)\Phi_0+\gamma\Phi_1$ is linear in $\gamma$, the scaling forms are readily expressed in terms of the parameters~\eqnref{hatgamma}. At fixed $\hat\gamma<1$ we find
\algn{\eqnlab{scalhat}
	\zeta_T^2\mu_{\Delta\lambda_1}d&\sim\hat\zeta_T^2(1-\hat\gamma)\Phi_0(\hat\kappa,\mbb{a})\,,\\
	\zeta_T^2\sigma^2_{\Delta\lambda_1}d^2&\sim\hat\zeta_T^2(3-2\hat\gamma)\Psi^2(\hat\kappa,\bar a)\,,
}
because the solenoidal contribution to the mean, $\hat\zeta_T^2\hat\gamma\Phi_1/(d-1)$, is smaller by a factor $1/d$. Mean and variance then scale in the same way for all $\hat\gamma<1$, and $\Delta P\sim\hat\zeta_T(1-\hat\gamma)\Phi_0/[\sqrt{2\pi(3-2\hat\gamma)}\,\Psi]$ tends to a finite, negative limit. At fixed force variance, weak disorder thus slows relaxation down on average in high-dimensional anisotropic backgrounds, for every $\hat\gamma<1$. Faster relaxation at large $d$, i.e., $\Phi_\gamma>0$ in \Eqnref{muscalmain}, requires the solenoidal force variance to grow with $d$, as it does at fixed $\gamma$.

In this section, we described the behaviour in anisotropic backgrounds, the left branch in the diagram in \Figref{flowdiagram}. To summarise, in an anisotropic background the unperturbed level $\lambda^0_1$ is non-degenerate, so that the disorder shifts $\lambda_1$ by a real value, whose mean is \emph{second} order in $\zeta_T$, i.e., $\langle\Delta\lambda_1\rangle_F\sim\langle\Delta \mu_1\rangle_F\sim\zeta_T^2$, while its fluctuations are of first order. At short correlation lengths the mean shift is universal, $\mu_{\Delta\lambda_1}\to-(1-\gamma)/d$ [\Eqnref{mushort}]. In this case, the particle samples many minima of the disorder potential, so relaxation is always slower, and the fluctuations are suppressed as $\kappa^d$. At long correlation lengths both mean and variance decay as $\kappa^{-4}$, and there relaxation can become \emph{faster}. Which of the two occurs is decided by the condition~\eqnref{prefac}, which depends on $\gamma$, on the stiffnesses and on $d$, but not on the correlation function, and which is satisfiable in any background once $\gamma>1/2$. For $d\geq d_\text{th}=4$, increasing $\gamma$ raises the mean and the variance whatever the background. This threshold disappears when the force variance is kept fixed. Since $\Delta P$ follows the sign of $\mu_{\Delta\lambda_1}$ at large $\kappa$, the same condition decides whether faster relaxation is more probable. The largest effect lies between the two limits, at correlation lengths comparable to the size $\ell_\text{rms}$, which corresponds to the typical diameter of the region the particle explores. In high dimensions, $\ell_\text{rms}$ becomes the only relevant scale, so that mean and variance collapse onto functions of $\hat\kappa=\ell_\text{c}/\ell_\text{rms}$, \Eqsref{muscalmain} and~\eqnref{sigscalmain}, with their optima at $\hat\kappa_{\text{max},\mu}$ and $\hat\kappa_{\text{max},\sigma}$ of order unity, i.e., at $\kappa\propto\sqrt d$. The shift $\mu_{\Delta\lambda_1}$ itself decays as $1/d$, while $\sigma^2_{\Delta\lambda_1}$ decays as either $d^{-1}$ or $d^{-2}$ for $\gamma>0$ and $\gamma=0$, respectively. Consequently, the probability shift $\Delta P$ takes a finite, negative limit for conservative forces, while it decays as $d^{-1/2}$ for non-conservative forces. At fixed force variance, mean and variance scale alike for all $\hat\gamma<1$, and disorder slows relaxation down on average. Simulations in $d=2$ confirm the perturbative predictions across the whole range of correlation lengths.

\section{Isotropic background}\seclab{iso}
\begin{figure}
	\includegraphics[width=\linewidth]{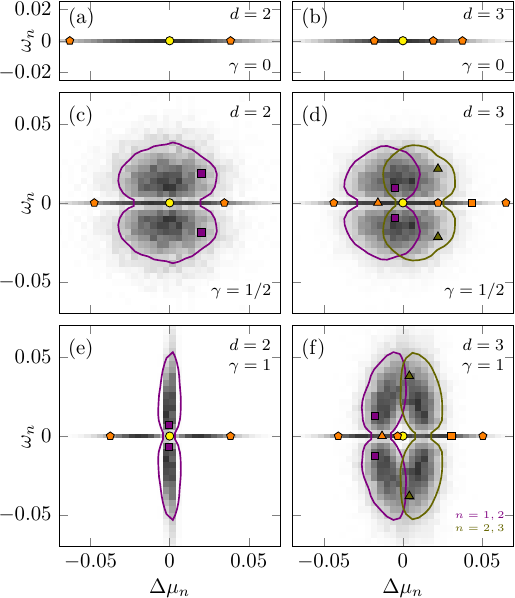}
	\caption{Numerically obtained shifts $\Delta\lambda_n$ of the $d$ lowest non-vanishing eigenvalues of the Fokker-Planck operator~\eqnref{FPop} in an \textit{isotropic} background, computed with the method of \Secref{numsim}, for $\gamma=0$ [(a),(b)], $\gamma=1/2$ [(c),(d)] and purely solenoidal forces, $\gamma=1$ [(e),(f)], in $d=2$ (left) and $d=3$ (right), correlation length $\kappa=1$, and disorder strength $\zeta_T=10^{-1}$. The simulation box is of size $L=20$ with $N=100$ ($d=2$) and $N=40$ ($d=3$) grid points per dimension, and $2.5\times10^4$ disorder realisations. Each panel shows the density of $\Delta\lambda_n$ in the complex plane, pooled over the branches, as a grey scale normalised to its maximum. At $\gamma=0$ all shifts are real, so that only the line on the real axis remains. Coloured markers show the shifts in single realisations, one of each type, with the shapes of \Figref{numsimaniso}: all shifts real (pentagons), $\lambda_1$ and $\lambda_2$ forming a complex-conjugate pair (squares) and, for $d=3$, $\lambda_2$ and $\lambda_3$ forming a pair (triangles). Real shifts are drawn in orange, the members of a pair in the colour of the corresponding contour. The violet line is the contour of the complex-conjugate pairs formed by $\lambda_1$ and $\lambda_2$ at a quarter of their maximum, the olive line that of the pairs formed by $\lambda_2$ and $\lambda_3$. The yellow bullet marks the unperturbed, degenerate eigenvalue at $\Delta\lambda=0$, onto which the density collapses as $\zeta_T\to0$.}
	\figlab{numsimiso}
\end{figure}
\begin{figure}
	\includegraphics[width=\linewidth]{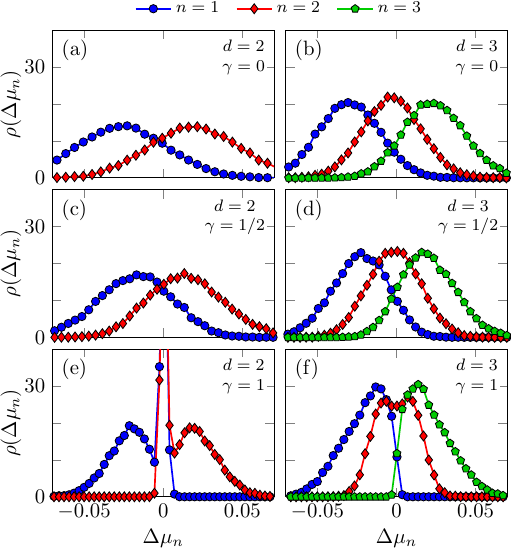}
	\caption{Marginal densities $\rho(\Delta\mu_n)$ of the real parts of the shifts in \Figref{numsimiso}, resolved by branch and labelled in order of increasing real part, $n=1$ (blue circles), $n=2$ (red diamonds) and $n=3$ (green pentagons), for $\gamma=0$ [(a),(b)], $\gamma=1/2$ [(c),(d)] and $\gamma=1$ [(e),(f)], in $d=2$ (left) and $d=3$ (right). The narrow peak at $\mathrm{Re}(\Delta\lambda)=0$ in (e) is clipped by the axis.}
	\figlab{numsimisore}
\end{figure}
Before developing a theory for the leading eigenvalue $\Delta\lambda_1$ for an isotropic background, we again illustrate the phenomenology with direct numerical simulations of~\Eqnref{Leig} for a large number ($\sim10^4$) of disorder realisations. When the background is isotropic, then the $d$ lowest non-vanishing levels of the unperturbed operator $\ms{L}_0^\dagger$ coincide exactly [\Figref{spectrum}(b)]. Consequently in the absence of disorder, this part of the spectrum consists of a single, $d$-fold degenerate point, marked by the yellow bullet at $\Delta\lambda=0$ in every panel of \Figref{numsimiso}.

Any realisation of the random force field breaks the isotropy of $\ms{L}^\dagger$ and thereby lifts this degeneracy, splitting the level into $d$ distinct eigenvalues. Over an ensemble of realisations, the degenerate point thus spreads into a continuous distribution of shifts $\Delta\lambda_n = \lambda_n - \lambda^0_n$ in the complex plane. Figure~\figref{numsimiso} shows this distribution for $\gamma=0$, $\gamma=1/2$ and purely solenoidal forces, $\gamma=1$, in $d=2$ [\Figref{numsimiso}(a), (c) and~(e)] and $d=3$ [\Figref{numsimiso}(b), (d), and~(f)]. These data are obtained from the full, non-perturbative eigenvalue problem by the method of \Secref{numsim}, at a disorder strength $\zeta_T=10^{-1}$.

\Figsref{numsimiso}(c)--(f) show that for $\gamma>0$ the distribution has two clearly separated components. It consists of a sharp line of real shifts on the real axis [$\mathrm{Im}(\Delta\lambda)=0$], and a diffuse cloud of complex eigenvalues. As we show below, such a separation is a generic property of finite-dimensional real random matrices. In particular, in the representation of \Secref{fpe} the discretised operator $\ms{L}^\dagger$ is real, so that its eigenvalues are either real or occur in complex-conjugate pairs, and each of these possibilities is realised with a finite probability. The markers in \Figref{numsimiso} show single realisations of each type. In $d=2$, the two shifts are either both real (pentagons) or they form a complex-conjugate pair (squares). In $d=3$, either all three shifts are real (pentagons), or one is real and the other two form a pair, which consists of $\lambda_1$ and $\lambda_2$ (squares) or of $\lambda_2$ and $\lambda_3$ (triangles). The type of a realisation is thus set by its number $R$ of real eigenvalues, $R=d$ or $R=d-2$. In the random-matrix description developed below, this is the decomposition~\eqnref{ev_dens_xi} of the eigenvalue density into sectors with a fixed number $R$ of real eigenvalues. For conservative forces, $\gamma=0$, $\ms{L}^\dagger$ can be transformed into a Hermitian form, so the complex component of the spectrum disappears, see \Figsref{numsimiso}(a) and~(b).

The second, vertical line of shifts close to the imaginary axis [$\Delta\mu_n\approx0$] in \Figref{numsimiso}(e) has no counterpart in $d=3$ or at $\gamma=1/2$ [\Figsref{numsimiso}(c) and~(d)]. It reflects a peculiarity of two dimensions at $\gamma=1$. As shown below, the perturbation matrix that determines the spectrum to leading order $\zeta_T$ is then traceless in every realisation, so that for $d=2$ its two eigenvalues are either both real and opposite in sign, or they form a purely imaginary pair, in which case both eigenvalues have vanishing real part. To leading order in $\zeta_T$, the density in \Figref{numsimiso}(e) therefore only lives on these two axes. The higher orders in $\zeta_T$, retained by the simulation shown here, affect the two lines differently. Eigenvalues that are real remain exactly real to high accuracy, so that the horizontal line survives almost unbroadened. The purely imaginary pairs, by contrast, acquire a small real part and are smeared out into the surrounding cloud. In $d=3$, tracelessness no longer pins the eigenvalues to the imaginary axis, and \Figref{numsimiso}(f) accordingly shows the real line together with a broad, almost featureless complex cloud.

Projecting these distributions onto the real axis and resolving the $d$ branches in order of increasing real part gives \Figref{numsimisore}, which makes the splitting of the degenerate level directly visible. In $d=2$ the level splits into two branches, in $d=3$ into three. In $d=2$ and at $\gamma=1$ both branches carry the sharp peak at $\Delta\mu=0$ discussed above [\Figref{numsimisore}(e)], and, because $\mbb{\tilde H}$ is traceless, they are mirror images of one another. In $d=3$ the peak is absent and the three branches are smooth and well separated [\Figref{numsimisore}(f)]. For $\gamma<1$ the peak is absent also in $d=2$, and the branches broaden as $\gamma$ decreases. At $\gamma=0$, where all shifts are real, they also lie furthest apart [\Figsref{numsimisore}(a) and~(b)]. Asymptotically, the leftmost branch, $n=1$, corresponds to $\Delta\tilde\mu_1$, which governs the long-time relaxation speed.

\subsection{Theory}
We now develop a theory to explain the behaviour in \Figref{numsimiso}. When the background is isotropic, all directions share the same stiffness, i.e., $\mbb{a}=\mbb{1}$ or $a_1=a_2=\ldots=a_d$. In this case, the lowest-lying unperturbed eigenvalue is $d$-fold degenerate. Therefore, the eigenvalue shift $\Delta\lambda_1$ of the smallest non-vanishing eigenvalue of $\ms{L}^\dagger$ is determined from the spectrum of the perturbation matrix $\mbb{H}$ with elements
\algn{\eqnlab{Hnmdef}
	H_{nm} = \langle \ve e_n| \ms{H}^\dagger_{F} |\ve e_m\rangle = \langle 0|c_n \ms{H}^\dagger_{F} c^\dagger_m|0\rangle\,.
}
The quenched-random forces $\ve F$ are Gaussian distributed, implying that $\mbb{H}$ is a Gaussian random matrix with zero-mean entries, $\langle H_{nm}\rangle_{F}=0$. Consequently, $\mbb{H}$ is completely determined by the covariances
\algn{\eqnlab{H1corr}
	\langle H_{nm}H_{op}\rangle_{F} = \Sigma^2(\kappa)\ms{C}_{nmop}(\gamma)\,,
}
where $\ms{C}_{nmop}$ is a rank-four tensor and $\Sigma(\kappa)$ is a covariance factor. From \Eqnref{Hnmdef} and \eqnref{HdagF}, we obtain an expression for $H_{nm}$ in terms of $\ve{\hat{F}}$,
\algn{
	H_{nm} = -i\kappa^{-2}\int_{\mbb{R}^d}\!\!\!\ed^d k\, \ee^{-\ve k^2/(2\kappa^2)}k_n \hat F_m(\ve k)\,,
}
which allows us to determine the right-hand side of \Eqnref{H1corr} from the correlation function
\begin{multline}
	\Sigma^2(\kappa)\ms{C}_{nmop}(\gamma) %
	= \kappa^{-4}\!\!\int_{\mbb{R}^d}\!\! \ed ^d\ve k\, \ee^{-\ve k^2/\kappa^2}k_nk_o\left[(1-\gamma)k_mk_p\right.\\
	\left. + \gamma (\delta_{mp} \ve k^2 -k_mk_p)\right]\hat C_F(\ve k)\,. \eqnlab{cint}
\end{multline}
Due to the isotropy of the random forces and the background, a generic quartic integral over $\ve k$ must have the form
\begin{multline}
	\kappa^{-4}\!\!\int_{\mbb{R}^d}\!\! \ed ^d\ve k\, \ee^{-\ve k^2/\kappa^2}k_nk_mk_ok_p\hat C_F(\ve k)	= \\\Sigma^2(\kappa)\left(\delta_{nm}\delta_{op} + \delta_{no}\delta_{mp} + \delta_{np}\delta_{mo}\right)\,, \eqnlab{cint2}
\end{multline}
which is independent of $\gamma$. Summing over $n=m=p=o$, and turning to spatial polar coordinates, we obtain a one-dimensional integral expression for $\Sigma^2(\kappa)$ in terms of $\rho(r^2)$:
\algn{
	\Sigma^2(\kappa)  &= \frac{2^{-(d+5)}}{\Gamma \left(\frac{d}{2}+2\right)} \!\!\int_{0}^\infty\!\!\!\!\ed r\, e^{-\frac{r^2}{4}}f_d(r)\rho\left(\frac{r^2}{\kappa^2}\right)\,, \eqnlab{sigint}
}
with the function
\algn{
	f_d(r) = r^{d-1}    \left[r^4 - 4(d+2)(r^2-d)\right]\,,
}
valid for $d\geq2$. In particular, for Gaussian correlations $\Sigma(\kappa)$ takes the explicit form
\algn{\eqnlab{SigGauss}
	\Sigma^2_G(\kappa) = \frac{\pi^2\kappa^d}{4(\pi + \kappa^2)^{\frac{d}2+2}}\,.
}
Furthermore, using \Eqsref{cint} and \eqnref{cint2}, we write the tensor $\ms{C}$ in \Eqnref{H1corr} as
\sbeqs{
\algn{
	\ms{C}_{nmop}(\gamma) &= \ms{A}\delta_{no}\delta_{mp} + \ms{B}\left(\delta_{np}\delta_{mo}+\delta_{nm}\delta_{op}\right) \,, \eqnlab{cexpr}
}
with
\algn{
	\ms{A} = 1+d\gamma\,,\qquad \ms{B} = 1-2\gamma \,.
}
}
To leading (linear) order in the disorder strength $\zeta_T$, the expressions \eqnref{sigint} and \eqnref{cexpr} for $\Sigma(\kappa)$ and $\ms{C}$ fully determine the statistical properties of the eigenvalue shift $\Delta\lambda_1$. Each realisation of the disordered force field $\ve F$ generates through \Eqnref{Hnmdef} a realisation of the Gaussian random matrix $\mbb{H}$, whose distinct eigenvalues determine the leading-order eigenvalue shifts compared with the unperturbed case. In particular, each realisation of $\ve F$ lifts the $d$-fold degeneracy of the unperturbed eigenvalue $\lambda^0_1$, which is split into the $d$ individual (random) eigenvalues of $\mbb{H}$.

In the long-time limit, the eigenvalue shift with the smallest real part dominates, meaning that for determining the asymptotic relaxation speed, we need to consider the eigenvalue of $\mbb{H}$ with the smallest real part. The eigenvalues of $\mbb{H}$ solve the characteristic equation
\algn{
	\det\left[\zeta_T\mbb{H}(\kappa,\gamma)-\Delta\lambda_1\mbb{1}	\right] = 0\,,
}
which we rewrite using \Eqnref{H1corr} in terms of rescaled quantities as
\algn{\eqnlab{Dlam1deg}
	\det\left[\mbb{\tilde H}(\gamma)-\Delta{\tilde\lambda}_1\mbb{1}	\right] = 0\,,
}
where
\algn{\eqnlab{Hresc}
	\mbb{\tilde H}(\gamma) = \frac{\mbb{H}}{\Sigma(\kappa)d}\,,\quad \Delta{\tilde\lambda}_1 = \frac{\Delta{\lambda}_1}{\Sigma(\kappa)d\zeta_T}\,.
}
Interestingly, after rescaling, $\mbb{\tilde H}(\gamma)$ solely depends on $\gamma$ and $d$, but is independent of the correlation length $\kappa$. In other words, for isotropic backgrounds, the dependence of the eigenvalue shift on the correlation length $\kappa$ is entirely encoded in $\Sigma(\kappa)$, which can be scaled out. Because $\Sigma(\kappa)>0$, this immediately rules out sign changes of the mean eigenvalue shift as a function of $\kappa$. Therefore, the asymptotic dynamics in isotropic and anisotropic backgrounds are fundamentally different.

After this rescaling, $\Delta{\tilde\lambda}_1$ is determined by the eigenvalue with the smallest real part of the random matrix $\mbb{\tilde H}(\gamma)$ with zero mean and correlation
\algn{\eqnlab{tHcorr}
	\langle \tilde H_{nm}\tilde H_{op}\rangle_{F} = \frac{1+d\gamma}{d^2}\left[\delta_{no}\delta_{mp} + \tau\delta_{np}\delta_{mo}+\tau\delta_{nm}\delta_{op}\right]\,,
}
where
\algn{\eqnlab{tau}
	\tau = \frac{1-2\gamma}{1+d\gamma}
}
denotes the ``correlation parameter'' which for fixed $d$ ranges between $\tau \in [\tau_\text{min},\tau_\text{max}]$, where $\tau_\text{min} = -1/(1+d)>-1$ and $\tau_\text{max} = 1$. These values are assumed for $\gamma=1$ and $\gamma=0$, respectively. The correlation parameter $\tau$ determines whether individual entries of the matrix are uncorrelated ($\tau = 0$), positively correlated ($\tau>0$), or negatively correlated ($\tau<0$). The extreme cases $\tau=+1$ and $\tau=-1$ correspond to symmetric and antisymmetric matrices, respectively. This explains why we have $\tau = \tau_\text{max}$ for $\gamma=0$: In this case the forces are entirely conservative and the dynamics fulfills detailed balance. Thus, the eigenvalues of $\mbb{\tilde H}$ and $\mbb{H}$ are real, which is guaranteed when $\mbb{\tilde H}$ is symmetric.

The correlation structure in \Eqnref{tHcorr} is that of a matrix from the Lehmann-Sommers Gaussian ensemble of matrices $\mbb{J}$ with correlation parameter $\tau$ treated in Ref.~\cite{Leh91}, plus an independent Gaussian identity contribution, i.e.,
\algn{\eqnlab{tHdec}
	\mbb{\tilde H} = \sqrt{\frac{1+d\gamma}{d}}(\mbb{J} + \sqrt{|\tau|}X \mbb{1})\,,
}
where $\langle J_{mn}X\rangle_F=0$ for all $m,n$, and
\algn{\eqnlab{Jcorr}
	\langle J_{nm}J_{op}\rangle_F = \frac1d\left(\delta_{no}\delta_{mp} + \tau\delta_{np}\delta_{mo}\right)\,,\quad \langle X^2\rangle_F = \frac{1}d\,.
}
For $\tau>0$, the additional identity contribution in \Eqnref{tHdec} leads to a random shift of all eigenvalues by the Gaussian random variable $X$, compared to the ensemble studied in Ref.~\cite{Leh91}. The distribution of $\Delta\tilde\lambda_1$ is thus obtained by a scaled convolution of the eigenvalue with the smallest real part of $\mbb{J}$ with a Gaussian distribution with zero mean and variance $|\tau|/d$. However, the parametrisation~\eqnref{tHdec} works only for $\tau\geq0$, i.e., $\gamma\leq1/2$. The reason is that for  $\tau<0$ ($\gamma>1/2$) the variance of the trace of $\mbb{\tilde H}$ is \textit{reduced} compared to that of $\mbb{J}$, which cannot be achieved by adding an independent Gaussian variable, as done in \Eqnref{tHdec}. Consequently, we cannot readily apply the results developed in Ref.~\cite{Leh91} to the current matrix ensemble but require an additional step.

This additional step involves first removing the trace from the Lehmann-Sommers ensemble altogether and then adding the correctly sized trace contribution, in a representation similar to \Eqnref{tHdec}. To this end, we first define the \textit{tracefree} Lehmann-Sommers ensemble as
\algn{\eqnlab{J0}
	\mbb{J}^0 = \mbb{J} - \frac{\tr (\mbb{J})}d \mbb{1}\,.
}
The correlations for $\mbb{J}^0$ read
\algn{
	\langle J^0_{nm}J^0_{op}\rangle_F = \frac1d\left[\delta_{no}\delta_{mp} + \tau\delta_{np}\delta_{mo} - \left(\frac{1+\tau}{d}\right)\delta_{nm}\delta_{op}\right] \,.
}
Now, starting from the tracefree Lehmann-Sommers ensemble, we are able to add an arbitrary trace term analogous to \Eqnref{tHdec} to incorporate the non-standard trace of our matrix ensemble, including for $\tau<0$. This parametrisation of $\tilde{\mbb{H}}$ then reads
\algn{\eqnlab{tHdec2}
	\mbb{\tilde H} = \sqrt{\frac{1+d\gamma}{d}}\left[\mbb{J}^0 + \sqrt{1+\tau(1+d)}Y \mbb{1}\right]\,,
}
where $\langle Y^2\rangle = 1/d^2$ and $\langle \mbb{J}^0 Y\rangle = 0$. Note that
\algn{
	1+\tau(1+d) = (1-\gamma)\frac{d+2}{1+d\gamma}
}
vanishes at $\gamma=1$, so that the trace term in \Eqnref{tHdec2} is absent and $\mbb{\tilde H}$ is \textit{exactly} traceless in every dimension. This is the matrix-level expression of the fact that a purely non-conservative force field is divergence-free.

To make use of the parametrisation~\eqnref{tHdec2} and compute the density of eigenvalues of $\mbb{\tilde H}$, we first determine the distribution of eigenvalues for the tracefree Lehmann-Sommers ensemble $\mbb{J}^0$.

Before doing so, we review the standard Lehmann-Sommers ensemble, for which the joint distribution of eigenvalues $\ve{\xi}=(\xi_1,\ldots,\xi_d)^{\sf T}$ is known~\cite{Leh91}. For a given number $R$ of real eigenvalues, the joint density of eigenvalues of $\mbb{J}$ was derived in Ref.~\cite{Leh91} as
\begin{multline}\eqnlab{ev_dens_xi}
	P^R_{\ve \xi}(\ve{\xi}) = K_d(\tau) \prod^d_{i>j}|\xi_i-\xi_j|\\
	\times\prod_{i=1}^R\ee^{-\frac{d\,\xi_i^2}{2(1+\tau)}} \prod_{j=1}^Q2\ee^{-\frac{d(\eta^2_j-\nu_j^2)}{1+\tau}}\erfc\left(\frac{\sqrt{2d}\,\nu_j}{\sqrt{1-\tau^2}}\right)\,,
\end{multline}
where $Q$ is the number of complex-conjugate pairs, so that $d= R + 2Q$, and $\xi_i = \eta_i +i\nu_i$. The constant $K_d(\tau)$ reads
\algn{
	K_d(\tau) = (1+\tau)^{-\frac{d}2}\left(\frac{d}{2}\right)^{\frac{d(d+1)}4}\prod_{n=1}^d\frac{1}{\Gamma\left(\frac{n}2\right)}\,.
}
The complete joint density of eigenvalues of $\mbb{J}$ is then given by the sum over all possible $0\leq R\leq d$:
\algn{
	P_{\ve \xi}(\ve{\xi}) = \sum_{R}P^R_{\ve \xi}(\ve{\xi})\,.
}
To pass to the density of the tracefree Lehmann-Sommers ensemble, we consider the joint distribution $P(\ve{\xi},T)$ of the eigenvalues $\ve{\xi}$ and the trace $T= \sum_{j=1}^d\xi_j$, which reads
\algn{
	P_{(\ve \xi,T)}(\ve{\xi},T) = P_{\ve \xi}(\ve{\xi})\delta\left(T-\sum_{i=1}^d \xi_i\right).
}
We now transform into the new coordinates $\xi^0_i$ given by
\algn{\eqnlab{x0trafo}
	\xi^0_i = \xi_i-\frac{T}d\,,
}
for $i=1,\ldots,d$, so that the sum of $\ve \xi^0$ vanishes. This transformation has a unit Jacobian and allows us to write $P(\ve \xi,T)$ in terms of $\ve \xi^0$ as the product of two contributions
\algn{\eqnlab{Pxi0}
	P_{(\ve \xi^0,T)}(\ve \xi^0,T) = P_{\ve \xi^0}(\ve \xi^0)\frac{\ee^{-\frac{T^2}{2(1+\tau)}}}{\sqrt{2\pi(1+\tau)}}\,,
}
where
\algn{
	P_{\ve \xi^0}(\ve \xi^0) = \sqrt{2\pi(1+\tau)}P(\ve \xi^0)\delta\left(\sum_{i=1}^d \xi^0_i\right)\,,
}
represents the density of eigenvalues of the \emph{traceless} Lehmann-Sommers ensemble $\mbb{J}^0$ given in \Eqnref{J0}. Hence, after the transformation~\eqnref{x0trafo}, the joint distribution $P_{(\ve \xi^0,T)}(\ve \xi^0,T)$ factorises into the distribution $P^0(\ve \xi^0)$ of the traceless Lehmann-Sommers ensemble and the Gaussian distribution of the trace. The factorisation of the joint probability distribution in \Eqnref{Pxi0} expresses the fact that the trace of a rotationally invariant ensemble can be manipulated by adding an independent Gaussian variable multiplied by the identity matrix, as in the decomposition~\eqnref{tHdec2}.

Using this, we now replace the distribution of the Lehmann-Sommers trace with variance $1+\tau$ in \Eqnref{Pxi0} by the complete trace with variance $1+\tau(d+1)$ and invert the transformation~\eqnref{x0trafo}. This way, we circumvent the previous problem with the decomposition~\eqnref{tHdec}, which did not work for $\tau<0$. After integrating out the variable $T$, we obtain the joint eigenvalue density of $\ve{\Delta\tilde \lambda}=(\Delta\tilde\lambda_1,\ldots,\Delta\tilde\lambda_d)^{\sf T}$
\algn{\eqnlab{joint_density}
	P_{\ve{\Delta\tilde\lambda}}(\ve{\Delta\tilde\lambda}) = \ms{N}(\tau) P_{\ve \xi}\left(\sqrt{\frac{d}{1+d\gamma}}\ve{\Delta\tilde \lambda}\right)\exp\left[\frac{d\left(\sum_{i=1}^d\Delta\tilde\lambda_i\right)^2}{2(1+d\gamma)\Delta\sigma^2}\right]\,,
}
where
\algn{
	\Delta\sigma^2 &= \frac{[1+\tau(d+1)](1+\tau)}{\tau d}\,,\\
	\ms{N}(\tau) &= \left(\frac{d}{1+d\gamma}\right)^{\frac{d}2}\sqrt{\frac{1+\tau}{1+\tau(d+1)}}\,.
}
Hence, by adjusting the trace of the Lehmann-Sommers ensemble, we have derived the joint density of eigenvalues of the rescaled perturbation matrix $\mbb{\tilde H}$.

In order to obtain the density of the smallest real part $\mu_1$, we need to marginalise the joint distribution~\eqnref{joint_density} over all remaining eigenvalues. This can be done explicitly in two dimensions $d=2$, where we have for \Eqnref{ev_dens_xi} the two cases $R=2$ and $R=0$ with
\algn{
	P^2_{\ve \xi}(\eta_1,\eta_2) &= K_2(\tau) |\eta_2-\eta_1|\ee^{-\frac{\eta_1^2+\eta_2^2}{1+\tau}}\,,	\\
	P^0_{\ve \xi}(\eta_1,\nu_1) &= 4K_2(\tau) |\nu_1|\ee^{-\frac{2(\eta^2_1-\nu_1^2)}{1+\tau}}\erfc\left(\frac{2\nu_1}{\sqrt{1-\tau^2}}\right)\,,
}
and $K_2(\tau) =1/[\sqrt{\pi}(1+\tau)]$. We change the trace of these contributions according to \Eqnref{joint_density} and marginalise over $\mu_2>\mu_1$ and $\omega_1$ [cf. \Eqnref{realimag}] to obtain
\algn{
	P^2_{\Delta\tilde\mu_1}(\Delta\tilde\mu_1) &= \frac{2\ee^{-\frac{\Delta\tilde\mu_1^2}{1-\gamma}}}{\sqrt{\pi}(3-2\gamma)}\sqrt{\frac{1-\gamma}{1+2\gamma}}\\\nn
	&\times\left(1-\frac{\sqrt{\pi}\Delta\tilde\mu_1\ee^{\frac{\Delta\tilde\mu_1^2}{(1-\gamma)(3-2\gamma)}}\erfc\left[\frac{\Delta\tilde\mu_1}{\sqrt{(1-\gamma)(3-2\gamma)}}\right]}{\sqrt{(1-\gamma)(3-2\gamma)}}\right)\,,\\
	P^0_{\Delta\tilde\mu_1}(\Delta\tilde\mu_1) &= \frac{\ee^{-\frac{\Delta\tilde\mu_1^2}{1-\gamma}}}{\sqrt{\pi(1-\gamma)}
	}\left(1-\frac{1}{\sqrt{1+2\gamma}}\right)\,.\eqnlab{P0d2}
}
Note that for $\tau=1$, $\mbb{J}$ is symmetric and the contribution $P^0_{\Delta\tilde\mu_1}$, which corresponds to the realisations with $Q=1$ complex-conjugate pair, vanishes identically.

\begin{figure}
	\includegraphics[width=\linewidth]{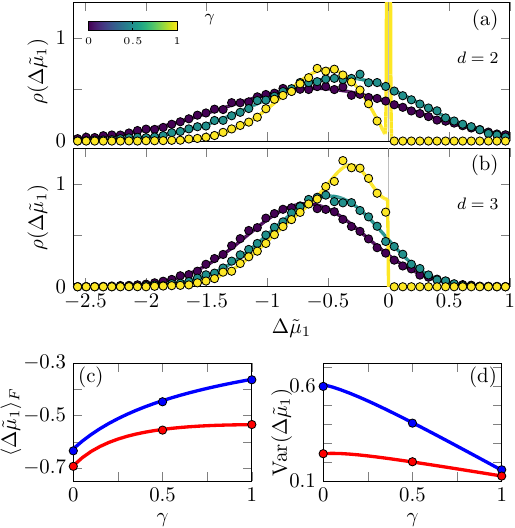}
	\caption{Probability density and first cumulants of $\Delta\tilde\mu_1$ in an isotropic background for different $\gamma$ (colour bar). The symbols are from numerical solutions of the Fokker-Planck equation with $2.5\cdot10^4$ realisations of the random forces, the lines from random-matrix theory. (a) Probability density $P(\Delta\tilde\mu_1)$ in $d=2$. The narrow peak at $\Delta\tilde\mu_1=0$ for $\gamma=1$ is an atom of weight $1-1/\sqrt3\simeq0.42$, discussed in the text. (b) As (a), for $d=3$, where no atom occurs and the density terminates at $\Delta\tilde\mu_1=0$. (c) Average $\langle\Delta\tilde\mu_1\rangle_F$ as a function of $\gamma$ for $d=2$ (blue) and $d=3$ (red). (d) Variance $\text{Var}(\Delta\tilde\mu_1)$ as a function of $\gamma$, same colours.}\figlab{prob3d}
\end{figure}
Beyond $d=2$ we have not been able to carry out the marginalisation in closed form. But \Eqnref{joint_density} remains useful, because we can marginalise it numerically to obtain $P(\Delta\tilde \mu_1)$ to high accuracy.  Figures~\figref{prob3d}(a) and~(b) show $P(\Delta\tilde\mu_1)$ obtained in this way for $d=2$ and $d=3$, together with the full numerical densities obtained from solving the eigenvalue problem~\eqnref{Leig} of the Fokker-Planck operator~\eqnref{FPop} for $\sim10^4$ realisations. In good agreement between the theory and simulations, we find that the distributions are quite broad, meaning that in low dimensions, the disorder shifts the relaxation rate by an amount that varies substantially between realisations. Although the mean is negative for every $\gamma$, a finite fraction of realisations relaxes \emph{faster} than the unperturbed background for $\gamma<1$. Panels~(c) and~(d) show the mean and variance as functions of $\gamma$. The mean rises towards zero as the forcing is made non-conservative, while the variance falls, in both dimensions.

Purely non-conservative forcing in $d=2$ is a special case. At $\gamma=1$ the matrix $\mbb{\tilde H}$ is traceless, so that a complex-conjugate pair of eigenvalues has \emph{both} real parts equal to zero. Whenever the two eigenvalues are complex, which for this ensemble happens with probability $1-1/\sqrt3\simeq0.42$, the shift therefore vanishes identically, and the density in \Figref{prob3d}(a) carries a delta function of that weight at $\Delta\tilde\mu_1=0$ in addition to its continuous part. This is the projection onto the real axis of the eigenvalue density concentrated on the imaginary axis in \Figref{numsimiso}(e). This implies that for purely solenoidal forces in two dimensions, the disorder leaves the relaxation rate unchanged at this order with probability $\approx0.42$. No such atom exists in $d=3$, where three traceless eigenvalues cannot all have vanishing real part, so that $\Delta\tilde\mu_1<0$ in every realisation and the density in \Figref{prob3d}(b) instead terminates abruptly at $\Delta\tilde\mu_1=0$.

The joint density~\eqnref{ev_dens_xi} also determines how the $d$ levels arrange relative to one another, that is, the level repulsion and attraction introduced in \Secref{perturbation}. Its factor $\prod_{i>j}|\xi_i-\xi_j|$ vanishes whenever two eigenvalues coincide. Two real eigenvalues therefore repel each other, and so do the two members $\xi$ and $\xi^*$ of a complex-conjugate pair, because $|\xi-\xi^*|=2|\nu|$. The factor involving the complementary error function, by contrast, suppresses complex-conjugate pairs as $\tau\to1$ and removes them altogether at $\gamma=0$. To see how these two effects compete, we consider the distance $s=\lambda_2-\lambda_1$ between the two lowest levels, ordered by their real parts. Because $s$ is unchanged when a multiple of the identity is added to $\mbb{\tilde H}$, its density follows from \Eqnref{ev_dens_xi} without the trace correction in \Eqnref{joint_density}. In $d=2$, integrating $P^2_{\ve\xi}$ and $P^0_{\ve\xi}$ over the centre of the two eigenvalues gives
\algn{\eqnlab{spacing2d}
	\rho(\mathrm{Re}\,\tilde s) &= \left(1-\frac{1}{\sqrt{1+2\gamma}}\right)\delta(x)+\frac{x\,\ee^{-x^2/2}}{\sqrt{1+2\gamma}}\,,\nn\\
	\rho(|\mathrm{Im}\,\tilde s|) &= \frac{\delta(y)}{\sqrt{1+2\gamma}}+\frac{y\,\ee^{y^2/2}}{\sqrt{1+2\gamma}}\,\erfc\left(y\sqrt{\frac{1+2\gamma}{4\gamma}}\right)\,,
}
with $x=\mathrm{Re}\,\tilde s$, $y=|\mathrm{Im}\,\tilde s|$ and $\tilde s=s/[\Sigma(\kappa)d\zeta_T]$ rescaled as in \Eqnref{Hresc}. The delta functions collect the realisations with a complex-conjugate pair, for which $\mathrm{Re}\,\tilde s=0$, and those with two real levels, for which $\mathrm{Im}\,\tilde s=0$. At $\gamma=0$ the second line reduces to $\delta(y)$, because all levels are real. When both levels are real, their distance thus follows the Wigner surmise of the Gaussian orthogonal ensemble~\cite{Meh04}, whose shape does not depend on $\gamma$. Non-conservative forces therefore do not weaken the repulsion between two real levels. Instead, they turn a growing fraction $1-1/\sqrt{1+2\gamma}$ of the realisations into complex-conjugate pairs, whose levels have attracted each other until their real parts coincide, $\mathrm{Re}\,s=0$. In the language of \Secref{perturbation}, these are the realisations with $4\tilde H_{12}\tilde H_{21}<-\delta^2$, in which the two levels lie beyond the exceptional point. Figure~\figref{numsimisospacing}(a) shows that the simulations agree with \Eqnref{spacing2d} for $d=2$ and, for $d=3$, with the densities obtained by sampling the ensemble~\eqnref{tHcorr}~\footnote{Note that in $d=3$ and for $\gamma>0$, the density $\rho(\mathrm{Re}\,\tilde s)$ remains finite as $\tilde s\to0$ [\Figref{numsimisospacing}(b)]. This is because in a fraction of the realisations, $\lambda_1$ is real while $\lambda_2$ and $\lambda_3$ form a pair, and the factor $|\xi_1-\xi_2||\xi_1-\xi_3|$ does not vanish when only the real parts coincide.}.
\begin{figure}
	\includegraphics[width=\linewidth]{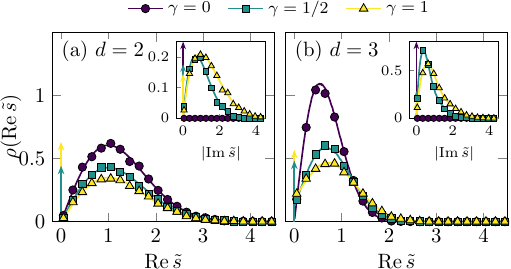}
	\caption{Repulsion and attraction of the levels in an \textit{isotropic} background: densities $\rho(\mathrm{Re}\,\tilde s)$ of the real part of the rescaled distance $\tilde s=(\lambda_2-\lambda_1)/[\Sigma(\kappa)d\zeta_T]$ between the two lowest levels for the data of \Figref{numsimiso}, in (a) $d=2$ and (b) $d=3$, for $\gamma=0$ (circles), $\gamma=1/2$ (squares) and $\gamma=1$ (triangles). The insets show the densities of $|\mathrm{Im}\,\tilde s|$. The lines show results of random-matrix theory, \Eqnref{spacing2d} for $d=2$ and sampling from~\eqnref{tHcorr} for $d=3$. Due to the formation of complex-conjugate pairs with $\text{Re}(\tilde s)=0$, and completely real eigenvalues with $\text{Im}(\tilde s)=0$, finite probabilities can be localised in delta functions, shown as arrows whose lengths give an estimate of the weight.}
	\figlab{numsimisospacing}
\end{figure}

In summary, in an isotropic background, the disorder generically \emph{slows} relaxation down already at \emph{first} order in $\zeta_T$. Because the unperturbed level is $d$-fold degenerate, the disorder lifts it already at leading order, which is why the shift is linear in $\zeta_T$ rather than quadratic as for an anisotropic background (\Secref{aniso}). But lifting a degeneracy splits the level in both directions, and the asymptotic relaxation rate is set by the branch with the smallest real part, so that the selected eigenvalue is systematically biased downwards. By this mechanism, $\langle\Delta\tilde\mu_1\rangle_F$ is negative for every $\gamma$, \Figref{prob3d}(c). Non-conservative forcing counteracts this mechanism through the phenomenon of ``level attraction''~\cite{Hei12} which creates complex-conjugate pairs: increasing $\gamma$ raises the mean and lowers the variance [\Figsref{prob3d}(c) and~(d)], and at $\gamma=1$ in $d=2$ it makes the shift vanish identically in a finite fraction of realisations. While it remains the exception, we note that individual realisations may still relax faster, as the weight at $\Delta\tilde\mu_1>0$ in \Figsref{prob3d}(a) and~(b) shows. In fact, as we show next, faster relaxation behaviour in an isotropic background becomes increasingly unlikely as the dimension $d$ grows.

\begin{figure}
	\includegraphics[width=\linewidth]{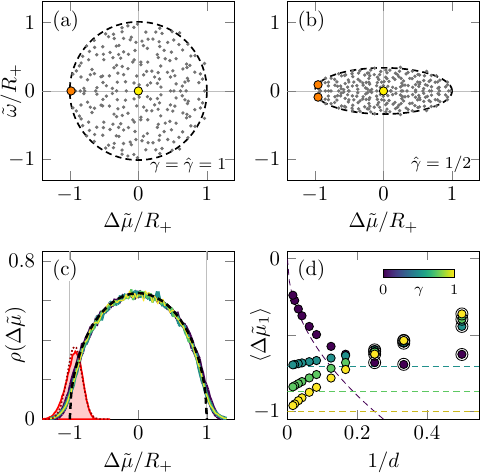}
	\caption{Spectrum of $\mbb{\tilde H}$ in high dimensions. (a) Eigenvalues $\Delta\tilde\lambda$ in the complex plane, $\Delta\tilde\mu=\mathrm{Re}\,\Delta\tilde\lambda$ and $\tilde\omega=\mathrm{Im}\,\Delta\tilde\lambda$ (grey), for a single realisation of the ensemble~\eqnref{tHcorr} with $d=256$ and $\gamma=1$, in units of the finite-$d$ edge $R_+$ of \Eqnref{ellipse}. The dashed line is the elliptic law~\eqnref{ellipse}, here a disc up to corrections of order $1/d$, the yellow bullet marks the unperturbed level and the orange bullet the eigenvalue with the smallest real part. (b) Same as (a), but at fixed force variance with $\hat\gamma=1/2$ [\Eqnref{hatgamma}], i.e., $\gamma=1/d$, where $\tau\approx1/2$ and the eigenvalues fill an ellipse. The smallest real part here belongs to a complex-conjugate pair. (c) Density $\rho(\Delta\tilde\mu)$ of the real parts of \textit{all} eigenvalues at $d=32$, in units of $R_+$, for $\gamma=0$, $1/2$, $3/4$ and $1$ [colour bar in (d)], drawn with decreasing line width. The dashed curve is the semicircle law. The red lines show the density of the smallest real part, divided by $10$, for $\gamma=1/2$ (filled) and $\gamma=0$ (dotted). (d) Mean $\langle\Delta\tilde\mu_1\rangle_F$ against $1/d$ for $\gamma=0$, $1/2$, $3/4$ and $1$ (colour bar). The dashed horizontal lines are the limits $-\sqrt\gamma$ in \Eqnref{edge2}, the dashed curve is $-2/\sqrt d$ of \Eqnref{edge2}. Filled symbols are from sampling, open circles from numerical integration of \Eqnref{joint_density}, where both are available.}\figlab{circle}
\end{figure}
\subsection{High-dimensional limit $d\gg1$}
We now consider the limit of high dimensions $d\gg1$. The correlation parameter~\eqnref{tau} which determines the correlations between the entries of the (rescaled) perturbation matrix $\mbb{\tilde H}$ then behaves as
\algn{
	\tau \sim \css{
	1\,, & \gamma=0\\
	\frac{1-2\gamma}{\gamma}\,\frac1d\,, & \gamma>0\,,}
}
showing that $\tau$ vanishes asymptotically as $d^{-1}$ for $\gamma>0$ while it remains at unity for all $d$ when $\gamma=0$. For $\gamma>0$ the entries of $\mbb{\tilde H}$ thus decorrelate with increasing dimension, and $\mbb{\tilde H}$ approaches a real Ginibre matrix~\cite{Gin65}, irrespective of the ratio of conservative and non-conservative parts. Only the strictly conservative case $\gamma=0$ escapes this, because there $\tau=1$ independently of $d$. The reason is that the disorder forces then derive from a potential so that $\mbb{\tilde H}$ is symmetric in every dimension $d$ and for every individual realisation.

The case $\gamma=1/2$ is also special, but in the sense that the entry correlations vanish, $\tau=0$, \textit{exactly} at every $d$. This can be seen from~\Eqnref{cexpr}, where the coefficient $\ms{B}=1-2\gamma$ multiplying the two correlated index pairings vanishes at $\gamma=1/2$, so that \Eqnref{tHcorr} reduces to
\algn{\eqnlab{tHcorrhalf}
	\langle \tilde H_{nm}\tilde H_{op}\rangle_{F} = \frac{d+2}{2d^2}\,\delta_{no}\delta_{mp}\,,
}
that is, the $d^2$ entries of $\mbb{\tilde H}$ are independent Gaussian variables of common variance $(d+2)/(2d^2)$, with no correlation between an entry and its transpose. This correlation structure corresponds to the real Ginibre ensemble~\cite{Gin65}, with spectral radius $\sqrt{d\,(d+2)/(2d^2)}=\sqrt{1/2+1/d}$. Consequently, when conservative and non-conservative forces contribute equally, $\mbb{\tilde H}$ becomes a real Ginibre matrix for all $d$ and not solely in the limit $d\to\infty$. %

For large dimensions $d$, the eigenvalue spectrum of $\mbb{\tilde H}$ therefore splits into two cases: For any $\gamma>0$ the matrix $\mbb{\tilde H}$ becomes a large real Ginibre matrix, whose eigenvalues fill a disc in the complex plane with uniform density~\cite{Gin65,Gir85,For10}. This disc is shown for $\gamma=1$ in \Figref{circle}(a) for a single realisation at $d=256$, where the eigenvalues fill it essentially uniformly. The eigenvalues on the real axis, $12$ of $256$ here, are the finite fraction of eigenvalues that remain real, a hallmark of the \emph{real} Ginibre ensemble at finite $d$ that the asymptotic laws do not resolve~\cite{Leh91}. The disc's radius is obtained from the prefactor in \Eqnref{tHcorr}~\cite{Gin65,Meh04} as $\sqrt{(1+d\gamma)/d}$, up to corrections of order $1/d$, which approaches $\sqrt{\gamma}$ as $d\to\infty$.

The circular law is a statement about the density of \emph{all} eigenvalues. Figure~\figref{circle}(c) shows the density of their real parts at $d=32$, measured in units of the finite-$d$ edge $R_+$, given in \Eqnref{ellipse} below. For every $\gamma>0$ it is close to a semicircle on $[-R_+,R_+]$, because integrating a uniform density over the imaginary direction returns a semicircle whatever the shape of the ellipse. The asymptotic relaxation speed, however, is set by the single eigenvalue with the smallest real part, whose density is the filled red curve in the same panel for $\gamma=1/2$. That density is strongly pinned to the left edge of the bulk, consistent with the location of the orange bullet of the single realisation in \Figref{circle}(a). Due to this pinning, $\langle\Delta\tilde\mu_1\rangle_F$ approaches $-\sqrt\gamma$, the limit of $-R_+$, as \Figref{circle}(d) shows.

We now consider the second case $\gamma=0$, where $\mbb{\tilde H}$ is symmetric for each realisation and at every $d$, because the corresponding forces derive from a potential. In this case, all eigenvalues are real, and the spectrum collapses onto an interval of the real axis, with the Wigner semicircle density~\cite{Wig55,Meh04} as the limiting distribution of eigenvalues. Its edge does not saturate but shrinks as $2/\sqrt d$. Measured in units of $R_+$, the density of the real parts falls onto the same semicircle as for $\gamma>0$, see \Figref{circle}(c). The smallest eigenvalue, dotted red in \Figref{circle}(c), is again pinned to its left edge $-R_+$, which now, however, shrinks with $d$, so that $\langle\Delta\tilde\mu_1\rangle_F$ converges to zero rather than to a finite value, see the $\gamma=0$ markers in \Figref{circle}(d).

The behaviours of both branches can be expressed as limiting cases of a single expression. Since $\tau$ interpolates between symmetric and antisymmetric matrices, the spectral density of $\mbb{\tilde H}$ at large but finite $d$ is given by the elliptic law~\cite{Som88,For10}: the eigenvalues fill, with uniform density, an ellipse in the complex plane centred at the origin with semi-axes
\algn{\eqnlab{ellipse}
	R_\pm = (1\pm\tau)\sqrt{\frac{1+d\gamma}{d}}
}
along the real and the imaginary direction, respectively. For $\tau\to0$ the ellipse becomes the circular-law disc, and for $\tau=1$ it collapses onto the real interval $[-R_+,R_+]$ with the Wigner semicircle density, consistent with the fact that $\mbb{\tilde H}$ is then symmetric. The dashed lines in \Figsref{circle}(a) and~(b) show this ellipse at $d=256$, for $\gamma=1$, where $\tau=-1/(d+1)$ makes it a disc up to corrections of order $1/d$, and for $\hat\gamma=1/2$, where $\tau=(d-2)/(2d)\approx1/2$. Because $\tau\to0$ and $\tau=1$ for $\gamma>0$ and $\gamma=0$, respectively, we use \Eqnref{ellipse} below only as a convenient way of treating both branches at once. The asymptotic statements about the eigenvalue densities are the circular and semicircle laws described above.

The eigenvalue with the smallest real part is located at the left-most point of the ellipse, so that
\algn{\eqnlab{edge}
	\Delta\tilde\mu_1 \sim -R_+ = -\frac{2+(d-2)\gamma}{\sqrt{d(1+d\gamma)}}\,,
}
where we used \Eqnref{ellipse} together with the definition of $\tau$. The limit~\eqnref{edge} is deterministic, i.e., $\Delta\tilde\mu_1=\langle\Delta\tilde\mu_1\rangle_F$, because the fluctuations of $\Delta\tilde\mu_1$ around its mean vanish quickly as $d$ grows. The reason is that the eigenvalues of $\mbb{\tilde H}$ are not independent but can be described as a strongly correlated Coulomb gas~\cite{Meh04}, in which each eigenvalue is confined by its neighbours. An eigenvalue close to the outer edge of the ellipse can therefore not deviate from the edge by significantly more than the local level spacing of size $d^{-1/2}$~\cite{Meh04}. In this respect, the high-dimensional case is very different from the low-dimensional case, where the distribution of $\Delta\tilde\mu_1$ is broad, see \Figref{prob3d}.

The limit~\eqnref{edge} implies the two qualitatively different asymptotic behaviours
\algn{\eqnlab{edge2}
	\Delta\tilde\mu_1 \sim\css{
	-\frac{2}{\sqrt d}\,,	& \gamma=0 \\
	-\sqrt{\gamma}\,,	& \gamma>0\,.}
}
Hence, for any non-vanishing non-conservative contribution, $\gamma>0$, $\Delta\tilde\mu_1$ tends to the finite negative value $-\sqrt\gamma$,
whereas for strictly conservative forces it shrinks as $d^{-1/2}$. The difference between the two cases originates from the coefficient $\ms{A}=1+d\gamma$ in \Eqnref{cexpr}, which sets the variance of the entries of $\mbb{\tilde H}$. It grows linearly with the dimension whenever $\gamma>0$, but is constant for conservative forces, $\gamma=0$. This growth reflects the $d-1$ solenoidal polarisation directions discussed in \Secref{dimensionless}. At fixed $\gamma$, each of them carries the same weight, so that the variance of the solenoidal forces grows with $d$ and the spectral radius saturates instead of shrinking.

To turn \Eqsref{edge} and \eqnref{edge2} into statements about the physical shift we undo the rescaling of \Eqnref{Dlam1deg}, $\Delta\lambda_1 = \Sigma(\kappa)d\zeta_T\Delta\tilde\mu_1$. The covariance factor $\Sigma(\kappa)$ depends on the dimension $d$. For Gaussian correlations it is given by \Eqnref{SigGauss}, which is the representative case, since by Schoenberg's theorem~\eqnref{schoenberg} every admissible isotropic correlation function becomes a superposition of Gaussians as $d\to\infty$~\cite{Yag86}.

At any \textit{fixed} correlation length, \Eqnref{SigGauss} decays exponentially in the dimension, $\Sigma^2_G\propto[\kappa^2/(\pi+\kappa^2)]^{d/2}$, so that the disorder-induced shift disappears in high dimensions unless the correlation length is scaled with $d$, analogous to the anisotropic background of \Secref{anisohighd}. In fact, this decay is governed by the same factor, because \Eqnref{Rmain} reduces at $\mbb{a}=\mbb{1}$ to $\hat{\ms R}=[1+\pi/\kappa^2]^{-d/2}$, in terms of which \Eqnref{SigGauss} reads
\algn{\eqnlab{SigR}
	\Sigma^2_G(\kappa) = \frac{\pi^2}{4}\,\frac{\hat{\ms R}}{(\pi+\kappa^2)^2}\,.
}
In isotropic backgrounds we have $\ell^2_\text{rms} = d\,\ell^2_\text{T}$, so that $\hat\kappa = \kappa/\sqrt d$ at $\mbb{a}=\mbb{1}$. The combination $\Sigma(\kappa)d$ that enters \Eqnref{Dlam1deg} then behaves for $d\to\infty$ as
\algn{\eqnlab{Siscal}
	\Sigma_G(\kappa)\,d \sim \Psi(\hat\kappa,1)\,,
}
which is the same scaling function $\Psi$ as in \Eqnref{sigscalmain}, evaluated for an isotropic background, $\bar a=1$. Correspondingly, as in \Eqsref{muscalmain} and \eqnref{sigscalmain}, the shift at fixed $\hat\kappa$ is independent of the dimension. The scaling function $\Psi$ has a single maximum,
\algn{\eqnlab{kapmax}
	\hat\kappa^2_{\text{max},\sigma} = \frac{\pi}{4}\,,\qquad \Psi(\hat\kappa_{\text{max},\sigma},1) = \frac2{\ee}\,,
}
where the variance attains its largest value. We hence obtain for $\langle\Delta\lambda_1\rangle_F$ at large $\kappa$ and $d$,
\algn{\eqnlab{Dlamopt}
	\langle\Delta\lambda_1\rangle_F \sim -\zeta_T \Psi(\hat\kappa,1)\times\css{\frac{2}{\sqrt{d}}\,,	& \gamma=0\\\sqrt{\gamma}\,, & \gamma>0\,.}
}
Equation~\eqnref{Dlamopt} is the central result of this section. At fixed $\gamma>0$, the disorder-induced slowdown \textit{saturates} at the finite value $-\zeta_T \Psi(\hat\kappa,1)\sqrt{\gamma}$, independent of the spatial dimension and determined entirely by $\gamma$, while for strictly conservative forces it vanishes as $d^{-1/2}$. High dimensionality alone therefore does not make the disorder more effective at slowing relaxation down. At fixed $\gamma$, however, breaking detailed balance ($\gamma>0$) in isotropic backgrounds does, with a slowdown that is bounded.

The saturation for $\gamma>0$ is a consequence of the growth of the solenoidal force variance with $d$ at fixed $\gamma$. In terms of the parameters~\eqnref{hatgamma}, which keep the force variance fixed, the correlation parameter~\eqnref{tau} and the edge~\eqnref{edge} read
\algn{\eqnlab{edgehat}
	\tau = \frac{d(1-\hat\gamma)-1}{d+2\hat\gamma-1}\,,\qquad \zeta_T R_+ = \hat\zeta_T\frac{2(d-1)-(d-2)\hat\gamma}{\sqrt{d(d-1)(d+2\hat\gamma-1)}}\,.
}
At fixed $\hat\gamma<1$ the correlation parameter tends to the finite value $\tau\to1-\hat\gamma$ as $d\to\infty$. The matrix $\mbb{\tilde H}$ then remains an elliptic random matrix with correlated entries~\cite{Som88,Fyo16}, see \Figref{circle}(b) for $\hat\gamma=1/2$, and instead of \Eqnref{Dlamopt} we obtain
\algn{\eqnlab{Dlamopthat}
	\langle\Delta\lambda_1\rangle_F \sim -\hat\zeta_T \Psi(\hat\kappa,1)\,\frac{2-\hat\gamma}{\sqrt{d}}\,,
}
for all $0\leq\hat\gamma\leq1$. At fixed force variance, the slowdown therefore vanishes as $d^{-1/2}$ for every $\hat\gamma$, and non-conservative forces reduce it, by at most a factor of two for purely solenoidal forces.%

\subsection{Approach to the high-dimensional limit}\seclab{isoapproach}
\begin{figure}
	\includegraphics[width=\linewidth]{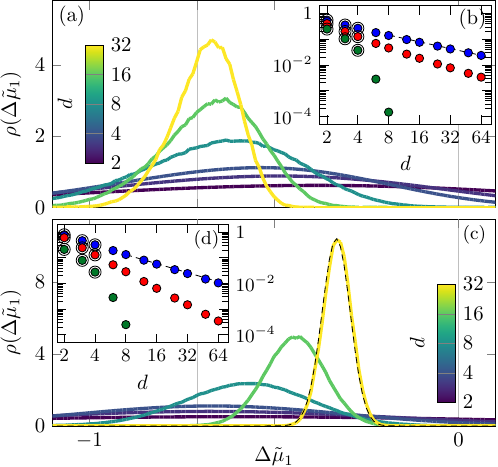}
	\caption{Approach of the density $P(\Delta\tilde \mu_1)$ to the spectral edge with increasing spatial dimension $d$ in an isotropic background, for $\gamma>0$, represented by $\gamma=1/2$ [panels (a), (b)], and for $\gamma=0$ [panels (c), (d)]. Throughout, lines for $d\leq4$ are obtained by numerical integration of the joint density~\eqnref{joint_density} and those for $d\geq8$ by sampling the random matrix ensemble~\eqnref{tHcorr}. (a) Probability density $P(\Delta\tilde\mu_1)$ at $\gamma=1/2$, for which $\tau=0$ and $\mbb{\tilde H}$ is a real Ginibre matrix at every $d$, for $d=2$ to $32$ (colour bar). The vertical line marks the asymptotic edge $-\sqrt\gamma=-1/\sqrt2$ of \Eqnref{edge2}. (b) Distance of the mean from the finite-$d$ edge, $\langle\Delta\tilde\mu_1\rangle_F+R_+$ (blue), variance $\text{Var}(\Delta\tilde\mu_1)$ (red), and probability $P(\Delta\tilde\mu_1>0)$ (green), as functions of $d$. Filled symbols from sampling, open circles from the numerical integration where both are available. The dashed line indicates the power law~\eqnref{edgeapp}. (c) As (a), but for $\gamma=0$, where $\mbb{\tilde H}$ is symmetric. Here the edge $-2/\sqrt d$ of \Eqnref{edge2} moves towards zero with $d$. (d) As (b), for $\gamma=0$.}\figlab{probedge}
\end{figure}

The previous section has shown how random-matrix theory allows us to deduce the density of the smallest eigenvalue at low dimensions and the behaviour of the mean in the high-dimensional limit. In the latter limit, the distribution of the smallest eigenvalue localises close to the spectral edge, making its variance vanish asymptotically.

We now consider in more detail how the density $P(\Delta\tilde\mu_1)$ behaves in the high-dimensional limit, and how known results from random matrix theory can be applied to analyse it. Figure~\figref{probedge} follows the approach to the limit~\eqnref{edge2} along the branches $\gamma=0$ and $\gamma>0$. Panels (a) and (c) show the density $P(\Delta\tilde\mu_1)$ for $d=2,\ldots,32$ at $\gamma=1/2$ and $\gamma=0$, respectively. In both cases the distribution sharpens as the dimension grows, so that the eigenvalue shift becomes self-averaging. As discussed in the previous section, in high dimensions disorder slows relaxation down by an amount that is essentially deterministic, in marked contrast to the broad distributions found for low dimensions in \Figref{prob3d}. The two branches differ in where the density accumulates. At $\gamma=1/2$ it migrates towards the fixed asymptotic edge $-1/\sqrt2$, marked by the vertical line in panel (a), while at $\gamma=0$ the edge $-2/\sqrt d$ itself drifts towards zero, and the density follows it. Panels (b) and (d) quantify these trends. Besides the distance of the mean from the edge and the variance, they show the probability $P(\Delta\tilde\mu_1>0)$ that the disorder makes relaxation \textit{faster}. This probability, which at $d=2$ equals $1/4$ for $\gamma=1/2$ and $(1-1/\sqrt3)/2\simeq0.21$ for $\gamma=0$, is suppressed exponentially with increasing $d$ and is already below $10^{-3}$ at $d=8$ for both values of $\gamma$. This highlights another difference compared with the anisotropic background of \Secref{aniso}: in high-dimensional isotropic backgrounds a disorder-induced speedup, while not forbidden, becomes \emph{exponentially unlikely}, as we now discuss in more detail.

The rate at which the edge~\eqnref{edge} is approached is a question about finite-$d$ corrections, and it is essential that the distance be measured from the \textit{finite}-$d$ edge $-R_+$. Measured numerically from $-R_+$, we find that the mean approaches the edge as a power law with
\algn{\eqnlab{edgeapp}
	\langle\Delta\tilde\mu_1\rangle_F + R_+ \propto d^{-\alpha}\,, \qquad \alpha = \css{ 1.20\pm0.04\,,	&	\gamma=0
	\\0.86\pm0.02 & \gamma>0}\,,
}
over the range $16\leq d\leq256$ at $\gamma=1/2$. We observe that the conservative $\gamma=0$ branch approaches its edge faster. Furthermore, the exponent $\alpha$ can be obtained from random matrix theory in this case: For $\gamma=0$ the matrix $\mbb{\tilde H}$ is symmetric and $\Delta\tilde\mu_1$ is the smallest eigenvalue of a Wigner matrix plus an independent Gaussian-random shift, see \Eqnref{tHdec2}. Its distribution at the soft semicircle edge is governed by the convolution of the Tracy-Widom law~\cite{Tra96} and a Gaussian distribution with zero mean and variance $1/d^2$. This distribution is shown for $d=32$ as the dashed line in \Figref{probedge}(c). Its mean lies a distance $\propto d^{-2/3}$ inside the edge measured in units of the semicircle radius. Since that radius itself decreases as $\sim 2/\sqrt d$, one obtains $\alpha=7/6\approx 1.17$, consistent with \Eqnref{edgeapp}.

The non-conservative branch, with $\alpha=0.86\pm0.02\approx7/8$, converges to its edge more slowly than the conservative one. The slower convergence for $\gamma>0$ is likely caused by a competition between the real and the complex eigenvalues of $\mbb{\tilde H}$. For a real Ginibre matrix the extreme real part is asymptotically attained by a complex eigenvalue. Refs.~\cite{Cip22,Ake14} show that it then lies a distance $\sqrt{\vartheta_d/4d}$ \textit{outside} the limiting spectrum, with
\algn{\eqnlab{gammad}
	\vartheta_d = \frac{\log d - 5\log\log d - \log(2\pi^4)}{2}
}
and Gumbel-distributed fluctuations. This scaling factor is negative for all $d\lesssim7\times10^8$, so the asymptotic regime is far beyond any dimension we can test numerically. At accessible $d$, the extreme is instead frequently attained by one of the $\sim\smash{\sqrt{2d/\pi}}$ exactly real eigenvalues, which lie on the real axis in \Figref{circle}(a), where one of them is the eigenvalue with the smallest real part. Indeed, we find numerically that in about half of all realisations at $d=64$ and still in a third at $d=512$ an exactly real eigenvalue is the smallest one. The exponent $\alpha\approx7/8$ in \Eqnref{edgeapp} therefore describes an intermediate regime, in which the two families of eigenvalues compete, and should not be extrapolated to $d\to\infty$.

A further difference between the branches concerns the variances. For $\gamma=0$ the density of $\Delta\tilde\mu_1$ is a convolution of the Tracy-Widom law with a Gaussian of width $1/d$ that adds to the trace. This width exceeds the Tracy-Widom width $d^{-7/6}$ by a factor $d^{1/6}$, so the Gaussian trace dominates the fluctuations, and $\text{Var}(\Delta\tilde\mu_1)\to1/d^2$ rather than the $\smash{d^{-7/3}}$ of the usual edge law~\cite{Tra96}. Numerically, we obtain $d^{-2.1}$, consistent with these considerations. The mean is unaffected, since the trace has zero mean, which is why the $\alpha=7/6$ scaling of the mean survives the convolution. For $\gamma>0$ the ordering is reversed: the trace variance, $(d+2)(1-\gamma)/d^3$, is again $O(d^{-2})$, but the edge fluctuations are now much larger, presumably due to the mentioned competition between the real and complex parts of the spectrum, and we find numerically that $\text{Var}(\Delta\tilde\mu_1)$ decays only as $d^{-1.25}$.

In this section, we described the behaviour in isotropic backgrounds, the middle branch in the diagram in \Figref{flowdiagram}. To summarise, the $d$-fold degeneracy of $\lambda^0_1$ makes the leading-order shift an eigenvalue of the Gaussian random matrix $\mbb{\tilde H}$, whose statistics depend only on $d$ and $\gamma$, while the correlation length enters only through $\Sigma(\kappa)$. Random-matrix theory explains why the shift is nevertheless non-Gaussian: unlike the single matrix element that sets the leading-order shift in anisotropic backgrounds [\Eqnref{ev1}], an eigenvalue depends nonlinearly on the Gaussian entries. The eigenvalues repel each other [\Eqnref{ev_dens_xi}], they can form complex-conjugate pairs for $\gamma>0$, and the relaxation rate is set by the one with the smallest real part, which biases the mean towards slower relaxation already at first order in $\zeta_T$. In low dimensions, the resulting distributions are broad, and for $\gamma<1$ a finite fraction of realisations relaxes faster, in agreement with simulations in $d=2$ and $d=3$ [\Figref{prob3d}]. In high dimensions, by contrast, the eigenvalues fill the ellipse~\eqnref{ellipse}, and since each eigenvalue is confined by its neighbours, $\Delta\tilde\mu_1$ is pinned to its left edge, \Eqnref{edge}. Its fluctuations are therefore suppressed, so that the shift becomes deterministic and faster relaxation exponentially unlikely [\Figref{probedge}]. This holds for conservative and non-conservative forces alike, whose variances decay as $d^{-2}$ and approximately as $d^{-1.25}$, respectively. The two cases differ, however, in the limiting ensemble and hence in the location of the edge. For $\gamma=0$, $\mbb{\tilde H}$ is symmetric and the edge shrinks as $2/\sqrt d$, so that the slowdown vanishes as $d^{-1/2}$. For $\gamma>0$, $\mbb{\tilde H}$ approaches a real Ginibre matrix, and the slowdown saturates at a finite value proportional to $\sqrt\gamma$ [\Eqnref{Dlamopt}]. At fixed force variance, by contrast, it vanishes as $d^{-1/2}$ for every $\hat\gamma$ [\Eqnref{Dlamopthat}].

\section{Quasi-isotropic background}\seclab{qiso}
Isotropic and anisotropic backgrounds are two limits of an intermediate regime, where the stiffnesses $a_{n\geq2}$ differ from $a_1=1$ by a value of the order of $\zeta_T$. In this intermediate regime, the background is \emph{quasi-isotropic}, i.e., its anisotropy is of the same order as the disorder strength. This quasi-isotropic regime connects the results of \Secref{aniso} and~\secref{iso}, which otherwise do not join smoothly. In particular, for the anisotropic backgrounds of \Secref{aniso}, the mean shift is quadratic in $\zeta_T$, but its coefficient~\eqnref{mu1} diverges as $a_n\to1$. By contrast, for the isotropic backgrounds in \Secref{iso} the shift is linear in $\zeta_T$ and independent of the $a_n$, because they are all equal to unity. The crossover between these two regimes occurs when $a_n-1\sim \zeta_T$.
\begin{figure}
	\centering
	\includegraphics[width=\linewidth]{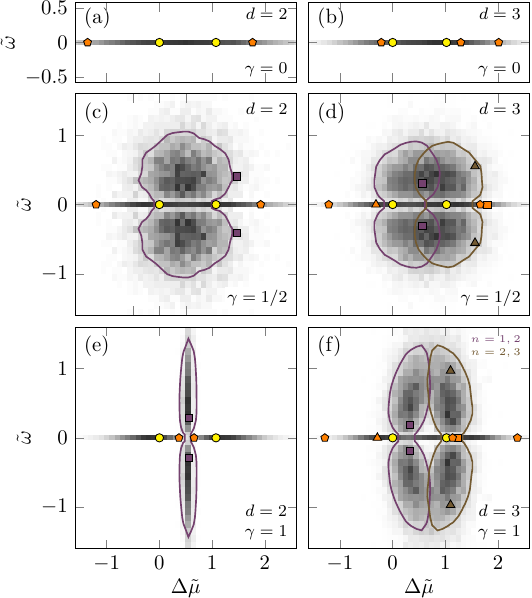}
	\caption{Density of eigenvalue shifts in the complex plane for a \textit{quasi-isotropic} background, shown as a grey scale normalised to its maximum. The densities were computed with the method of \Secref{numsim} for $2.5\times10^4$ disorder realisations at correlation length $\kappa=1$, disorder strength $\zeta_T=10^{-1}$ and stiffnesses $\mbb{a}=\text{diag}(1,1.04)$ for $d=2$ and $\mbb{a}=\text{diag}(1,1.04,1.04)$ for $d=3$, corresponding to the rescaled offsets $\delta\hat a=1.07$ and $1.02$ of \Eqnref{dahat}, for $\gamma=0$ [(a),(b)], $\gamma=1/2$ [(c),(d)] and $\gamma=1$ [(e),(f)], in $d=2$ (left) and $d=3$ (right). All shifts are measured from $\lambda^0_1$ and rescaled by $\zeta_T\Sigma(\kappa)d$, so that the unperturbed levels sit at $0$ and at $\delta\hat a$ (yellow bullets), respectively. Markers show single realisations of each type, with shapes and colours as in \Figref{numsimiso}. The violet line shows the contour of the complex-conjugate pairs formed by $\lambda_1$ and $\lambda_2$ at a quarter of their maximum, the olive line that of the pairs formed by $\lambda_2$ and $\lambda_3$.}
	\figlab{numsimqiso}
\end{figure}

In a quasi-isotropic background, we therefore write the stiffnesses as $a_n=1+\zeta_T\delta\tilde a_n$ with $\delta\tilde a_1=0$ and $\delta\tilde a_n=O(1)$. As discussed in \Secref{perturbation}, the anisotropy then enters the degenerate perturbation theory at the same order as the disorder, and the first-order shifts are the eigenvalues of $\zeta_T(\delta\tilde{\mbb{a}}+\mbb{H})$, where $\delta\tilde{\mbb{a}}=\text{diag}(\delta\tilde a_1,\ldots,\delta\tilde a_d)$ and $\mbb{H}$ is the isotropic perturbation matrix~\eqnref{Hnmdef}. In other words, in the quasi-isotropic case, we are now dealing with the problem of determining the eigenvalues of the random matrix $\mbb{H}$ plus the deterministic diagonal matrix $\delta \mbb{\tilde a}$. This type of problem has a long history~\cite{May72}, with closed-form solutions typically known either for $d=2$ or $d\to\infty$~\cite{For23}. Applying the rescaling~\eqnref{Hresc} of the isotropic case, the characteristic equation~\eqnref{Dlam1deg} becomes
\algn{\eqnlab{qisoeig}
	\det\left[\mbb{\tilde H}(\gamma)+\delta\hat{\mbb{a}}-\Delta\tilde\lambda_1\mbb{1}\right]=0\,,
}
with the diagonal matrix $\delta\hat{\mbb{a}}=\text{diag}(0,\delta\hat a_2,\ldots,\delta\hat a_d)$ of rescaled stiffness offsets
\algn{\eqnlab{dahat}
	\delta\hat a_n=\frac{\delta\tilde a_n}{\Sigma(\kappa)d}=\frac{a_n-1}{\zeta_T\Sigma(\kappa)d}\,.
}
\begin{figure}
	\centering
	\includegraphics[width=\linewidth]{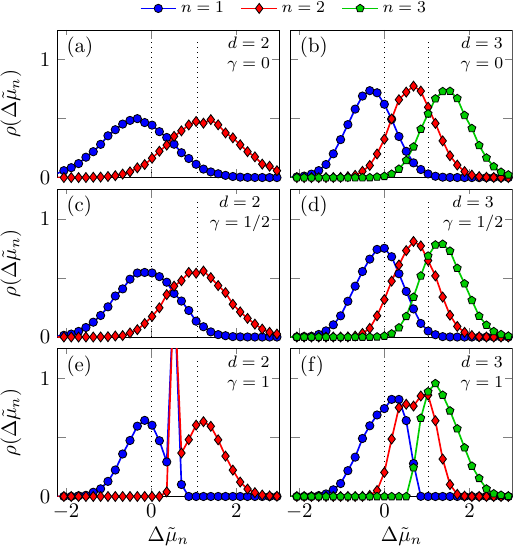}
	\caption{Repulsion and attraction of the levels in a \textit{quasi-isotropic} background: densities $\rho$ of the real parts of the shifts in \Figref{numsimqiso} resolved by branch, $n=1$ (blue), $n=2$ (red) and $n=3$ (green), for $\gamma=0$ [(a),(b)], $\gamma=1/2$ [(c),(d)] and $\gamma=1$ [(e),(f)], in $d=2$ (left) and $d=3$ (right). The dotted lines mark the unperturbed levels. The narrow peak at $\delta\hat a/2$ in (e) collects the coalesced realisations and is clipped by the axis.}
	\figlab{numsimqisore}
\end{figure}

The random part $\mbb{\tilde H}(\gamma)$ is the same matrix as in \Secref{iso}, with correlations~\eqnref{tHcorr}. The distribution of $\Delta\tilde\mu_1=\mathrm{Re}\,\Delta\tilde\lambda_1$ therefore depends on $\gamma$, $d$ and the offsets $\delta\hat a_n$. The correlation length, the disorder strength and the anisotropy of the background enter only through the combination~\eqnref{dahat}. For the real part $\Delta\mu_1$ of the eigenvalue shift, this gives
\algn{\eqnlab{qisoscal}
	\Delta\mu_1\sim\zeta_T\Sigma(\kappa)d\,\Delta\tilde\mu_1\,.
}
The shift is linear in $\zeta_T$ as in the isotropic case, provided that the offsets $a_n-1$ are scaled proportionally to $\zeta_T$. Its mean and standard deviation are therefore both linear in $\zeta_T$. Due to the scaling~\eqnref{qisoscal}, numerical data taken at different $\kappa$, $\zeta_T$ and $a_n$ must collapse onto a single curve when plotted against $\delta\hat a_n$. In most of this section, we take the stiffer directions to be equal, $a_{n\geq2}=a_2$, so that $\langle\Delta\tilde\mu_1\rangle_F$ depends on the single variable $\delta\hat a=\delta\hat a_2$. This simplifies the theoretical analysis in particular, because then $\delta\mbb{\hat a}=\delta \hat a (\mbb{1}-\ve e_1\ve e_1^{\sf T})$ becomes proportional to the identity matrix minus the rank-one matrix $\ve e_1\ve e_1^{\sf T}$. For rank-one perturbations of random matrices, much more is known, especially in the high-dimensional limit~\cite{Tao13,Pot21,For23}.

To analyse the quasi-isotropic regime, as before, we first look at direct numerical solutions of \Eqnref{Leig}. Figure~\figref{numsimqiso} shows the distributions of the rescaled shifts $\Delta\tilde\lambda_n= \Delta\tilde\mu_n + i\tilde\omega_n$ in the complex plane at $\zeta_T=10^{-1}$ for a weakly anisotropic background in $d=2,3$. The background is given by $a_{n\geq2}=1.04$, so the unperturbed levels are separated by an amount comparable to the width of the densities in \Figref{numsimiso}.  To compare the lowest eigenvalues $\lambda_n$, $n=1,\ldots,d$ on one axis, we measure all shifts from the same unperturbed level $\lambda^0_1=1$, so \Figref{numsimqiso} shows the density of all $d$ branches together.
\begin{figure}
	\centering
	\includegraphics[width=\linewidth]{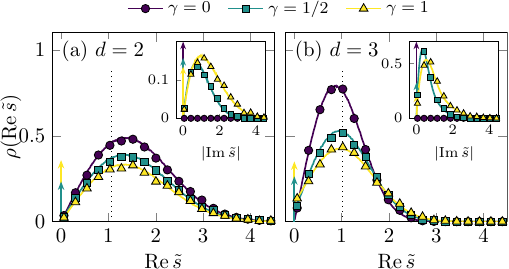}
	\caption{Repulsion and attraction of the levels in a \textit{quasi-isotropic} background: densities $\rho(\mathrm{Re}\,\tilde s)$ of the real part of the rescaled distance $\tilde s$ between the two lowest levels for the data of \Figref{numsimqiso}, in (a) $d=2$ and (b) $d=3$, for $\gamma=0$ (circles), $\gamma=1/2$ (squares) and $\gamma=1$ (triangles). The insets show the densities of $|\mathrm{Im}\,\tilde s|$. The lines show first-order theory, obtained by sampling of \Eqnref{qisoeig}, and the dotted lines mark the unperturbed distance $\delta\hat a$. The delta functions at $\tilde s=0$ have the weights $P(\mathrm{Re}\,\tilde s=0)=0.21$ and $0.33$ at $\gamma=1/2$ and $1$ in $d=2$, and $0.24$ and $0.32$ in $d=3$ (main panels), and $P(\mathrm{Im}\,\tilde s=0)=1$, $0.79$ and $0.67$ at $\gamma=0$, $1/2$ and $1$ in $d=2$, and $1$, $0.47$ and $0.31$ in $d=3$ (insets).}
	\figlab{numsimqisospacing}
\end{figure}

At $\gamma=0$ the spectrum of $\mc{L}^\dagger$ is real. Therefore, the densities in \Figsref{numsimqiso}(a) and (b) condense on the real axis. By contrast, for $\gamma=1/2$, shown in \Figref{numsimqiso}(c),(d), the densities have a sharp component on the real axis, and a more diffuse cloud in the complex plane, with a significant weight located at $\Delta\tilde\mu_n\approx\delta\hat a/2$, i.e., with real parts between the unperturbed levels (yellow bullets). These parts result from individual sets of eigenvalues that are either completely real (pentagons) or contain complex conjugate pairs (squares and triangles). For $\gamma=1$, when the forces are divergence-free, this behaviour is the strongest. In particular for $d=2$ [\Figref{numsimqiso}(e)], the density of complex-conjugate realisations forms a narrow ridge at $\Delta\tilde\mu_n=\delta\hat a/2$. In $d=3$, the behaviour is similar, yet less pronounced. The pairs formed by $\lambda_1$ and $\lambda_2$ (violet contour) have real parts between the unperturbed levels that are no longer pinned to $\delta\hat a/2$, while the pairs formed by $\lambda_2$ and $\lambda_3$ within the degenerate upper level (olive contour) are centred close to $\delta\hat a$.

Figure~\figref{numsimqisore} shows the densities of the real parts $\Delta\tilde\mu_n$ of the shifts for the same data as in \Figref{numsimqiso}, resolved by branch $n=1,\ldots,d$ in different colours. At $\gamma=0$ [\Figref{numsimqisore}(a),(b)] the real parts of the branches repel each other, so the mean of $\Delta\tilde\mu_1$ is pushed toward negative values. For $\gamma=1/2$ and $\gamma=1$ [\Figref{numsimqisore}(c)--(f)], the real eigenvalues still repel, but the formation of complex-conjugate eigenvalue pairs, whose two members share a common real part, pushes the densities of the real parts together. This mechanism of ``level attraction'' through the creation of complex-conjugate pairs is a property of non-Hermitian operators~\cite{Yu24}, here due to the presence of non-conservative forces.  In the most extreme case $\gamma=1$ and $d=2$, shown in \Figref{numsimqisore}(e), the vertical ridge observed in \Figref{numsimqiso}(e) collapses onto $\Delta\tilde \mu_n=\delta\hat a/2$, forming a strong peak that increases the mean of $\Delta\tilde \mu_1$.

The distance $\tilde s$ between the two lowest levels in \Figref{numsimqisospacing} shows how anisotropy shifts the balance between level repulsion and attraction. Nearby levels still repel each other, but this effect is weaker than in the isotropic background [\Figref{numsimisospacing}] and only slightly shifts the bulk of $\rho(\tilde s)$ beyond the unperturbed distance $\delta \hat a$ (dotted line). Level attraction, by contrast, is still significant, as seen from the delta functions in \Figref{numsimqisospacing}, and it now plays a more important role, because it shifts $\Delta\tilde\mu_1$ toward the centre between the two levels.

\subsection{\texorpdfstring{Scaling collapse and faster relaxation}{Scaling collapse and faster relaxation}}\seclab{qisoscaling}
To test the scaling collapse predicted by \Eqsref{dahat} and~\eqnref{qisoscal}, we first evaluate $\langle\Delta\tilde \mu_1\rangle_F$ by sampling $\mbb{\tilde H}$ from the ensemble~\eqnref{tHcorr} and diagonalising \Eqnref{qisoeig}. The solid lines in \Figref{qisoscal}(a) and (b) show the resulting $\langle\Delta\tilde \mu_1\rangle_F$ as a function of $\delta\hat a$ for $d=2$ and $d=3$ and different $\gamma$. The corresponding data of the full Fokker-Planck operator at different $\kappa$ and $\zeta_T$ and rescaled according to \Eqnref{Hresc} are shown as the markers. Each simulation marker corresponds to one combination of $\kappa$, $\zeta_T$ and $a_2=a_{n\geq2}$, where $\zeta_T$ and $a_2-1$ vary over two and three decades, respectively. Despite this wide range of parameters, we observe in \Figref{qisoscal}(a) and (b) that the data collapse onto the random-matrix curves over four decades of $\delta\hat a$, in $d=2$ as well as in $d=3$. This confirms that the correlation length, the disorder strength and the anisotropy enter only through the combination~\eqnref{dahat}. For $\zeta_T\leq10^{-2}$ the simulations agree well with the random-matrix result, while systematic deviations occur at the largest disorder strengths $\zeta_T\sim10^{-1}$.
\begin{figure}
	\centering
	\includegraphics[width=\linewidth]{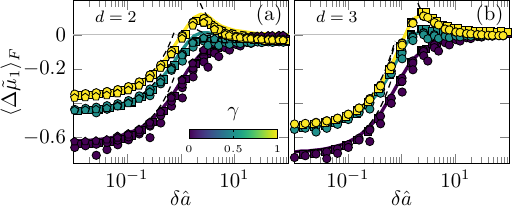}
	\caption{Rescaled mean $\langle\Delta\tilde\mu_1\rangle_F$ of the real eigenvalue shift~\eqnref{qisoscal} in a quasi-isotropic background with equal stiffer directions, $a_{n\geq2}=a_2$, for $\gamma=0$, $1/2$ and $1$ (colour bar), in (a) $d=2$ and (b) $d=3$. Lines are obtained by sampling the ensemble~\eqnref{tHcorr} with $2\times10^5$ matrices per point. Markers show direct numerical solutions of \Eqnref{Leig}, computed with the method of \Secref{numsim} for $10^3$ disorder realisations, at $\kappa=1$ (circles), $\kappa=2$ (squares) and, for $d=2$, $\kappa=4$ (pentagons). Each marker corresponds to one of eleven disorder strengths $10^{-3}\leq\zeta_T\leq10^{-1}$ and five stiffnesses $10^{-4}\leq a_2-1\leq0.12$, and is placed at its rescaled offset~\eqnref{dahat}. The grids are $L=30$, $N=120$ for $d=2$ and $L=20$, $N=40$ for $d=3$. The dashed lines show the first-order result~\eqnref{qisosmall} at small $\delta\hat a$ and the large-$\delta\hat a$ expansion~\eqnref{qisotail4}.}
	\figlab{qisoscal}
\end{figure}

Since the scaling collapse predicted by \Eqnref{qisoscal} holds true for the entire distribution of $\Delta\mu_1$, we confirm it in Appendix~\secref{collapse} for the variance and for the probabilities that $\Delta\mu_1>0$ and $\omega_1\neq0$.

With the collapse confirmed, let us now turn to the contents of \Figsref{qisoscal}(a) and (b). In agreement between the two- and three-dimensional cases, we find that $\langle \Delta\tilde \mu_1\rangle_F$ is negative for small values of $\delta\hat a$, as expected from the result of the isotropic background in \Secref{iso}, and that it initially increases linearly with $\delta\hat a$, as we show in \Secref{isomatch}. For $\gamma=0$ it stays negative and approaches zero from below as $\delta\hat a^{-1}$, see \Secref{qisomatch}, and the same holds for $\gamma=1/4$ (not shown). However, for larger $\gamma$ it can become \emph{positive} for $\delta \hat a\sim 1$. For $\gamma=1$ it changes sign at $\delta\hat a$ of order unity, reaches a positive maximum of about $0.12$ at $\delta\hat a$ between $2$ and $3$, in both $d=2$ and $d=3$, and then decays to zero. By \Eqnref{qisoscal}, the largest disorder-induced speedup at a given disorder strength is thus reached for $a_2-1\approx2.5\,\zeta_T\Sigma(\kappa)d$ and amounts to $\langle\Delta\mu_1\rangle_F\approx0.1\,\zeta_T\Sigma(\kappa)d$. In contrast to the anisotropic background, this speedup is linear in $\zeta_T$.

The onset of faster relaxation lies slightly below the threshold value $\gamma=1/2$ of the anisotropic background in \Secref{aniso}, at which the large-$\delta\hat a$ tail of $\langle\Delta\tilde\mu_1\rangle_F$ changes sign, as we show in \Secref{qisomatch}. Numerically, we find the onset at $\gamma^*\approx0.44$ in both $d=2$ and $d=3$. For smaller $\gamma$, we find no positive values for $\langle \Delta\tilde \mu_1\rangle_F$. Faster relaxation at linear order in $\zeta_T$ therefore requires a non-conservative fraction $\gamma$ of about one half, independently of the correlation function and of $\kappa$. This differs from the anisotropic background, where faster relaxation at second order is possible also for $\gamma<1/2$ if the background is sufficiently anisotropic, see \Figref{statediag}(a).

As anticipated earlier, the positive mean is carried by the realisations in which $\lambda_1$ forms a complex-conjugate pair with the next level. To make this explicit, we consider the two-dimensional case $d=2$. Here, the real part of a complex pair is half the trace of the $2\times2$ matrix in \Eqnref{qisoeig}, $(\tilde H_{11}+\tilde H_{22}+\delta\hat a)/2$. At $\gamma=1$ the trace $\tilde H_{11}+\tilde H_{22}$ vanishes identically, so that every complex pair has real part exactly $\delta\hat a/2$, which is the ridge in \Figref{numsimqiso}(e). The sum $\tilde H_{11}+\tilde H_{22}$ has zero mean and is independent of the condition for complex eigenvalues, which involves only $\tilde H_{11}-\tilde H_{22}$ and $\tilde H_{12}\tilde H_{21}$. The complex realisations therefore contribute $\delta\hat a/2>0$ on average, which increases with $\delta\hat a$. However, the probability of a complex pair decreases with $\delta\hat a$, from about $0.4$ at $\delta\hat a=0$ for $\gamma=1$ to $\approx 0.2$ at $\delta\hat a=2$, see \Figref{qisocollapse}(d) in Appendix~\secref{collapse}. The repulsion of the real parts is strongest in the isotropic case and decreases with $\delta\hat a$. Hence, the positive maximum of $\langle\Delta\tilde\mu_1\rangle_F$ for $\gamma>\gamma^*$ results from a competition between eigenvalue repulsion and attraction. For $\gamma=0$ all eigenvalues are real and only level repulsion plays a role. Physically, disorder speeds up the relaxation of a quasi-isotropic background by merging the slowest mode with its neighbour into an oscillating pair, cf.~\Eqnref{oscill}, whose decay rate lies between the two unperturbed rates.
\subsection{Crossover to the isotropic background}\seclab{isomatch}
For $\delta\hat a\to0$, it is clear from \Eqnref{qisoeig} that the distribution of $\Delta\tilde\mu_1$ reduces to that of the isotropic case, where the mean is negative for all $\gamma$, see \Secref{iso}. For small $\delta\mbb{\hat a}$, we may treat $\delta\hat{\mbb a}$ in \Eqnref{qisoeig} as a perturbation of the isotropic matrix $\mbb{\tilde H}$ of \Secref{iso}, whose eigenvalues are distinct for each realisation.

To first order, the eigenvalue $\Delta\tilde \lambda_n$ of $\mbb{\tilde H}+\delta\hat{\mbb a}$ reads
\algn{
	\Delta\tilde\lambda_n \sim \Delta\tilde\lambda^{\mbb{a}=\mbb{1}}_n + \ve l^{\sf T}_n\cdot \delta\mbb{\hat a}\cdot \ve r_n\,,
}
where $\Delta\tilde\lambda^{\mbb{a}=\mbb{1}}_n$ is the isotropic eigenvalue of \Secref{iso}, and $\ve l_n$ and $\ve r_n$ are the corresponding left and right eigenvectors of $\mbb{\tilde H}$, normalised such that $\ve l_n\cdot \ve r_m = \delta_{nm}$. The ensemble~\eqnref{tHcorr} is invariant under rotations $\mbb{\tilde H}\to\mbb{O}\mbb{\tilde H}\mbb{O}^{\sf T}$, so that the eigenvectors are uniformly oriented with respect to the coordinate axes, independently of the eigenvalues. Upon averaging, each eigenvalue is therefore shifted by the mean $\tr(\delta\hat{\mbb a})/d$ of the offsets, and we obtain for $\langle\Delta\tilde\mu_n\rangle_F$
\algn{\eqnlab{qisosmall}
	\langle\Delta\tilde\mu_n\rangle_F\sim\langle\Delta\tilde\mu^{\mbb{a}=\mbb{1}}_n\rangle_F+\frac1d\tr\delta\mbb{\hat a}\,.
}
For equal offsets $\delta \hat a_{n\geq2} = \delta\hat a$, $\langle\Delta\tilde\mu_1\rangle_F$ thus rises by $(d-1)\delta\hat a/d$ for every $\gamma$, see the dashed lines at small $\delta\hat a$ in \Figsref{qisoscal}(a) and~(b). For a general level $n=1,\ldots,d$, the eigenvalues associated with a $\delta\hat a_n$ smaller than the mean offset $\delta\bar a = d^{-1}\tr\delta\mbb{\hat a}$ increase, the ones above it, decrease, compared with the levels $\delta\hat a_n$ in the absence of disorder.

Physically, the disorder mixes the $d$ nearly degenerate modes, whose orientations are random, so they sample all stiffnesses with equal weight. To first order, the background looks isotropic, with mean stiffness $\tr(\mbb{a})/d>1$. The perturbation theory also gives an indication of why the densities in \Figsref{numsimqiso} and \figref{numsimqisore} are centred around the mean offset $\delta\bar a$, that is, $\delta\hat a/2 \approx 0.53$ ($d=2$) and $2\delta\hat a/3\approx 0.68$ ($d=3$).

For $\delta\hat a\to\infty$ the lowest level separates from the others, and we expect to recover the anisotropic case of \Secref{aniso}. We now show that this is indeed the case and that the matching fixes the sign of $\langle\Delta\tilde\mu_1\rangle_F$ at large $\delta\hat a$.
\subsection{Crossover to the anisotropic background}\seclab{qisomatch}
For $\delta\hat a_n\gg1$ we may treat the off-diagonal entries of $\mbb{\tilde H}$ in \Eqnref{qisoeig} as a perturbation of $\delta\hat{\mbb{a}}$. Non-degenerate perturbation theory for the lowest level, whose unperturbed value is $0$, gives to second order
\algn{\eqnlab{qisoPT}
	\Delta\tilde\lambda_1\sim\tilde H_{11}-\sum_{n=2}^d\frac{\tilde H_{1n}\tilde H_{n1}}{\delta\hat a_n}\,.
}
The first term has zero mean. The correlation~\eqnref{tHcorr} of an entry with its transpose is $\langle\tilde H_{1n}\tilde H_{n1}\rangle_F=\tau(1+d\gamma)/d^2=\ms{B}/d^2$, with $\ms{B}=1-2\gamma$ from \Eqnref{cexpr}. Hence
\algn{\eqnlab{qisotail}
	\langle\Delta\tilde\mu_1\rangle_F\sim-\frac{1-2\gamma}{d^2}\sum_{n=2}^d\frac1{\delta\hat a_n}\,,\qquad \delta\hat a_n\gg1\,.
}
Undoing the rescaling with \Eqnref{dahat}, the disorder strength $\zeta_T$ and the factor $\Sigma(\kappa)d$ reappear squared, and we obtain
\algn{\eqnlab{qisotailphys}
	\langle\Delta\mu_1\rangle_F\sim-\zeta_T^2\Sigma^2(\kappa)(1-2\gamma)\sum_{n=2}^d\frac1{a_n-1}\,.
}
This is a second-order shift that diverges as $a_n\to1$. Interestingly, this divergence matches the divergent part of the anisotropic result~\eqnref{mu1}. There, the terms with $n\geq2$ carry the factor $s^{a_n-2}$, whose integral near $s=0$ grows as $(a_n-1)^{-1}$. At $s=0$ the remaining $\ve k$ integral is $\kappa^{-4}\int\ed^dk\,\ee^{-\ve k^2/\kappa^2}\hat C_F(\ve k)(1-2\gamma)k_1^2k_n^2$, which by \Eqnref{cint2} equals $(1-2\gamma)\Sigma^2(\kappa)$. This reproduces \Eqnref{qisotailphys} for an arbitrary correlation function.

For the variance, we take the average of the square of the first term in \Eqnref{qisoPT}, $\langle\tilde H_{11}^2\rangle_F=[3+\gamma(d-4)]/d^2$, which gives $\sigma^2_{\Delta\lambda_1}=\Sigma^2(\kappa)[3+\gamma(d-4)]$. This result is recovered from \Eqnref{sigmain1} evaluated at $\mbb{a}=\mbb{1}$ and using \Eqnref{cint}. The quasi-isotropic regime thus regularises the divergence of \Secref{aniso}: as the background approaches isotropy the second-order shift grows as $(a_n-1)^{-1}$ until $a_n-1\sim\zeta_T\Sigma(\kappa)d$, where it turns into a shift that is linear in $\zeta_T$.

The sign of \Eqnref{qisotailphys} can be interpreted in the light of the previous discussion on the importance of eigenvalue attraction for faster relaxation. The coefficient $\ms{B}=1-2\gamma$ measures the correlation between the coupling $\tilde H_{1n}$ of the slowest mode to the mode $n$ and the reverse coupling $\tilde H_{n1}$. For predominantly conservative forces, $\gamma<1/2$, the coupling is predominantly symmetric. The two levels then repel each other, the lower one is pushed down, and relaxation becomes slower. For predominantly non-conservative forces, $\gamma>1/2$, the coupling is predominantly antisymmetric. The real parts of the levels then attract each other, the lower one moves up, and relaxation becomes faster. If the attraction is strong enough, the two levels merge into a complex-conjugate pair whose real part lies between the unperturbed levels. The same threshold $\gamma=1/2$ appears in the anisotropic case, see the discussion below~\Eqnref{prefac} and the state diagram in \Figref{statediag}(a).

At $\gamma=1/2$ the second-order term~\eqnref{qisotail} vanishes, and the next order decides the sign. The third-order contribution to the mean vanishes, because it contains an odd number of Gaussian entries. The fourth-order contribution follows from Wick's rule (Appendix~\secref{qiso4}), and we obtain
\algn{\eqnlab{qisotail4}
	\langle\Delta\tilde\mu_1\rangle_F\sim-\frac{1-2\gamma}{d^2}\sum_{n=2}^d\frac1{\delta\hat a_n}+\frac{(1+d\gamma)\left[(d+4)\gamma-1\right]}{d^4}\sum_{n=2}^d\frac1{\delta\hat a_n^3}\,.
}
The fourth-order term is positive for $\gamma>1/(d+4)$, and in particular at $\gamma=1/2$ in every dimension. There $\langle\Delta\tilde\mu_1\rangle_F$ approaches zero from above as $\delta\hat a^{-3}$, and relaxation is faster on average. This suggests that for $\gamma\geq1/2$, $\langle \Delta\tilde \mu_1\rangle_F$ is positive at large $\delta\hat a$ --- relaxation is faster on average --- in a quasi-isotropic background independently of the disorder correlation function. As mentioned earlier, we find numerically that relaxation can be faster on average even for $\gamma$ values smaller than $1/2$, down to $\gamma^*\approx0.44$. This positive window is consistent with \Eqnref{qisotail4}: for $\gamma$ slightly below $1/2$, the positive fourth-order term outweighs the small negative second-order term at moderate offsets, for instance for $\delta\hat a\lesssim2.8$ ($d=2$) and $\delta\hat a\lesssim2.4$ ($d=3$) at $\gamma=0.45$, where the expansion is, however, only qualitative.
\subsection{High-dimensional limit $d\gg1$}\seclab{qisohighd}
\begin{figure}
	\centering
	\includegraphics[width=\linewidth]{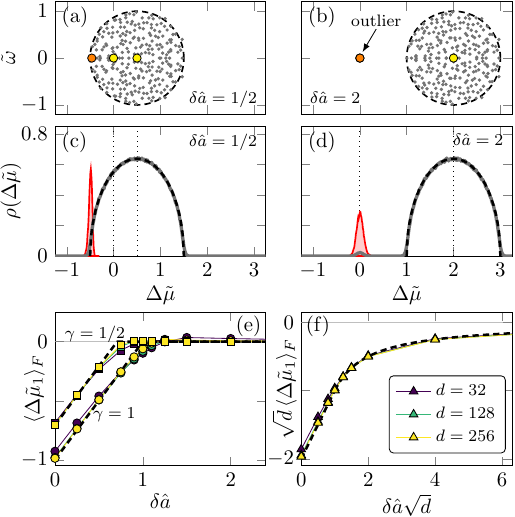}
	\caption{Quasi-isotropic background in high dimensions, with equal stiffer directions. (a) Eigenvalues $\Delta\tilde\lambda$ of \Eqnref{qisoeig} in the complex plane, $\Delta\tilde\mu=\mathrm{Re}\,\Delta\tilde\lambda$ and $\tilde\omega=\mathrm{Im}\,\Delta\tilde\lambda$ (grey), for a single realisation of the ensemble~\eqnref{tHcorr} with $d=256$, $\gamma=1$ and $\delta\hat a=1/2$, using the same $\mbb{\tilde H}$ as in \Figref{circle}(a). The dashed circle is the edge of the limiting disc of radius $\sqrt\gamma$ around $\delta\hat a$, the yellow bullets mark the unperturbed levels $0$ and $\delta\hat a$, and the orange bullet the eigenvalue with the smallest real part. (b) Same realisation as (a), but for $\delta\hat a=2$, where the eigenvalue with the smallest real part is an outlier outside the disc. (c) Densities of the real parts of all eigenvalues (grey) and of the smallest real part alone (red, divided by $20$) for the ensemble of (a), from $2\times10^4$ samples, on the axis of (a). The dashed curve is the projection of the limiting disc, a semicircle of radius $\sqrt\gamma$ around $\delta\hat a$, and the dotted lines mark the unperturbed levels. (d) Same as (c), for the ensemble of (b), where the smallest real part is the outlier near $0$. (e) Mean $\langle\Delta\tilde\mu_1\rangle_F$ obtained by sampling the ensemble~\eqnref{tHcorr} for $d=32$, $128$ and $256$ (colours, over $4\cdot 10^3$, $10^3$ and $4\cdot10^2$ matrices, respectively), as a function of $\delta\hat a$ for $\gamma=1$ (circles) and $\gamma=1/2$ (squares). Dashed lines show the limit~\eqnref{qisohighd}, with kinks at $\delta\hat a=\sqrt\gamma$. (f) Conservative forces, $\gamma=0$, with $\sqrt d\,\langle\Delta\tilde\mu_1\rangle_F$ as a function of $\delta\hat a\sqrt d$, colours as in (e). The dashed line shows \Eqnref{qisohighd0}.}
	\figlab{qisohighd}
\end{figure}
In high dimensions the random part of \Eqnref{qisoeig} approaches the limiting ensembles of \Secref{iso}, and the deterministic part has a simple structure. In the decomposition of $\delta\mbb{\hat a} = \delta\hat a(\mbb{1}-\ve e_1\ve e_1^{\sf T})$ introduced in the beginning of this section, the first term shifts the whole spectrum of $\mbb{\tilde H}$ by $\delta\hat a$. The second term is a rank-one perturbation of strength $-\delta\hat a$ along the slowest direction. For large random matrices it is known~\cite{Tao13,ORo14} that a perturbation of bounded rank creates an isolated eigenvalue, an outlier, at the position of its own eigenvalue whenever that position lies outside the limiting spectrum. For $\gamma>0$ the limiting spectrum of $\delta\hat a\mbb{1}+\mbb{\tilde H}$ is a disc of radius $\sqrt\gamma$ centred at $\delta\hat a$ (\Secref{iso}), and the potential outlier is located at $\Delta\tilde\lambda=0$. An actual outlier forms once $\Delta\tilde\lambda=0$ lies outside the disc, that is for $\delta\hat a>\sqrt\gamma$. Otherwise the smallest real part is attained at the left edge of the disc.

Figures~\figref{qisohighd}(a) and~(b) show the mechanism for a single realisation of the ensemble with $d=256$ and $\gamma=1$, with the same matrix $\mbb{\tilde H}$ for $\delta\hat a=1/2$ and $\delta\hat a=2$. For $\delta\hat a=1/2<\sqrt\gamma$ the origin lies inside the disc, and the eigenvalue with the smallest real part is located at the left edge of the disc. For $\delta\hat a=2>\sqrt\gamma$ the disc lies to the right of the origin, and the spectrum contains a single isolated real outlier. Figures~\figref{qisohighd}(c) and~(d) show the densities of the real parts over the ensemble. For $\delta\hat a=1/2$ the smallest real part forms a narrow peak at the left edge of the bulk. For $\delta\hat a=2$ the smallest real part is the outlier, and its density is approximately Gaussian as in \Secref{aniso} and separated from the bulk.

This implies that in the quasi-isotropic case \Eqnref{edge2} generalises to
\algn{\eqnlab{qisohighd}
	\langle\Delta\tilde\mu_1\rangle_F\to\min\left(\delta\hat a-\sqrt\gamma,\,0\right)\,,\qquad \gamma>0\,,
}
for $d\to\infty$, with corrections of order $1/d$. For $\delta\hat a\gg 1$, these corrections are given by \Eqnref{qisotail}, which reads $-(1-2\gamma)/(d\,\delta\hat a)$ for large $d$. Figure~\figref{qisohighd}(e) compares \Eqnref{qisohighd} (dashed line) with numerical results obtained by sampling the ensemble~\eqnref{tHcorr} for $d$ up to $256$. Away from the transition point at $\delta\hat a = \sqrt{\gamma}$, the sampled numerical results agree with the limit already at $d=32$, with slower convergence at the transition, when the outlier coincides with the disc's left edge.

In physical units, $\Sigma(\kappa)d\to\Psi(\hat\kappa,1)$ by \Eqnref{Siscal}, and \Eqnref{qisohighd} states that $\langle\Delta\mu_1\rangle_F\to\min(\delta a-\zeta_T\sqrt\gamma\,\Psi,0)$, where $\delta a=a_2-1$. The disorder spreads the $d-1$ levels at $a_2$ into a disc of radius $\zeta_T\sqrt\gamma\,\Psi$, while the level $\lambda^0_1= a_1 =1$ becomes an isolated outlier once it lies outside this disc, so that the asymptotic relaxation rate is set by whichever of the two lies further to the left. As long as the anisotropy is smaller than the disorder-induced broadening, $\delta a<\zeta_T\sqrt\gamma\,\Psi$, relaxation is slowed down as in the isotropic case, with a shift that decreases linearly with $\delta a$. Beyond this point, the background is effectively anisotropic, and the shift is of order $1/d$, as in \Secref{anisohighd}. The positive values of $\langle\Delta\tilde\mu_1\rangle_F$ found in low dimensions are part of this $1/d$ correction and vanish as $d\to\infty$. At fixed force variance, the disc is replaced by the ellipse of \Eqnref{edgehat}, whose size decreases as $d^{-1/2}$ for every $\hat\gamma$. The crossover then occurs at $\delta a\sim\hat\zeta_T\Psi/\sqrt d$, as for conservative forces below.

For conservative forces, $\gamma=0$, the matrix $\mbb{\tilde H}$ is symmetric with a semicircle of radius $2/\sqrt d$. For a rank-one perturbation of a symmetric random matrix the outlier separates from the semicircle only when the perturbation exceeds half the radius, and it is then displaced from the perturbation's eigenvalue towards the bulk~\cite{Fer07,ORo14}. Here this displacement is given by $-1/(d\,\delta\hat a)$, in agreement with \Eqnref{qisotail} at $\gamma=0$. We obtain
\algn{\eqnlab{qisohighd0}
	\langle\Delta\tilde\mu_1\rangle_F\to\css{\delta\hat a-\frac{2}{\sqrt{d}}\,, & \delta\hat a<\frac{1}{\sqrt d}\,,\\ -\frac1{d\,\delta\hat a}\,, & \delta\hat a>\frac{1}{\sqrt d}\,,}
}
which is negative everywhere. The crossover between the linear behaviour in $\delta\hat a$ and the asymptotic approach of $\langle\Delta\tilde \mu_1\rangle_F\to0$ now occurs at $\delta\hat a\sim1/\sqrt{d}$ or $\delta a\sim\zeta_T\Psi/\sqrt d$. Figure~\figref{qisohighd}(f) shows how $\sqrt{d}\langle\Delta\tilde\mu_1\rangle_F$ as a function of $x = \sqrt{d} \delta\hat a$ collapses onto $x-2$ for $x<1$ and $-1/x$ for $x>1$, with the transition at $x=1$. The agreement between the asymptotic form and the numerical data obtained from sampling the ensemble~\eqnref{tHcorr} for $\gamma=0$ is quite good already at $d=32$.

In this section, we described the behaviour in quasi-isotropic backgrounds, the right branch in the diagram in \Figref{flowdiagram}. To summarise, a background whose anisotropy is of the order of the disorder strength interpolates between the isotropic and the anisotropic cases. The mean shift $\langle \Delta\mu_1\rangle_F$ remains linear in $\zeta_T$ and depends on the correlation length, the disorder strength and the anisotropy only through the rescaled offsets $\delta\hat a_n$ of \Eqnref{dahat}, so that the rescaled means $\langle\Delta\tilde\mu_1\rangle_F$ collapse onto a single curve. For small offsets, the disorder mixes the nearly degenerate modes, and the negative isotropic mean is shifted by the mean offset $\tr(\delta\hat{\mbb a})/d$, \Eqnref{qisosmall}. For large offsets, it matches the second-order result of \Secref{aniso} and removes its divergence at $a_n\to1$, with a sign set by $1-2\gamma$, i.e., by whether level repulsion or level attraction dominates. In between, relaxation is faster on average for $\gamma>\gamma^*\approx0.44$ in $d=2$ and $d=3$, by up to about $0.1\,\zeta_T\Sigma(\kappa)d$. This speedup is linear in $\zeta_T$ and thus parametrically larger than that of anisotropic backgrounds. It is carried by level attraction, which merges the slowest mode with its neighbour into an oscillating complex-conjugate pair. It is largest at offsets $\delta\hat a\approx2$ to $3$, where the levels are close enough to merge in a sizeable fraction of the realisations, but far enough apart for the merged pair to relax substantially faster than the unperturbed slowest mode. The stiffening of the background, the other mechanism of faster relaxation, is of second order in $\zeta_T$ and therefore subdominant here. In high dimensions, $\langle\Delta\tilde\mu_1\rangle_F\to\min(\delta\hat a-\sqrt\gamma,0)$, which implies that anisotropy only ends the isotropic slowdown once it exceeds the disorder-induced broadening, and the speedup reduces to a correction of order $1/d$.
\section{Conclusions}\seclab{conclusions}
In this work, we have determined the asymptotic relaxation rate of a Brownian particle in a harmonic background potential and a weak quenched-random Gaussian force field in dimensions $d\geq2$, generalising the one-dimensional analysis of Ref.~\cite{Mei25a} to forces with conservative and non-conservative components. Combining perturbation theory for the slowest eigenvalue of the Fokker-Planck operator with random-matrix theory and numerical solutions of the eigenvalue problem in $d=2$ and $d=3$, we have characterised not only the mean of the disorder-induced shift $\Delta\lambda_1$, but also its fluctuations, and its full distribution. Three factors decide whether disorder slows relaxation down or speeds it up: the geometry of the background, which sets the degeneracy of the slowest unperturbed mode, the correlation length $\kappa$, and the fraction $\gamma$ of non-conservative forces.

In anisotropic backgrounds, the slowest mode is non-degenerate, and the mean shift is of second order in $\zeta_T$, with near-Gaussian fluctuations of first order. At short correlation lengths it approaches the universal value $-(1-\gamma)/d$ of \Eqnref{mushort}, which corresponds to an effectively softer background, \Eqnref{Aeff0}. At long correlation lengths it decays as $\kappa^{-4}$, with a sign set by the condition~\eqnref{prefac}. The resulting state diagram in \Figref{statediag}(a) separates the region in which disorder only slows relaxation down from the region in which it can also speed it up, independently of the functional form of the correlation function. For $\gamma>1/2$, relaxation at long correlation lengths is faster in every anisotropic background, because the attraction of the slowest mode to the faster levels then outweighs their repulsion. For $d\geq d_\text{th}=4$, increasing $\gamma$ speeds relaxation up at long correlation lengths, whatever the background. This threshold dimension refers to a fixed weight of each solenoidal polarisation direction and disappears when the total force variance is kept fixed.

In isotropic backgrounds, the disorder lifts the $d$-fold degeneracy of the slowest mode already at first order in $\zeta_T$. The shift is then set by the eigenvalue with the smallest real part of a Gaussian random matrix, whose statistics depend only on $d$ and $\gamma$, while the correlation length enters only through the factor $\Sigma(\kappa)$ and therefore cannot change the sign of the shift. Because the relaxation rate selects the slowest of $d$ mutually repelling eigenvalues, the mean shift is negative and its distribution strongly non-Gaussian, so that relaxation is generically slowed down. In low dimensions, non-conservative forces counteract this slowdown through level attraction, which creates complex-conjugate pairs. In high dimensions, the eigenvalues fill the ellipse~\eqnref{ellipse}, and the slowest one is pinned to its edge. Consequently, the shift becomes deterministic and faster relaxation exponentially unlikely. At a fixed fraction $\gamma>0$ of non-conservative forces, the slowdown saturates at a value proportional to $\sqrt\gamma$, \Eqnref{Dlamopt}, whereas for conservative forces, and at fixed total force variance for any mixture, it vanishes as $d^{-1/2}$, \Eqnref{Dlamopthat}.
\begin{figure}
	\centering
	\includegraphics[width=\linewidth]{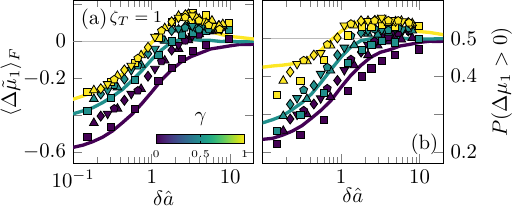}
	\caption{Quasi-isotropic background at strong disorder, $\zeta_T=1$, in $d=2$. (a) Rescaled mean $\langle\Delta\tilde\mu_1\rangle_F$ and (b) probability $P(\Delta\mu_1>0)$ of faster relaxation against $\delta\hat a$, for $\gamma=0$, $1/2$ and $1$ (colour bar). Lines: random-matrix result, as in \Figsref{qisoscal}(a) and~\figref{qisocollapse}. Markers: direct numerical solutions of \Eqnref{Leig} at $\kappa=2$ (squares), $\kappa=2.5$ (triangles), $\kappa=3$ (diamonds), $\kappa=3.5$ (inverted triangles) and $\kappa=4$ (pentagons), each with $4\times10^3$ pairs of realisations $\pm\ve F$, for seven to nine stiffnesses $0.05\leq a_2-1\leq3.2$, see Appendix~\secref{numerics}.}
	\figlab{qisostrong}
\end{figure}

The quasi-isotropic regime, in which the anisotropy is of the same order as the disorder strength, connects the two cases. The shift remains linear in $\zeta_T$ and depends on the correlation length, the disorder strength and the anisotropy only through the rescaled offsets~\eqnref{dahat}. For small offsets, the disorder mixes the nearly degenerate modes, so that the anisotropy enters only through its mean, \Eqnref{qisosmall}. For large offsets, the second-order result of the anisotropic case is recovered, and its divergence at vanishing anisotropy is removed. In between, relaxation becomes faster on average for $\gamma>\gamma^*\approx0.44$ in $d=2$ and $d=3$, by an amount that is linear in $\zeta_T$ and thus parametrically larger than in anisotropic backgrounds. This speedup is carried by level attraction, which merges the slowest mode with its neighbour into an oscillating complex-conjugate pair. In high dimensions, it reduces to a correction of order $1/d$.

Taken together, our results refine the intuition that disorder slows relaxation down. Higher dimensions do generate more ways to be slow: the slowest of many split levels is selected, which makes the slowdown in isotropic backgrounds generic and, for non-conservative forces, robust in high dimensions. A speedup, by contrast, arises from two distinct mechanisms. The first is the stiffening of the background known from one dimension~\cite{Mei25a}: at long correlation lengths, the disorder acts as an additional linear force, which on average stiffens the background around its displaced minimum. It does not require non-conservative forces, but for predominantly conservative forces it leads to faster relaxation only when the stiffer directions are stiff enough to make the repulsion from the faster levels weak. The second mechanism, level attraction, has no counterpart in one dimension. Non-conservative forces make the coupling of the slowest mode to its neighbours antisymmetric on average, so that the slowest level is attracted towards faster ones. In anisotropic backgrounds, this is a correction of second order in the disorder strength, analogous to the acceleration of Gaussian diffusions by non-reversible drifts~\cite{Hwa93}. When the anisotropy of the background is comparable to the disorder, it acts at first order and can merge the slowest mode with its neighbour into an oscillating complex-conjugate pair. This mechanism is therefore most effective at intermediate anisotropy, where the levels are close enough to merge, yet far enough apart for the merger to raise the relaxation rate substantially. In anisotropic and isotropic backgrounds alike, the disorder is most effective when its correlation length is comparable to the root-mean-square size $\ell_\text{rms}$ of the thermal fluctuations, which grows as $\sqrt d$.

We now summarise possible future directions. First, we have focused on the asymptotic relaxation rate. In isotropic and quasi-isotropic backgrounds, the slowest levels are split only at order $\zeta_T$. In high-dimensional isotropic backgrounds, moreover, the $d$ split levels are spread over a region of fixed size, so that their spacing closes as $d^{-1/2}$. The time needed to reach the asymptotic regime then grows as the disorder weakens and with the dimension. On shorter time scales, all nearly degenerate levels contribute to the decay, and for non-conservative forces the non-normality of the perturbation matrix can amplify transients before the asymptotic decay sets in~\cite{Cha98}. Such eigenvector non-orthogonality affects transient Langevin dynamics, fluctuation-dissipation relations, and entropy production even at fixed spectrum~\cite{Fyo25}. Relaxation on these intermediate time scales is not captured by the slowest mode alone and requires a separate analysis than the one carried out here. Second, we focussed on weak disorder because it is accessible analytically, but our numerics suggest that the mechanisms for relaxation speedups and slowdowns survive at stronger disorder. In $d=2$, the regions of faster relaxation predicted for weak disorder persist up to $\zeta_T=1$ (\Figref{statemap}), although for non-conservative forces in anisotropic backgrounds the region recedes to larger correlation lengths. Figure~\figref{qisostrong} shows simulations in $d=2$ at $\zeta_T=1$ in the quasi-isotropic case with an anisotropy of the order of the disorder. Although the data deviate from the leading-order random-matrix curves, the rescaled mean for $\gamma=1$ retains its positive maximum near $\delta\hat a\approx3$ for $2\leq\kappa\leq4$, and faster relaxation remains more likely than slower relaxation. For $\kappa=2$, the mean relaxation rate exceeds its unperturbed value by about $5\%$. Third, the model can be realised with colloidal particles in optical traps combined with laser speckles or synthetic random light fields~\cite{Bew16a,Vol14,Vos26}, while non-conservative forces arise, for instance, in rotating laser traps~\cite{Bli07b}. Since the long-time decay of the position correlation function $C_x(t)$ can be measured from stationary trajectories (\Secref{relspeed}), the disorder-induced speedup in quasi-isotropic backgrounds should be accompanied by damped oscillations of $C_x(t)$, cf.~\Eqnref{oscill}. Finally, a crucial next step is to apply the insights from the present minimal model to more complicated systems of interacting particles, to time-dependent landscapes~\cite{Mak26}, or to the rugged loss landscapes of neural-network learning~\cite{Win25}.

\section*{Appendix}
\appendix
\section{Near-Gaussian statistics in anisotropic backgrounds}\seclab{gaussian}
In this appendix, we derive the distribution of the eigenvalue shift $\Delta\lambda_1$ in an anisotropic background to next-to-leading order, and from it the probability shift $\Delta P$ of \Eqnref{DP}. We write the second-order expansion~\eqnref{ev1} as $\Delta\lambda_1\sim\zeta_TX_1+\zeta_T^2X_2$ with
\algn{\eqnlab{X1X2}
	X_1 = \langle\ve e_1|\ms{H}^\dagger_F|\ve e_1\rangle\,,\qquad X_2 = -\sum_{\ve m\neq\ve e_1}\frac{A_{\ve m}B_{\ve m}}{\ve a\cdot\ve m-a_1}\,,
}
where $A_{\ve m}=\langle\ve e_1|\ms{H}^\dagger_F|\ve m\rangle$ and $B_{\ve m}=\langle\ve m|\ms{H}^\dagger_F|\ve e_1\rangle$. All matrix elements are linear in the Gaussian force field $\ve F$, so that $X_1$, $A_{\ve m}$ and $B_{\ve m}$ are jointly Gaussian with zero mean. As in the main text, we denote by $\sigma^2_{\Delta\lambda_1}=\langle X_1^2\rangle_F$ the variance of \Eqnref{sigmain1} and by $\mu_{\Delta\lambda_1}=\langle X_2\rangle_F$ the mean of \Eqnref{mu1}.

The distribution function of the rescaled shift $\Delta\lambda_1/\zeta_T$ follows by expanding the indicator function of the event $X_1+\zeta_TX_2<u$ in $\zeta_T$,
\algn{\eqnlab{cdfDlam}
	P(\Delta\lambda_1<\zeta_Tu) &\sim P(X_1+\zeta_TX_2<u)\nn\\
	&\sim G\left(\frac {u}{\sigma_{\Delta\lambda_1}}\right)-\zeta_Tp_{X_1}(u)\langle X_2|X_1=u\rangle_F\,,
}
where $p_{X_1}(u)=\ee^{-u^2/(2\sigma_{\Delta\lambda_1}^2)}/(\sqrt{2\pi}\sigma_{\Delta\lambda_1})$ is the density of $X_1$, $G$ the standard Gaussian cumulative distribution function,
\algn{
	G(x) = \int_{-\infty}^x\!\!\!\ed x'\,\frac{\ee^{-\frac{x'^2}2}}{\sqrt{2\pi}}\,,
}
 and $\langle X_2|X_1=u\rangle_F$ the mean of $X_2$ over the realisations with $X_1=u$. To leading order, $\Delta\lambda_1$ is thus Gaussian with zero mean and variance $\zeta_T^2\sigma_{\Delta\lambda_1}^2$~\cite{Mei25a}.

The conditional mean follows from the joint Gaussian statistics. We decompose $A_{\ve m}=\alpha_{\ve m}X_1+A'_{\ve m}$ and $B_{\ve m}=\beta_{\ve m}X_1+B'_{\ve m}$ with $\alpha_{\ve m}=\langle X_1A_{\ve m}\rangle_F/\sigma_{\Delta\lambda_1}^2$ and $\beta_{\ve m}=\langle X_1B_{\ve m}\rangle_F/\sigma_{\Delta\lambda_1}^2$. The remainders $A'_{\ve m}$ and $B'_{\ve m}$ are uncorrelated with $X_1$ and Gaussian, and therefore independent of $X_1$. Hence $\langle A_{\ve m}B_{\ve m}|X_1=u\rangle_F=\langle A_{\ve m}B_{\ve m}\rangle_F+\alpha_{\ve m}\beta_{\ve m}(u^2-\sigma_{\Delta\lambda_1}^2)$, and
\algn{\eqnlab{condmean}
	\langle X_2|X_1=u\rangle_F = \mu_{\Delta\lambda_1}-\frac{u^2-\sigma_{\Delta\lambda_1}^2}{\sigma_{\Delta\lambda_1}^4}S\,,
}
where
\algn{\eqnlab{Sdef}
	S & =\sum_{\ve m\neq\ve e_1}\frac{\langle X_1A_{\ve m}\rangle_F\langle X_1B_{\ve m}\rangle_F}{\ve a\cdot\ve m-a_1}\,.
}
Differentiating \Eqnref{cdfDlam} with respect to $u$ and using \Eqnref{condmean} gives the density of $\Delta\lambda_1/\zeta_T$,
\algn{\eqnlab{pdfDlam}
	p(u)\sim p_{X_1}(u)\left[1+\zeta_T\left(\frac{\mu_{\Delta\lambda_1} u}{\sigma_{\Delta\lambda_1}^2}-\frac{S}{\sigma_{\Delta\lambda_1}^3}\mathrm{He}_3\!\left(\frac {u}{\sigma_{\Delta\lambda_1}}\right)\right)\right]\,,
}
with the Hermite polynomial $\mathrm{He}_3(x)=x^3-3x$. Equation~\eqnref{pdfDlam} contains the first terms of a so-called Edgeworth series~\cite{Kol06} in terms of the cumulants of $\Delta\lambda_1/\zeta_T$, where the mean of $\Delta\lambda_1/\zeta_T$ is $\zeta_T\mu_{\Delta\lambda_1}$, and its third cumulant is $3\zeta_T\mathrm{Cov}(X_2,X_1^2)=-6\zeta_TS$ to leading order in $\zeta_T$. To this order, the distribution of $\Delta\lambda_1$ is therefore determined by $\mu_{\Delta\lambda_1}$, $\sigma_{\Delta\lambda_1}^2$ and $S$ alone. The probability shift follows from \Eqnref{cdfDlam} at $u=0$,
\algn{\eqnlab{DPcorr}
	\Delta P &=\frac12-P(\Delta\lambda_1<0)\sim\zeta_Tp_{X_1}(0)\langle X_2|X_1=0\rangle_F\nn\\
	&=\frac{\zeta_T}{\sqrt{2\pi}\sigma_{\Delta\lambda_1}}\left(\mu_{\Delta\lambda_1}+\frac S{\sigma_{\Delta\lambda_1}^2}\right)\,,
}

To evaluate $S$, we proceed as for the mean in \Secref{aniso}, using \Eqnref{amint}, the matrix elements~\eqnref{element1} and~\eqnref{element2}, and the generating sum over $\ve m$. This gives
\algn{\eqnlab{Sint}
	S=\int_0^1\!\!\ed s\,s^{-2}\left[H(s)-s\sigma_{\Delta\lambda_1}^4\right]\,,
}
with
\begin{multline}\eqnlab{Hs}
	H(s)=\kappa^{-8}\sum_{n=1}^ds^{a_n}\!\!\int_{\mbb{R}^d}\!\!\ed^dk\int_{\mbb{R}^d}\!\!\ed^dq\,\hat C_F(\ve k)\hat C_F(\ve q)\\
	\times\ms{G}_{1n}(\ve k)\,\ms{G}_{11}(\ve q)\,k_1q_1q_n(k_1+s\,q_1)\,\ee^{-\ve y^2-\ve y_q^2-\ve y^{(s)}\cdot\ve y_q}\,,
\end{multline}
where $\ve y$ is given by \Eqnref{xydef} and $\ve y_q=\kappa^{-1}\mbb{a}^{-1/2}\ve q$. The second term in \Eqnref{Sint} removes $\ve m=\ve e_1$, for which $A_{\ve e_1}=B_{\ve e_1}=X_1$, so that the integrand remains finite for $s\to0$.

For the Gaussian correlation function $\hat C_F(\ve k)=\pi^{-d}\ee^{-\ve k^2/\pi}$ the pairs $(k_n,q_n)$ are bivariate Gaussian and independent for different $n$, so that $H(s)$ is a sum of moments of bivariate normal distributions, and only the integral over $s$ in \Eqnref{Sint} remains to be done numerically. Figure~\figref{prob} shows the resulting probability shift for $d=2$. For the parameters shown there, $S/\sigma_{\Delta\lambda_1}^2$ changes $\Delta P$ noticeably at intermediate $\kappa$, while it is negligible compared with $\mu_{\Delta\lambda_1}$ for $\kappa\to\infty$ and, for $\gamma<1$, for $\kappa\to0$.

In the high-dimensional limit of \Secref{anisohighd}, $S$ is negligible. Evaluating \Eqnref{Sint} at fixed $\hat\kappa=1/2$, $1$ and $2$ for $\mbb{a}=\text{diag}(1,a_2,\ldots,a_2)$ with $a_2=5/4$ and $a_2=10$ and in dimensions up to $d=10^4$, we find that $S/\sigma^4_{\Delta\lambda_1}$ tends to a negative constant for $\gamma=0$ and decays as $d^{-1}$ for $\gamma>0$. With \Eqnref{sigscalmain}, $S/\sigma^2_{\Delta\lambda_1}$ thus decays as $d^{-2}$ for every $\gamma$, and also at fixed $\hat\gamma<1$. Since $\mu_{\Delta\lambda_1}$ decays only as $d^{-1}$ [\Eqnref{muscalmain}], the relative correction $S/(\sigma^2_{\Delta\lambda_1}\mu_{\Delta\lambda_1})$ to \Eqnref{probshift} vanishes as $d^{-1}$, like the deviation of $d\mu_{\Delta\lambda_1}$ from the scaling function $\Phi_\gamma$ at finite $d$.

\section{Linear forces}\seclab{linforce}
As a first step we consider linear forces in the non-degenerate case in the form $\ve{F}^\text{lin}(\ve x)=\mbb{b}\ve x$ with Fourier transform
\algn{\eqnlab{linfourier}
	\ve{\hat{F}}^\text{lin}(\ve k) =i\mbb{b}\nabla_{\ve k}\delta(\ve k)\,.
}
The matrix $\mbb{b}$ can be either entirely deterministic or quenched-random. Linear conservative forces correspond to symmetric $\mbb{b}$, while non-conservative forces are represented by an antisymmetric matrix $\mbb{b}$. For the Taylor expansion of the force field in Appendix~\secref{kapexp}, we allow $\mbb{b}$ itself to be an expansion in $\zeta_T$ of the form
\algn{\eqnlab{bexp}
	\mbb{b} = \mbb{b}^{(0)} + \zeta_T \mbb{b}^{(1)} + O(\zeta_T^2)\,.
}
Since $\mbb{b}$ enters the dynamics multiplied by $\zeta_T$, its terms of order $\zeta_T^2$ affect the eigenvalues only at third order. We split $\mbb{b}$ into its symmetric and antisymmetric parts,
\algn{
	\mbb{b} = \mbb{b}^\text{s} + \mbb{b}^\text{a}\,,\qquad \mbb{b}^\text{s} = \frac12\left(\mbb{b}+\mbb{b}^{\sf T}\right)\,,\qquad \mbb{b}^\text{a} = \frac12\left(\mbb{b}-\mbb{b}^{\sf T}\right)\,,
}
and denote the elements of $\mbb{b}$, $\mbb{b}^\text{s}$ and $\mbb{b}^\text{a}$ by $b_{nm}$, $b^\text{s}_{nm}$ and $b^\text{a}_{nm}$. To compute $\Delta\lambda_1$ for the linear forces \eqnref{linfourier}, we apply \Eqnref{Dlam_N}. To leading order in $\zeta_T$, we obtain
\algn{
	-\frac{\zeta_T}{\kappa}\int_{\mbb{R}^d}\!\!\ed^dk\, \hat F^{\text{lin}}_1(\ve k)(iy_1)\ee^{-\frac{\ve y^2}2}= -\frac{\zeta_T}{\kappa^2}b_{11}= -\frac{\zeta_T}{\kappa^2}b^\text{s}_{11}\,,
}
where the antisymmetric part vanishes because of its vanishing diagonal elements. With \Eqnref{bexp}, this term contributes $-\zeta_T\kappa^{-2}b^{(0)}_{11}$ at first order and $-\zeta_T^2\kappa^{-2}b^{(1)}_{11}$ at second order.

The second-order term of \Eqnref{Dlam_N} requires only $\mbb{b}^{(0)}$ and has the structure
\begin{multline}
	\sum_{j,l,m=1}^d\sqrt{a_m}b^{(0)}_{1j}b^{(0)}_{ml}\int_{\mbb{R}^d}\!\!\ed^dk\int_{\mbb{R}^d}\!\!\ed^dk'\, \partial_{k_j}\delta(\ve k) \partial_{k'_l} \delta(\ve k')\\
	\times y_m^{(s)}\partial_{y'_1}\left(\ee^{-\ve y^{(s)}\cdot\ve y'}-\delta_{1m} \right)\ee^{-\frac12\left({\ve y}^2+{\ve y'}^2\right)}\,.
\end{multline}
Using integration by parts we write this expression as
\begin{multline}
	\sum_{j,l,m=1}^d\sqrt{a_m}\frac{b^{(0)}_{1j}b^{(0)}_{ml}}{\kappa^2\sqrt{a_j a_l}}\partial_{y_j}\partial_{y'_l}\partial_{y'_1}\left[y_m^{(s)}\left(\ee^{-\ve y^{(s)}\cdot\ve y'}\right.\right.\\
	\left.\left.-\delta_{1m}\right)\ee^{-\frac12\left({\ve y}^2+{\ve y'}^2\right)}	\right]\bigg|_{\ve y=\ve y'=0}\!\!\!\!\!\!\!\!\!=-\sum_{m=1}^ds^{a_m}\frac{b^{(0)}_{1m}b^{(0)}_{m1}}{\kappa^2}(1-\delta_{1m})\,.
\end{multline}
Multiplying by $s^{-2}$ and integrating over $s$, we obtain $\Delta\lambda_1$ to second order in $\zeta_T$ for linear forces. Repeating these steps for the level $a_n$ of the state $\ve e_n$ instead of $\ve e_1$, we find for any non-degenerate level
\algn{\eqnlab{detforcecorr}
	\Delta\lambda_n 		&\sim-\frac{\zeta_T}{\kappa^2}b^{(0)}_{nn}-\frac{\zeta_T^2}{\kappa^2}b^{(1)}_{nn} -\frac{\zeta_T^2}{\kappa^4}\sum_{m\neq n}^d\frac{b^{(0)}_{nm}b^{(0)}_{mn}}{a_m-a_n}\,,\\
								&\sim-\frac{\zeta_T}{\kappa^2}b^\text{s}_{nn}-\frac{\zeta_T^2}{\kappa^2}b^{(1)}_{nn} +\frac{\zeta_T^2}{\kappa^4}\sum_{m\neq n}^d\frac{(b^\text{a}_{nm})^2-(b^\text{s}_{nm})^2}{a_m-a_n}\,,
}
where $\mbb{b}^\text{s}$ and $\mbb{b}^\text{a}$ in the second line are the parts of $\mbb{b}^{(0)}$. When $\mbb{b}$ is purely antisymmetric, we have $b_{m1}=-b_{1m}$, and since $a_m>1$ for $m\geq2$, $\Delta\lambda_1$ is always positive in this case.  This means that adding a weak non-conservative force consistently increases the rate of relaxation towards the stationary state. The opposite is true for a purely symmetric $\mbb{b}=\mbb{b}^\text{s}$, which only replaces the stiffness matrix $\mbb{a}$ of the background potential by $\mbb{a}-\zeta_T\kappa^{-2}\mbb{b}^\text{s}$. The smallest eigenvalue of this matrix lies below its diagonal element $1-\zeta_T\kappa^{-2}b^\text{s}_{11}$, so that the sum in \Eqnref{detforcecorr} is negative. We shall see below that generically, this sort of behaviour persists for non-linear quenched-random forces.

For linear forces, the dynamics is an Ornstein-Uhlenbeck process, and the eigenvalues of the Fokker-Planck operator are sums, with non-negative integer coefficients, of the eigenvalues of the matrix of the total linear restoring force~\cite{Met02}. Equation~\eqnref{detforcecorr} is therefore the shift, to second order in $\zeta_T$, of the eigenvalue of this matrix closest to $a_n$, and the relaxation rate depends on its spectrum alone. The non-orthogonality of its eigenvectors, which for non-conservative forces shapes transients, fluctuation-dissipation relations and entropy production even at fixed spectrum~\cite{Fyo25}, does not enter.

Note also that the correction in \Eqnref{detforcecorr} becomes large as $a_m\to a_n$, i.e., as the stiffnesses of the background potential become equal. In this case, the spectrum of the Fokker-Planck operator becomes degenerate and the non-degenerate perturbation theory that we have set up here does not apply. We shall discuss the degenerate case in \Secref{iso}.

\section{Anisotropic backgrounds at short correlation lengths}\seclab{short}
Repeating the steps that lead to \Eqnref{mu1} for the level $a_n$ of the state $\ve e_n$ instead of $\ve e_1$, we obtain its mean shift
\begin{multline}\eqnlab{mulevn}
	\langle\Delta\lambda_n\rangle_F \sim -\frac{\zeta_T^2}{\kappa^4}\sum_{m=1}^d\int_0^1\!\!\ed s\, s^{a_m-a_n-1}\!\!\int_{\mbb{R}^d}\!\!\ed^dk\, \hat C_F(\ve k)\ms{G}_{nm}(\ve k)k_nk_m\\
	\times\ee^{-\ve y^2}\left[(1-s^{a_n})\ee^{\ve y^{(s)}\cdot\ve y}-\delta_{nm}\right]\,.
\end{multline}
The limit $\kappa\to0$ corresponds to $\ve y \to \infty$. In this limit, the $s$ integral in \Eqnref{mulevn} is dominated by the behaviour around $s=1$. Transforming $\tilde s = 1-s$ and expanding to lowest order in $\tilde s$ we thus obtain
\begin{multline}
	\int_0^1\!\!\ed s\, s^{a_m-a_n-1}\ee^{-\ve y^2}\left[(1-s^{a_n})\ee^{\ve y^{(s)}\cdot\ve y}-\delta_{nm}\right]\\
	\sim a_n\!\!\!\int_{0}^1\!\!\!\ed \tilde s\, \tilde s\, \ee^{-\ve k^2 \tilde s/\kappa^2} \sim \frac{a_n \kappa^4}{k^4}\,,
\end{multline}
which is independent of the index $m$. Summing over $m$ and using that $\sum_{m=1}^d \ms{G}_{nm}k_m = (1-\gamma)k_n {\ve k}^2$ projects out the non-conservative contributions, we obtain
\algn{\eqnlab{smallkapN}
	\langle \Delta\lambda_n\rangle_{F} \sim -\zeta_T^2a_n(1-\gamma)\int_{\mbb{R}^d}\!\!\ed^dk\ \hat C_F(\ve k)\frac{k^2_n}{k^2} = -a_n\frac{1-\gamma}{d}\zeta_T^2\,.
}
In particular, for $n=1$ we obtain in the limit $\kappa\to0$ the simple formula
\algn{\eqnlab{smallkap}
	\langle \Delta\lambda_1\rangle_{F}  \sim -\frac{1-\gamma}{d}\zeta_T^2\,,
}
independently of the correlation function. This expression generalises the one-dimensional result $\langle \Delta\lambda_1\rangle_F\sim-\zeta_T^2$ obtained in \cite{Mei25a} ($\gamma=0$ in one dimension).
\section{Replica calculation}\seclab{replica}
The result~\eqnref{smallkapN} can be recovered by a simple replica calculation. The idea is that when $\kappa\to0$ the disorder is so irregular that each individual relaxing particle experiences the entire statistics of the disorder ensemble within a single relaxation process. One can then compute the disorder-averaged free energy of the particle according to
\algn{
	\langle \ms{F}\rangle_{F} = -\langle \ln Z\rangle_{F}\,,
}
where the partition function reads
\algn{
	Z = \int_{\mbb{R}^d}\!\!\!\ed^d x \ee^{-V_0(\ve x) - \zeta_T\sqrt{1-\gamma}\,V(\ve x/\kappa)}\,,
}
where $V(\ve x)$ is the random potential of \Eqnref{force}, which generates the conservative part of the random forces $\ve F(\ve x/\kappa)$. The logarithm is evaluated using the identity
\algn{
	\langle\ln Z\rangle_{F} = \langle\DD{r}Z^r\big|_{r=0}\rangle_F\,.
}
Swapping the order of derivative and average, we need to compute the average $\langle Z^r\rangle_F$, where $r$ is the number of so-called replicas. This average is written as
\algn{
	\langle Z^r \rangle_{F} = \int_{\mbb{R}^{d r}}\!\!\!\ed^{d r} x\ \ee^{-\sum_{l=1}^rV_0(\ve x^{(l)})}  \langle\ee^{- \zeta_T\sqrt{1-\gamma}\sum_{l=1}^rV(\ve x^{(l)}/\kappa)}\rangle_F\,.
}
The Gaussian average over $V$ can be evaluated as
\algn{
	\langle\ee^{- \zeta_T\sqrt{1-\gamma}\sum_{l=1}^rV(\ve x^{(l)}/\kappa)}\rangle_{F} = \ee^{\frac{(1-\gamma)\zeta_T^2}2\sum_{l,l'=1}^rC_F(|\ve x^{(l)}-\ve x^{(l')}|/\kappa)}\,,
}
where we have used that $\langle V\rangle_{F}=0$. For $\kappa\to0$ all off-diagonal terms in the sum over correlation functions vanish and with $C_F(0)=1$, we have
\algn{
	\sum_{l,l'=1}^rC_F(|\ve x^{(l)}-\ve x^{(l')}|/\kappa) \sim r\,.
}
The remaining integral is Gaussian and we find
\algn{
	\langle Z^r \rangle_{F} = \left[\frac{(2\pi)^d}{\prod_{n=1}^d \ee^{-(1-\gamma)\zeta_T^2/d}a_n }\right]^{\frac{r}2}\,.
}
Taking the $r$ derivative and evaluating at $r=0$ gives
\algn{
	\langle \ms{F}\rangle_{F} = -\frac12\ln\left[\frac{(2\pi)^d}{\prod_{n=1}^d \ee^{-(1-\gamma)\zeta_T^2/d}a_n }\right]\,,
}
which we identify as the equilibrium free energy of a single particle in a harmonic potential with the decreased stiffness
\algn{
	a^\text{eff}_n = \ee^{-(1-\gamma)\zeta_T^2/d}a_n\,.
}
This shows that relaxation at $\kappa\to0$ is effectively relaxation in a harmonic potential with disorder-dependent stiffness. Using that $\langle\Delta\lambda_n\rangle_{F}\sim (a_n^\text{eff}-a_n)$ and expanding for small $\zeta_T$, we recover the perturbative result~\eqnref{smallkapN}.
\section{Anisotropic backgrounds at large correlation lengths}\seclab{long}
In the opposite limit of large correlation lengths $\kappa\to\infty$, i.e., $y_i\to0$, a similar series of simplifications occurs. Also in this case, the $s$ integral in \Eqnref{mulevn} simplifies considerably:
\begin{multline}
	\int_0^1\!\!\ed s\, s^{a_m-a_n-1}\ee^{-\ve y^2}\left[(1-s^{a_n})\ee^{\ve y^{(s)}\cdot\ve y}-\delta_{nm}\right]\\
	\sim \int_0^1\!\!\ed s\, s^{a_m-a_n-1}\left[1-s^{a_n}-\delta_{nm}\right] = \frac{1-\delta_{nm} }{a_m-a_n }-\frac{1}{a_m}\,,%
\end{multline}
we then obtain for $\langle \Delta\lambda_n\rangle_F$ the expression
\begin{multline}
	\langle \Delta\lambda_n\rangle_F \sim-\frac{\zeta^2_T}{\kappa^4}\sum_{m=1}^d\int_{\mbb{R}^d}\!\!\ed^dk\ \hat C_F(\ve k)\ms{G}_{nm}(\ve k)k_nk_m\\
	\times\left(\frac{1-\delta_{nm} }{a_m-a_n}-\frac{1}{a_m}\right)\,,
\end{multline}
which, after reinstalling the force correlation function and an integration by parts, gives
\algn{\eqnlab{largekap}
	\langle \Delta\lambda_n\rangle_F \sim\frac{\zeta^2_T}{\kappa^4}\sum_{m=1}^d\left(\frac{1-\delta_{nm} }{a_m-a_n}-\frac{1}{a_m}\right)\partial_n\partial_m\langle F_n(\ve x)F_m(0)\rangle_F|_{\ve x=0}\,.
}
Again, in one dimension ($d=1$, $\gamma = 0$), we recover $\langle \Delta\lambda_1 \rangle_F \sim \zeta^2_T C_F^{(4)}(0)/\kappa^4$~\cite{Mei25a}.
As in the small-$\kappa$ case, we obtain an essentially universal result, which, however, still depends on the local properties of the correlation function. In this sense, it is somewhat less universal than \Eqnref{smallkap}.
\section{Second-order shift for Gaussian correlations}\seclab{gauss2}
For the Gaussian correlation function $\rho(r^2)=\ee^{-\pi r^2/4}$ of \Secref{model}, whose Fourier transform is $\hat C_F(\ve k)=\pi^{-d}\ee^{-\ve k^2/\pi}$, the integrals in \Eqnref{mu1} can be carried out in closed form for every $\kappa$ and every $d$. Collecting the exponents in \Eqnref{mu1}, the Gaussian weight, the factor $\ee^{-\ve y^2}$ and the factor $\ee^{\ve y^{(s)}\cdot\ve y}$ combine into a single anisotropic Gaussian,
\algn{\eqnlab{gaussexp}
	\hat C_F(\ve k)\,\ee^{-\ve y^2}\ee^{\ve y^{(s)}\cdot\ve y} = \frac1{\pi^d}\exp\left[-\sum_{n=1}^d\frac{k_n^2}{\pi r_n^2(s)}\right]\,,
}
where we defined the dimensionless widths
\sbeqs{\eqnlab{rdef}
\algn{
	r_n(s) = \frac1{\sqrt{1+\frac{\pi\left(1-s^{a_n}\right)}{a_n\kappa^2}}}\,,\quad \hat r_n = r_n(0) = \frac1{\sqrt{1+\frac{\pi}{a_n\kappa^2}}}\,.
}
}
The subtracted term in \Eqnref{mu1}, which carries no factor $\ee^{\ve y^{(s)}\cdot\ve y}$, is obtained from the same expression by replacing $r_n(s)\to\hat r_n$. Both $r_n$ and $\hat r_n$ increase monotonically towards unity with increasing $\kappa$.

The remaining $\ve k$ integrals are Gaussian moments with
\algn{\eqnlab{Rdef}
	\ms{R}(s) = \prod_{n=1}^d r_n(s)\,, \qquad \hat{\ms{R}} = \ms{R}(0)\,,
}
the normalisation and the fourth moments read
\sbeqs{
\algn{
	\frac1{\pi^d}\int_{\mbb{R}^d}\!\!\ed^dk\,\ee^{-\sum_n\frac{k_n^2}{\pi r_n^2}} &= \ms{R}\,,\\
	\frac1{\pi^d}\int_{\mbb{R}^d}\!\!\ed^dk\,\ee^{-\sum_n\frac{k_n^2}{\pi r_n^2}}k_n^2k_m^2 &= \frac{\pi^2}4\ms{R}\,r_n^2r_m^2\left(1+2\delta_{nm}\right)\,.
}
}
It is convenient to collect the two index structures of $\ms{G}_{1n}$ into
\algn{\eqnlab{Qdef}
	\ms{Q}_n(s) = \ms{R}(s)\,r_1^2(s)\,r_n^2(s)\left(1+2\delta_{1n}\right)\,,
}
and
\algn{
	 w_1 = 1-\gamma\,,\quad w_{n\geq2} = \gamma\,,
}
so that the $n=1$ contribution to \Eqnref{mu1}, which contains both $k_1^4$ and $k_1^2(\ve k^2-k_1^2)$, becomes $\sum_n w_n\ms{Q}_n$, while the contributions with $n\geq2$ are proportional to $\ms{Q}_n$ alone. Performing the $\ve k$ integrals, \Eqnref{mu1} becomes
\begin{multline}\eqnlab{muGauss}
	\mu_{\Delta\lambda_1} = -\frac{\pi^2}{4\kappa^4}\Bigg\{\int_0^1\!\!\!\ed s\, s^{-1}\sum_{n=1}^dw_n\left[(1-s)\ms{Q}_n(s)-\ms{Q}_n(0)\right]\\
	+ (1-2\gamma)\sum_{n=2}^d\int_0^1\!\!\ed s\, s^{a_n-2}(1-s)\,\ms{Q}_n(s)\Bigg\}\,.
\end{multline}
Equation~\eqnref{muGauss} is the exact perturbative expression for $\mu_{\Delta\lambda_1}$ with Gaussian correlation functions. Both $s$ integrals converge: as $s\to0$ one has $r_n(s)\to\hat r_n$, so that the bracket in the first line vanishes linearly in $s$, and the integrand of the second line behaves as $s^{a_n-2}$ with $a_n>1$.

To check \Eqnref{muGauss}, we evaluate it at large $\kappa$, where $r_n\to1$ for all $n$ and $\ms{R}\to1$, so that $\sum_nw_n\ms{Q}_n\to3(1-\gamma)+\gamma(d-1)$ and $\ms{Q}_n\to1$ for $n\geq2$. The first integral then evaluates to $-[3(1-\gamma)+\gamma(d-1)]$ and the second to $\sum_{n\geq2}[a_n(a_n-1)]^{-1}$, so that
\algn{
	\mu_{\Delta\lambda_1}\to\frac{\pi^2}{4\kappa^4}\left[3+\gamma(d-4)-\sum_{n=2}^d\frac{1-2\gamma}{a_n(a_n-1)}\right]\,,
}
which is \Eqnref{mulong} with $\rho''(0)=\pi^2/16$.

We now take $d$ large at fixed $\hat\kappa$. In this regime $\pi/(a_n\kappa^2)$ is small, so that each individual $r_n$ tends to unity, but the product~\eqnref{Rdef} does not, because it contains $d$ such factors:
\algn{\eqnlab{Rasym}
	\hat{\ms{R}}\to\ee^{-\frac{\pi}{2\hat\kappa^2}}\,,\qquad \ms{R}(s)\to\ee^{-\frac{\pi}{2\hat\kappa^2}\left[1-g(s)\right]}\,,
}
where the background enters only through the function $g(s)$ of \Eqnref{gdefmain}, whose slope at $s=1$ is the harmonic mean of the stiffnesses,
\algn{\eqnlab{abardef}
	\bar a = g'(1) = \frac{d}{\sum_{n=1}^d a_n^{-1}} = \frac{d\,\ell^2_\text{T}}{\ell^2_\text{rms}}\,.
}
Moreover, $\ms{Q}_n(s)\to \ms{R}(s)$ and $\ms{Q}_n(0)\to\hat{\ms{R}}$ both become independent of the index $n$. Retaining only the terms that grow with $d$ in \Eqnref{muGauss}, both sums contribute a factor $d$, and we obtain the scaling form~\eqnref{muscalmain}, with $\Phi_\gamma$ given by \Eqnref{muscalmain2} and the integrals $\ms{J}$ and $\ms{K}$ of \Eqnref{Jdefmain}.

For $\hat\kappa\to\infty$ one has $\ms{J}\to-1$ and $\ms{K}\to d^{-1}\sum_{n\geq2}[a_n(a_n-1)]^{-1}$, so that $\Phi_\gamma/d$ reduces to the large-$d$ limit of \Eqnref{mulong}. For $\hat\kappa\to0$ both integrals in \Eqnref{Jdefmain} are dominated by the integrand at $s\to1$, where $1-g(s)\to\bar a(1-s)$ by \Eqnref{abardef}, so that $\ms{J},\ms{K}\to[2\hat\kappa^2/(\bar a \pi)]^{2}$ and $\Phi_\gamma\to-(1-\gamma)$, so $\Phi_\gamma/d$ then recovers the small-$\kappa$ limit in \Eqnref{mushort}. Hence $\Phi_\gamma$ as a function of $\hat\kappa$ interpolates between both known limiting cases for large $d$.
\section{\texorpdfstring{Taylor expansion of the force field}{Taylor expansion of the force field}}\seclab{kapexp}
The result~\eqnref{largekap} can also be obtained directly from the Langevin equation~\eqnref{laneqn} by expanding in large $\kappa$. With summation over repeated indices, we have
\algn{
	\dot{x_i} \sim{}& -a_i x_i + \frac{\zeta_T}{\kappa} F_i(0) + \frac{\zeta_T}{\kappa^2} \partial_jF_i(0)x_j\nn\\
	&+ \frac12\frac{\zeta_T}{\kappa^3} \partial_k\partial_jF_i(0)x_jx_k + \xi_i(t)\,.\eqnlab{lanexp}
}
We first determine the noise-free fixed point $\ve x^*$ of $\ve x$ to lowest order in $\kappa^{-1}$. This gives
\algn{
	x_i^* \sim \frac{\zeta_T}{\kappa}\frac{F_i(0)}{a_i} +\frac{\zeta_T^2}{\kappa^3}\frac{\partial_j F_i(0)F_j(0)}{a_ia_j}\,,
}
which shows that the fixed point is shifted in the presence of the random force. Therefore, we express \Eqnref{lanexp} in terms of the shifted variable $\ve{\tilde x} = \ve x -\ve x^*$, leading us to a linear equation for $\ve{\tilde x}$:
\algn{
	\dot{\tilde x}_i \sim\left[-a_i\delta_{ij} +  \frac{\zeta_T}{\kappa^2} \partial_jF_i(0)+\frac{\zeta^2_T}{\kappa^4} \frac{\partial_k\partial_jF_i(0)F_k(0)}{a_k}\right]\tilde{x}_j + \xi_i(t)\,.
}
The matrix in brackets is $\mbb{b}^\text{eff}_{\kappa\to\infty}$ of \Eqnref{linforce}. It has the form $-\mbb{a}+\zeta_T\kappa^{-2}\mbb{b}$ of Appendix~\secref{linforce}, with the expansion~\eqnref{bexp} given by $b^{(0)}_{ij}=\partial_jF_i(0)$ and $b^{(1)}_{ij}=\kappa^{-2}\sum_k\partial_k\partial_jF_i(0)F_k(0)/a_k$. Averaging \Eqnref{detforcecorr} over the disorder, we find
\begin{multline}\eqnlab{lamNsums}
	\langle\Delta\lambda_n\rangle_{F}\sim -\frac{\zeta_T}{\kappa^2}\langle \partial_n F_n(0)\rangle_{F} \\- \frac{\zeta_T^2}{\kappa^4}\Bigg[\sum_{m\neq n}^d\frac{\langle \partial_n F_m(0)\partial_m F_n(0)\rangle_F}{a_m-a_n}\\
	{}+ \sum_{m=1}^d\frac{\langle \partial_m\partial_n F_n(0)F_m(0)\rangle_F}{a_m}\Bigg]\,.
\end{multline}
The first term vanishes upon averaging, while we can move both derivatives on the $F_n(0)$ upon adding a minus sign. This way, we recover \Eqnref{largekap}.

We can simplify the expression~\eqnref{largekap} by using the parametrisation~\eqnref{Fcorr2} in terms of $\rho(r^2)$. We find the simple expression for the correlation function
\begin{multline}\eqnlab{ddFFNM}
	\langle \partial_n\partial_m F_n(\ve x)F_m(0)\rangle_F|_{\ve x=0} =\\
	 -4\left[\delta_{nm}\left[2+\gamma(d-2)\right]+1-2\gamma\right]\rho''(0)\,,
\end{multline}
which gives for \Eqnref{largekap}
\algn{
	\langle \Delta\lambda_n\rangle_F \sim4\frac{\zeta^2_T}{\kappa^4}\rho''(0)\left[\frac{3+\gamma(d-4)}{a_n}- \sum_{m\neq n}^d\frac{(1-2\gamma)a_n}{a_m(a_m-a_n)}\right]\,.
}
If we specify $n=1$, we find
\algn{
	\langle \Delta\lambda_1\rangle_F \sim4\frac{\zeta^2_T}{\kappa^4}\rho''(0)\left[3+\gamma(d-4)- \sum_{m= 2}^d\frac{1-2\gamma}{a_m(a_m-1)}\right]\,.
}
For the correlation function $C_F$ to be positive definite, we require that $(-1)^n\rho^{(n)}(0)\geq0$, which shows that the prefactor is non-negative, independently of the correlation function. This shows that, unless $\gamma=1$, there are at least two regimes if the bracket is positive, one at small $\kappa$, where relaxation is always slower than in the unperturbed case, and one at large $\kappa$, where it is faster.

This can be studied systematically in the simple case when $1=a_1<a_2 = a_3 = a_4=\ldots=a_d$. Figure~\figref{statediag} shows the regions where relaxation is asymptotically faster for $\kappa\to\infty$ in different dimensions. We observe that there are always regions in parameter space where relaxation is faster in the anisotropic case.

The two sums in \Eqnref{lamNsums} separate the mean of the linear force~\eqnref{linforce} from its fluctuations. Averaging \Eqnref{linforce} over realisations, the first-order term vanishes, and \Eqnref{Fcorr2} gives
\algn{\eqnlab{ddFF}
	\langle\partial_k\partial_jF_i(0)F_k(0)\rangle_F=-4\rho''(0)\,\delta_{ij}\left[1-2\gamma+(2+\gamma(d-2))\,\delta_{ik}\right]\,,
}
which contains \Eqnref{ddFFNM} as the case $i=j=n$, $k=m$. The mean background matrix is therefore diagonal, $\langle\mbb{b}^\text{eff}_{\kappa\to\infty}\rangle_F=-\mbb{a}^\text{eff}_{\kappa\to\infty}$, with effective stiffness matrix
\begin{multline}\eqnlab{Aeffinf}
	\mbb{a}^\text{eff}_{\kappa\to\infty}=\mbb{a}+\frac{4\rho''(0)\zeta_T^2}{\kappa^4}\Big\{\left[2+\gamma(d-2)\right]\mbb{a}^{-1}\\
	{}+(1-2\gamma)\tr(\mbb{a}^{-1})\,\mbb{1}\Big\}\,.
\end{multline}
Potential forces stiffen every direction. Solenoidal forces, by contrast, are divergence free and leave $\tr\mbb{a}^\text{eff}_{\kappa\to\infty}=\tr\mbb{a}$ unchanged, so that they only make the background more isotropic, which still stiffens its slowest direction. The second sum in \Eqnref{lamNsums} gives the resulting shift $(\mbb{a}^\text{eff}_{\kappa\to\infty})_{nn}-a_n$ of the level $a_n$. The first sum is the level coupling due to the fluctuations of $\partial_jF_i(0)$, with $\langle\partial_nF_m(0)\partial_mF_n(0)\rangle_F=4(1-2\gamma)\rho''(0)$ for $m\neq n$. Together,
\algn{\eqnlab{lamNsplit}
	\langle\Delta\lambda_n\rangle_F\sim(\mbb{a}^\text{eff}_{\kappa\to\infty})_{nn}-a_n-\frac{4\rho''(0)\zeta_T^2}{\kappa^4}\sum_{m\neq n}^d\frac{1-2\gamma}{a_m-a_n}\,,
}
which reduces to \Eqnref{mulongsplit} for $n=1$.
\section{Probability shift}\seclab{probapp}
This appendix collects the full $\kappa$-dependence of the probability shift
$\Delta P$ summarised in \Secref{aniso}, together with the comparison to
numerical simulations.
\subsection{Short correlation lengths}
At small $\kappa$, $\sigma_{\Delta\lambda_1}^2$ becomes small while
$\mu_{\Delta\lambda_1}$ converges to a finite, negative value for
$\gamma\neq1$. By \Eqnref{probshift}, $\zeta^{-1}_T\Delta P$ therefore takes
large negative values, growing as $\kappa^{-1}$ in $d=2$. The prefactor of the
$\kappa^{-d/2}$ scaling in \Eqnref{DPsmall} follows from the ratio of the
small-$\kappa$ scalings of $\mu_{\Delta\lambda_1}$ and $\sigma_{\Delta\lambda_1}$.
Equation~\eqnref{DPsmall} is a nonlinear function of $\gamma$, but it is
straightforward to show that $\DD{\gamma}\Delta P>0$ for all $\gamma$ and $d$,
which is the universal statement quoted in the main text.
\subsection{Large correlation lengths}
In the opposite limit of large correlation lengths, $\kappa\gg1$, inserting the asymptotics of $\mu_{\Delta\lambda_1}$ from \Eqnref{mulong} and of $\sigma^2_{\Delta\lambda_1}$ from \Eqnref{siglong} into \Eqnref{probshift} gives
\algn{\eqnlab{DPlarge}
	\zeta_T^{-1}\Delta P \sim \sqrt{\frac{2\rho''(0)}{\pi}}\frac{3+\gamma(d-4)- \sum_{m= 2}^d\frac{1-2\gamma}{a_m(a_m-1)}}{\sqrt{3+(d-4)\gamma}\;\kappa^2}\,.
}
Since $\sigma^2_{\Delta\lambda_1} > 0$, the sign of $\Delta P$ at large $\kappa$ is determined entirely by the sign of $\mu_{\Delta\lambda_1}$, as is already apparent from \Eqnref{probshift}.
Consequently, the parameter regions in which faster relaxation becomes more probable ($\Delta P > 0$) coincide precisely with those in which the mean relaxation speed is enhanced ($\mu_{\Delta\lambda_1} > 0$), i.e., the yellow-hatched region of the state diagram in \Figref{statediag}(a).

The effect of non-conservative forces on $\Delta P$ is captured by the sign of $\partial_\gamma \Delta P$.
Taking the derivative of \Eqnref{DPlarge} with respect to $\gamma$ and determining when it is positive, one finds that increasing $\gamma$ increases $\Delta P$ if and only if the expression
\algn{\eqnlab{sign}
	(d-4)\left[3+\gamma(d-4)\right] + \left[8+d+2\gamma(d-4)\right]\sum_{m= 2}^d\frac{1}{a_m(a_m-1)}
}
is positive.
Since $a_m > 1$ for all $m$, the sum is strictly positive, and for $d \geq d_\text{th} = 4$ both terms in \Eqnref{sign} are non-negative for all $\gamma \in [0,1]$.
Hence $\partial_\gamma \Delta P > 0$ for $d \geq d_\text{th}$, i.e., non-conservative forces increase the probability of faster relaxation for $d\geq4$, regardless of the background stiffnesses or the correlation function. This is the same threshold $d_\text{th} = 4$ found for $\mu_{\Delta\lambda_1}$ and $\sigma^2_{\Delta\lambda_1}$, because $\Delta P$ at large $\kappa$ is the ratio of these two quantities. At fixed force variance [\Eqnref{hatgamma}], increasing $\hat\gamma$ increases $\Delta P$ if and only if $(2d+1-d\hat\gamma)\sum_{m=2}^d[a_m(a_m-1)]^{-1}>(d-1)(3-2\hat\gamma)$, which depends on the background in every dimension.

For $d = 2$ and $d = 3$, the sign of \Eqnref{sign} depends on $\gamma$ and the $a_m$.
Specialising to $a_2 = a_3 = \cdots = a_d$, the boundary $a_2^*$ at which $\partial_\gamma\Delta P$ changes sign satisfies
\begin{multline}
	[8+d+2\gamma(d-4)](d-1) \\+ (d-4)[3+\gamma(d-4)]\,a_2^*(a_2^*-1)=0\,.
\end{multline}
Solving for $a_2^*$ as a function of $\gamma$, one finds that $a_2^*$ ranges between approximately $1.88$ and $2.30$ for $d=2$, and between approximately $3.25$ and $3.54$ for $d=3$.
For $a_2 < a_2^*$, increasing $\gamma$ increases $\Delta P$, and for $a_2 > a_2^*$ it decreases it.
In both cases the boundary is set by the competition between the two terms in \Eqnref{sign}: the background-dependent sum, which always favours non-conservative forcing, and the $(d-4)$ prefactor, which is negative for $d < 4$ and can dominate for stiff backgrounds.
\subsection{Comparison with numerical simulations}
Figure~\figref{prob} compares $\zeta_T^{-1}\Delta P$ from \Eqnref{DPcorr} (solid lines) and \Eqnref{probshift} (dashed lines) with values computed from the eigenvalue problem in $d=2$ for $10^3$ disorder realisations (markers), for two background stiffnesses.

Three features carry the physics. First, $\zeta_T^{-1}\Delta P$ is large and negative at small $\kappa$ for every $a_2$ and $\gamma$, as required by the $\kappa^{-1}$ divergence of \Eqnref{DPsmall} in $d=2$: a dense landscape of local minima traps the particle, making slower relaxation overwhelmingly likely. Second, $\Delta P$ changes sign at a $\kappa$ of order unity whenever $a_2$ or $\gamma$ is large enough, close to where $\mu_{\Delta\lambda_1}$ changes sign. The covariance $S$ of \Eqnref{Sdef} shifts this sign change to larger $\kappa$, for example from $\kappa\approx1.7$ to $\kappa\approx2.1$ for $a_2=10$ and $\gamma=0$, and the simulations follow this shift. Third, as the insets show, the ordering with $\gamma$ at large $\kappa$ inverts between the two backgrounds: larger $\gamma$ increases $|\Delta P|$ for $a_2=1.25$ and decreases it for $a_2=10$, as predicted by \Eqnref{sign}. The magnitude of $|\Delta P|$ grows as $a_2$ approaches unity, so nearly isotropic backgrounds offer the most favourable conditions for a large probability of faster relaxation.
\begin{figure}
	\includegraphics[width = \linewidth]{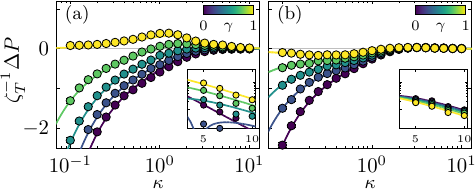}
	\caption{Scaled probability shift $\zeta_T^{-1}\Delta P$ in an anisotropic background in $d=2$ from theory (solid lines, \Eqnref{DPcorr}) and from simulations of the eigenvalue equation (markers) for $10^3$ realisations, or $2\times10^3$ pairs of fields $\pm\ve F$ for $a_2=10$ at $\kappa\geq2.5$ and $a_2=1.25$ at $\kappa=10$, shown as a function of $\kappa$ for different $\gamma$ [colour bars in panels (a) and (b)]. The thin dashed lines show \Eqnref{probshift}, which neglects the dependence between $X_1$ and $X_2$. The simulation values are $p_{X_1}(0)\langle X_2|X_1=0\rangle_F$ of \Eqnref{DPcorr}, evaluated with the coefficients $X_1$ and $X_2$ that we extract for each realisation from its eigenvalues at several $\zeta_T\leq0.02$, with error bars of one standard error. (a) $a_2 = 1.25$. (b) $a_2 = 10$. The insets show $|\zeta_T^{-1}\Delta P|$ at large $\kappa$ on double-logarithmic axes, the abscissa again being $\kappa$.}\figlab{prob}
\end{figure}

The simulations agree with \Eqnref{DPcorr} within their statistical errors. They deviate from \Eqnref{probshift} by up to about ten standard errors at intermediate $\kappa$, where $S/\sigma^2_{\Delta\lambda_1}$ is comparable to $\mu_{\Delta\lambda_1}$.
\section{\texorpdfstring{Variance and probabilities in quasi-isotropic backgrounds}{Variance and probabilities in quasi-isotropic backgrounds}}\seclab{collapse}
The scaling form~\eqnref{qisoscal} applies to the entire distribution of $\Delta\tilde\mu_1$, not only to its mean. \Figref{qisocollapse} tests it for the variance of $\Delta\tilde\mu_1$ and for the probabilities $P(\Delta\tilde\mu_1>0)$ and $P(\omega_1\neq0)$, using the same simulations as in \Figref{qisoscal}. As for the mean, the data collapse onto the random-matrix curves.
\begin{figure}
	\centering
	\includegraphics[width=\linewidth]{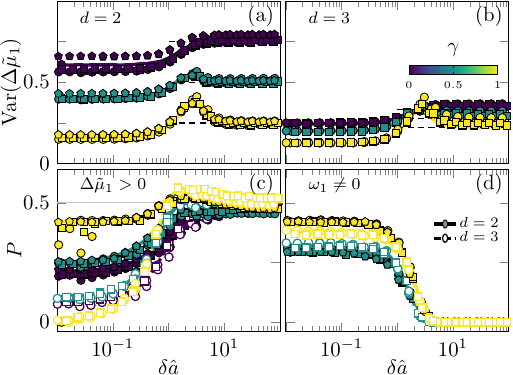}
	\caption{Variance and probabilities of the rescaled shift $\Delta\tilde\mu_1$ in a quasi-isotropic background with equal stiffer directions, $a_{n\geq2}=a_2$, for $\gamma=0$, $1/2$ and $1$ (colour bar). Lines and markers are obtained from the same random-matrix samples and simulations as in \Figref{qisoscal}. (a) Variance $\text{Var}(\Delta\tilde\mu_1)$ for $d=2$. The dashed lines show its large-$\delta\hat a$ limit $\langle\tilde H_{11}^2\rangle_F=[3+\gamma(d-4)]/d^2$. (b) Same as (a), but for $d=3$. (c) Probability $P(\Delta\tilde\mu_1>0)$ of faster relaxation for $d=2$ (solid lines, filled markers) and $d=3$ (dashed lines, open markers). The grey line marks $P=1/2$. (d) Probability $P(\omega_1\neq0)$ that $\lambda_1$ is part of a complex-conjugate pair, for $\gamma=1/2$ and $1$, with line styles and markers as in (c). For $\gamma=0$ all eigenvalues are real, so that $P(\omega_1\neq0)=0$.}
	\figlab{qisocollapse}
\end{figure}

The variance of $\Delta\tilde\mu_1$ in \Figsref{qisocollapse}(a) and~(b) increases from its isotropic value at $\delta\hat a=0$ towards the variance $\langle\tilde H_{11}^2\rangle_F=[3+\gamma(d-4)]/d^2$ of the first-order term $\tilde H_{11}$ in \Eqnref{qisoPT}, which dominates the shift at large $\delta\hat a$. For $\gamma>0$ it overshoots this limit near $\delta\hat a\approx2$ to $3$, where also $\langle\Delta\tilde\mu_1\rangle_F$ has its maximum, and for $\gamma=1$ it reaches $0.38$ in $d=2$ and $0.33$ in $d=3$, compared with the limits $1/4$ and $2/9$. The simulations agree with the random-matrix results to within about ten per cent. This is of the order of the statistical error of a variance estimated from $10^3$ realisations, about five per cent, and, as for the mean, it is correlated between markers that share the same random numbers.

Figure~\figref{qisocollapse}(c) shows the probability $P(\Delta\tilde\mu_1>0)$ of faster relaxation, which is related to the probability shift~\eqnref{DP} by $\Delta P=P(\Delta\tilde\mu_1>0)-1/2$. In the isotropic limit it is small, $0.21$ for $\gamma=0$ and $d=2$. For $\gamma=1$ in $d=2$ it vanishes at $\delta\hat a=0$, where the slowest mode has $\Delta\tilde\mu_1\leq0$ in every realisation (\Secref{iso}), but jumps to $0.42$ for any $\delta\hat a>0$. The reason is that the complex pairs, whose real part is exactly zero in the isotropic case, acquire the positive real part $\delta\hat a/2$ discussed in \Secref{qisoscaling}. In the simulations this jump is smeared out. At finite $\zeta_T$, the real part of a pair acquires corrections of relative order $\zeta_T$, which compete with $\delta\hat a/2$, so that $P(\Delta\tilde\mu_1>0)$ reaches $0.42$ only once $\delta\hat a$ exceeds a value of the order of $\zeta_T$. Apart from this, the simulated probabilities in \Figsref{qisocollapse}(c) and~(d) agree with the random-matrix results to within the statistical error of about $0.02$. With increasing $\delta\hat a$ it approaches $1/2$, because at large $\delta\hat a$ the shift is dominated by the symmetric first-order term $\tilde H_{11}$ in \Eqnref{qisoPT}. Faster relaxation becomes the more probable outcome, $\Delta P>0$, only for $\gamma>1/2$, with a largest value $P(\Delta\tilde\mu_1>0)\approx0.55$ at $\gamma=1$ for both $d=2$ and $d=3$. For $\gamma^*<\gamma<1/2$ the mean is positive while $\Delta P$ remains slightly negative, which means that the positive mean is carried by a minority of realisations with large positive shifts.

\Figref{qisocollapse}(d) shows the probability $P(\omega_1\neq0)$ that $\lambda_1$ is part of a complex-conjugate pair, whose decrease with $\delta\hat a$ limits the speedup discussed in \Secref{qisoscaling}. The coalesced pairs in \Figref{numsimqiso} are of this kind. At its parameters, $\delta\hat a=1.07$ for $d=2$ and $1.02$ for $d=3$, $\lambda_1$ is part of a complex-conjugate pair in a fraction $0.33$ and $0.32$ of the realisations at $\gamma=1$, respectively, in agreement with $P(\omega_1\neq0)=0.32$ from the random-matrix ensemble [\Figref{qisocollapse}(d)].
\section{Fourth-order shift in quasi-isotropic backgrounds}\seclab{qiso4}
Here we derive the expansion~\eqnref{qisotail4}. We separate the slowest direction in the matrix of \Eqnref{qisoeig} and write
\algn{
	\mbb{\tilde H}+\delta\hat{\mbb{a}}=\pmat{\tilde H_{11} & \ve u^{\sf T}\\ \ve w & \mbb{A}+\mbb{\tilde H}'}\,,
}
with $u_n=\tilde H_{1n}$ and $w_n=\tilde H_{n1}$ for $n\geq2$, the diagonal matrix $\mbb{A}=\text{diag}(\delta\hat a_2,\ldots,\delta\hat a_d)$, and the remaining block $\mbb{\tilde H}'$. The eigenvalue $\Delta\tilde\lambda_1$ that tends to zero for $\mbb{\tilde H}\to0$ solves the Schur-complement equation
\algn{
	\Delta\tilde\lambda_1=\tilde H_{11}+\ve u^{\sf T}\left(\Delta\tilde\lambda_1-\mbb{A}-\mbb{\tilde H}'\right)^{-1}\ve w\,.
}
We expand the inverse in powers of $\mbb{A}^{-1}$ and iterate. With $\mbb{X}=\mbb{\tilde H}'-\tilde H_{11}\mbb{1}$ this gives, up to fourth order in the entries of $\mbb{\tilde H}$,
\begin{multline}
	\Delta\tilde\lambda_1\sim\tilde H_{11}-\ve u^{\sf T}\mbb{A}^{-1}\ve w+\ve u^{\sf T}\mbb{A}^{-1}\mbb{X}\mbb{A}^{-1}\ve w\\
	+\left(\ve u^{\sf T}\mbb{A}^{-1}\ve w\right)\left(\ve u^{\sf T}\mbb{A}^{-2}\ve w\right)-\ve u^{\sf T}\mbb{A}^{-1}\mbb{X}\mbb{A}^{-1}\mbb{X}\mbb{A}^{-1}\ve w\,.
\end{multline}
By \Eqnref{tHcorr}, each pair $(u_n,w_n)$ has variance $v=(1+d\gamma)/d^2$ and covariance $\langle u_nw_n\rangle_F=v\tau$, and it is independent of the other pairs, of $\tilde H_{11}$ and of $\mbb{\tilde H}'$. The first and third terms have zero mean, because they contain an odd number of Gaussian entries. The second term gives $-v\tau S_1$, with $S_k=\sum_{n\geq2}\delta\hat a_n^{-k}$, which is \Eqnref{qisotail}. For the fourth-order terms, Wick's rule gives
\sbeqs{
\algn{
	\left\langle\left(\ve u^{\sf T}\mbb{A}^{-1}\ve w\right)\left(\ve u^{\sf T}\mbb{A}^{-2}\ve w\right)\right\rangle_F&=v^2(1+2\tau^2)S_3+v^2\tau^2(S_1S_2-S_3)\,,\\
	\left\langle\ve u^{\sf T}\mbb{A}^{-1}\mbb{X}\mbb{A}^{-1}\mbb{X}\mbb{A}^{-1}\ve w\right\rangle_F&=v^2\tau\left[2(1+\tau)S_3+\tau(S_1S_2-S_3)\right]\,,
}
}
where we used $\langle X_{nn}^2\rangle_F=2v(1+\tau)$ and $\langle X_{nm}X_{mn}\rangle_F=v\tau$ for $n\neq m$. The terms proportional to $S_1S_2$ cancel, and the fourth-order mean is
\algn{
	v^2(1-2\tau)S_3=\frac{(1+d\gamma)\left[(d+4)\gamma-1\right]}{d^4}\sum_{n=2}^d\frac1{\delta\hat a_n^3}\,.
}
The fifth order again has zero mean, so the next correction is of order $\delta\hat a^{-5}$. We have checked this result against sampling of the ensemble for $2\leq d\leq8$, for equal and unequal offsets, where the difference decreases as $\delta\hat a^{-2}$, as expected.
\section{Numerical simulations}\seclab{numerics}
In this appendix, we describe the method used to numerically solve the eigenvalue equation~\eqnref{Heig}.
\paragraph{Discretisation.} We represent $\ms{H}^\dagger$ on a periodic box with $N$ points per dimension, $x^{(j)} = jL/N-L/2$ with $j=0,\ldots,N-1$, using a Fourier pseudospectral scheme~\cite{Tre00}. Along direction $m$, the side $L$ is replaced by $L_m=L/\sqrt{a_m}$ here and below, so that the box is shorter along the stiffer directions, in proportion to the width of the eigenfunctions. This gives the same resolution in every direction. In isotropic backgrounds, the box is cubic, $L_m=L$. In this scheme the first and second derivatives along one coordinate are represented by the circulant matrices
\sbeqs{\eqnlab{diffmat}
\algn{
	(\mbb{D}_1)_{jl} &= \frac{\pi}{L}(-1)^{j-l}\cot\left[\frac{\pi(j-l)}{N}\right]\,,\\
	(\mbb{D}_2)_{jl} &= -\frac{2\pi^2}{L^2}\frac{(-1)^{j-l}}{\sin^2\left[\pi(j-l)/N\right]}\,,
}
}
for $j\neq l$, with $(\mbb{D}_1)_{jj}=0$ and $(\mbb{D}_2)_{jj}=-(4\pi^2/L^2)(N^2/12+1/6)$. These matrices differentiate any function with finite frequency range on the grid exactly~\cite{Tre00}. Writing $\mbb{X}=\text{diag}(x^{(0)},\ldots,x^{(N-1)})$, the one-dimensional blocks of the ladder operators~\eqnref{ladders} and of the unperturbed Hamiltonian~\eqnref{Ham} read
\algn{\eqnlab{blocks}
	\mbb{c}_m = \frac{\sqrt{a_m}}2\mbb{X} + \frac{1}{\sqrt{a_m}}\mbb{D}_1\,,\qquad
	\mbb{Q}_m = -\mbb{D}_2 + \frac{a_m^2}4\mbb{X}^2\,.
}
We embed each of them into the full $N^d$-dimensional space by the Kronecker product $\mbb{O}^{(m)} = \mbb{1}_{N^{d-m}}\otimes\mbb{O}\otimes\mbb{1}_{N^{m-1}}$, so that $\ms{H}^\dagger_0\mapsto\sum_{m}\mbb{Q}^{(m)}_m-\tr\mbb{a}/2$ and $c_m\mapsto\mbb{c}^{(m)}_m$. The domain must be large enough that the eigenfunctions, which decay on the scale of the harmonic background, are negligible at the boundary, and the spacing $L/N$ small enough to resolve both that scale and the correlation length of the disorder.

\paragraph{Random force field.} For each realisation, the potentials $V$ and $\Phi_{ij}$ that generate $\ve F$ in \Eqnref{force} are drawn in Fourier space on the same grid, with $k_m = 2\pi n_m/L_m$ and integer $n_m\in[-N/2,N/2)$. Writing
\algn{\eqnlab{window}
	W(\ve k) = \sqrt{\hat C_F(\ve k)}\,,%
	\qquad p_d = 2^{\frac{d+1}2}\kappa^{\frac d2}\Big(\prod_{m=1}^d L_m\Big)^{-\frac12}\,,
}
we draw independent standard complex Gaussian coefficients $\eta^{(s)}_{\ve k}$, with $\langle\eta^{(s)}_{\ve k}\eta^{(s')*}_{\ve k'}\rangle_F=\delta_{ss'}\delta_{\ve k\ve k'}$ and $\langle\eta^{(s)}_{\ve k}\eta^{(s')}_{\ve k'}\rangle_F=0$, and form the complex fields
\algn{\eqnlab{gfield}
	g^{(s)}(\ve x) = p_d\sum_{\ve k}W(\ve k)\,\eta^{(s)}_{\ve k}\,\ee^{i\ve k\cdot\ve x}\,.
}
Because $W$ is real and even and the set of wavevectors is symmetric, the real and imaginary parts of $g^{(s)}$ are two \textit{independent} real Gaussian random fields, each with covariance $C_F(|\ve x-\ve y|/\kappa)$. The prefactor $p_d$ enforces the normalisation $C_F(0)=1$. The sum in \Eqnref{gfield} represents the random fields faithfully only if the box is large and the grid spacing small compared with the correlation length, $L_m\gg\kappa\gg L_m/N$, in every direction. We therefore use different grids for different ranges of $\kappa$, see Table~\ref{tab:numerics}. Since its real and imaginary parts are independent, a single complex field supplies both potentials needed in $d=2$, the scalar potential $V$ and the stream function $\phi$, and two complex fields supply the four needed in $d=3$, namely $V$ and the three components of the vector potential $\ve A$, see \Eqsref{potentials}. The force is then assembled according to \Eqnref{force} with all derivatives applied as multiplications by $i\ve k$ before the inverse Fourier transform. Since these fields have correlation length $\kappa$ in $\ve x$, differentiating them yields $\kappa^{-1}\ve F(\ve x/\kappa)$ directly, which is the combination that enters \Eqsref{laneqn} and~\eqnref{HdagF}.

\paragraph{Eigenvalue solver.} The discretised operator has dimension $N^d$, of order $10^4$ in $d=2$ and $10^5$ in $d=3$ at the grids used here. To save memory, the corresponding matrix is never formed explicitly. Instead, the method only considers its action on a vector, where the derivative matrices enter through sparse Kronecker products and $\ve F$ enters as a diagonal.  We compute the $d+1$ eigenvalues with the smallest real part, that is, the stationary state $\lambda_{\ve 0}=0$ together with the first $d$ leading eigenvalues, by an implicitly restarted Arnoldi iteration with residual tolerance $10^{-10}$. For isotropic backgrounds, these $d+1$ eigenvalues contain the $d$-fold degenerate level of the unperturbed operator $\ms{H}_0^\dagger$, see \Figref{spectrum}(b). For $\gamma>0$ the operator is non-Hermitian, so that its eigenvalues may occur in complex-conjugate pairs that we order by real part. The iteration is limited to $10^3$ steps, which was sufficient for all realisation of the data shown here. It requests two ($d=2$) or three ($d=3$) eigenvalues beyond the $d+1$ that are needed, as a margin for convergence.

\paragraph{Reference spectrum.} It turns out that the spectrum of the discretised $\ms{H}^\dagger_0$ does not coincide exactly with the continuum levels $\ve a\cdot\ve m$. On the grids used here, the first excited level instead lies slightly above its continuum value, which is a property of the discretisation and is independent of the disorder. However, this shift is not necessarily small compared with the disorder-induced shift itself. We therefore diagonalise the discretised operator once per grid at $\zeta_T=0$, obtaining its own levels $\lambda^0_n$, and define the shifts as $\Delta\lambda_n = \lambda_n-\lambda^0_n$, so that this discretisation offset is removed.

\paragraph{Extraction of the perturbative coefficients.} Comparing the numerical results with the perturbative predictions is not entirely trivial, because the predictions are the coefficients of the leading term in $\zeta_T$, whereas the simulation returns the full $\zeta_T$ dependence. We therefore compute $\Delta\lambda_1$ for a logarithmically spaced grid of $11$ disorder strengths, $10^{-4}\leq\zeta_T\leq10^{-2}$ for \Figsref{mean}, \figref{var} and~\ref{fig:prob} and $10^{-3}\leq\zeta_T\leq10^{-1}$ for \Figsref{qisoscal} and~\ref{fig:qisocollapse}, keeping the \textit{same} realisation of $\ve F$ across the $\zeta_T$ grid. This makes the residual sampling noise common to all $\zeta_T$ and thereby tightens the fits. Furthermore, it allows each numerical eigenvalue solve to be started from the eigenvectors converged at the preceding, smaller $\zeta_T$, while the sweep is chosen to start from the exactly known $\zeta_T=0$ solution. We fit each series $\Delta\lambda_1(\zeta_T)$ by a quadratic polynomial in $\zeta_T$ (for \Figref{prob}, by a cubic polynomial without constant term) and extract the coefficient of the appropriate leading power. This isolates the perturbative contribution from the higher-order terms that become noticeable towards the largest $\zeta_T$. Taking averages and variances of these fitted coefficients over realisations, we obtain our numerical approximations of the leading-order perturbative contributions. The leading power is $\zeta_T^2$ in the anisotropic case and $\zeta_T$ in the isotropic case. In the quasi-isotropic case the shift is not a power of $\zeta_T$ at fixed $a_2$, so no fit is performed. Instead, the shift at each $\zeta_T$ is divided by $\zeta_T\Sigma(\kappa)d$ and plotted against the corresponding offset~\eqnref{dahat}. The fit also serves as a diagnostic, because a non-vanishing linear coefficient in the anisotropic case would indicate that the assumed scaling is incorrect, while a quadratic coefficient that is large compared with the linear one in the isotropic case would indicate that the fitting window extends beyond the perturbative regime. Where Table~\ref{tab:numerics} lists pairs, each realisation $\ve F$ is instead solved together with $-\ve F$ at two disorder strengths, $\zeta_T$ and $2\zeta_T$. Since $\Delta\lambda_1(-\ve F,\zeta_T)=\Delta\lambda_1(\ve F,-\zeta_T)$, the odd and even parts of the pair, divided by $\zeta_T$ and $\zeta_T^2$, give the linear and quadratic coefficients up to corrections of order $\zeta_T^2$, which cancel in the combination $[4f(\zeta_T)-f(2\zeta_T)]/3$ of the values $f$ at the two disorder strengths. Unlike the quadratic fit, this leaves no contribution of the cubic term in the quadratic coefficient.

\paragraph{Strong disorder.} For \Figsref{statemap} and~\figref{qisostrong}, $\Delta\lambda_1$ is computed at fourteen logarithmically spaced disorder strengths $0.05\leq\zeta_T\leq1$, keeping the same realisation of $\ve F$ across the $\zeta_T$ grid. At $\zeta_T$ of order one, we found that the periodic box produces spurious eigenvalues localised at its edges, where the ladder operators $\mbb{c}_m$ jump. We therefore switch the disorder off smoothly within $2.5/\sqrt{a_m}$ of the edges, where the eigenfunctions are negligible for $L\geq16$, so that $\lambda_1$ is unaffected. Each realisation $\ve F$ is paired with $-\ve F$, which has the same statistics, and each parameter set uses at least $10^3$ such pairs with independent random numbers. The maps in \Figref{statemap} are interpolated between the simulated parameters and $\zeta_T=0$, where the shift vanishes, by a modified Akima interpolation in $\ln\kappa$ and $\zeta_T$. The level lines and the dotted lines in \Figref{statemap} are contours of this interpolation. Near $\kappa^*$ in \Figref{statemap}(e), the mean is small compared with its fluctuations, so that up to $9\times10^4$ pairs are used there. They locate the dotted line at $\zeta_T=0.05$ at $\kappa=0.39\pm0.01$, in agreement with $\kappa^*=0.40$.

\paragraph{Parameters and statistics.} Table~\ref{tab:numerics} lists the parameters of all simulations. The means and variances in \Figsref{mean} and~\ref{fig:var} are based on up to $10^3$ realisations, or up to $2\times10^3$ pairs $\pm\ve F$, per parameter set. For $10^3$ realisations, the standard error of $\mu_{\Delta\lambda_1}$ is typically $0.005$ and at most $0.02$, and the relative standard error of $\sigma^2_{\Delta\lambda_1}$ is $\sqrt{2/10^3}\approx5\%$. All parameter sets on the same grid use the same random numbers, so that their statistical errors are correlated. The pairs use independent random numbers for each $\kappa$. The densities in \Figsref{numsimaniso}, \figref{numsimiso}, \figref{numsimisore}, \figref{prob3d}, \figref{numsimqiso} and~\figref{numsimqisore} are obtained at the three disorder strengths $\zeta_T=10^{-3}$, $10^{-2}$ and $10^{-1}$ only, from $25$ independent batches of $10^3$ realisations each that are pooled at every parameter point.

\begin{table*}
\caption{Parameters of the numerical simulations: dimension $d$, stiffnesses $a_{n\geq2}=a_2$, correlation length $\kappa$, fraction $\gamma$ of non-conservative forces, disorder strength $\zeta_T$, box parameter $L$ and number $N$ of grid points per dimension (Appendix~\secref{numerics}), and number of disorder realisations per parameter set. Entries $x/y$ refer to $d=2$ and $d=3$, respectively. Ranges of $\kappa$ are sampled logarithmically with ten values per decade, ranges of $\zeta_T$ with eleven logarithmically spaced values. In \Figsref{qisoscal} and~\figref{qisocollapse}, $a_2-1=10^{-4}$, $10^{-3}$, $10^{-2}$, $0.04$ and $0.12$. For the pairs $\pm\ve F$ with $L=12$, the field is drawn with the wavevectors $k_m=2\pi n_m/\max(L_m,6\kappa)$ of a larger periodic box and evaluated on the grid. For the pairs in $d=3$ at $\kappa\leq0.5$, $N=84$, $68$, $54$ and $48$ at $\kappa=0.2$, $0.25$, $0.32$ and $\kappa\geq0.4$, with $300$ pairs at $\kappa=0.2$. In \Figref{qisostrong}, $a_2-1=0.05$, $0.1$, $0.2$, $0.4$, $0.6$, $0.8$ and $1.1$ for all $\kappa$, together with $1.6$ and $3.2$ for $\kappa=2$, $1.6$ and $2.2$ for $\kappa=2.5$, $1.6$ for $\kappa=3$ and $1.4$ for $\kappa=3.5$. In \Figref{statemap}, $\zeta_T$ takes fourteen logarithmically spaced values, and $\kappa=0.28$, $0.4$, $0.56$, $0.8$, $1$, $1.4$, $2$, $2.5$, $3$, $3.5$, $4$, $5.6$ and $8$, with $N=172$, $120$, $86$, $60$, $80$, $58$, $60$, $70$, $76$, $86$, $90$, $126$ and $176$, and with $L=24$, $28$, $30$, $34$ and $36$ for $2\leq\kappa\leq4$. Each parameter set has $10^3$ pairs, or $4\times10^3$ pairs for $a_2>1$ and $2\leq\kappa\leq4$. In \Figref{statemap}(e), $\kappa=0.33$ and $0.47$ are added, with $N=146$ and $104$, and the parameter sets with $\kappa\leq0.8$ have $5\times10^3$ to $9\times10^4$ pairs.}\label{tab:numerics}
\begin{ruledtabular}
\begin{tabular}{lllllllll}
Figure & $d$ & $a_2$ & $\kappa$ & $\gamma$ & $\zeta_T$ & $L$ & $N$ & Realisations\\
\hline
\ref{fig:statemap} & $2$ & $1,\,1.1,\,2.1$ & $0.28$--$1.4$ & $0,\,1$ & $0.05$--$1$ & $16$ & $172$--$58$ & $10^3$--$9\times10^4$ pairs\\
 & & & $2$--$4$ & & & $24$--$36$ & $60$--$90$ & \\
 & & & $5.6$, $8$ & & & $50$, $70$ & $126$, $176$ & \\
\ref{fig:numsimaniso} & $2/3$ & $1.25$ & $1$ & $1$ & $10^{-1}$ & $20$ & $100/40$ & $2.5\times10^4$\\
\ref{fig:mean}, \ref{fig:var}, \ref{fig:prob} & $2$ & $1.25$ & $0.1$--$0.4$ & $0,\,\tfrac14,\,\tfrac12,\,\tfrac34,\,1$ & $10^{-4}$--$10^{-2}$ & $10$ & $150$ & $\leq10^3$\\
 & & & $0.5$--$4$ & & & $20$ & $100$ & \\
 & & & $5$--$8$ & & & $30$ & $150$ & \\
 & & & $10$ & & $10^{-2}$, $2\times10^{-2}$ & $12$ & $48$ & $2\times10^3$ pairs\\
 & & $10$ & $0.1$--$0.4$ & & $10^{-4}$--$10^{-2}$ & $10$ & $150$ & $\leq10^3$\\
 & & & $0.5$--$2$ & & & $20$ & $100$ & \\
 & & & $2.5$--$10$ & & $10^{-2}$, $2\times10^{-2}$ & $12$ & $48$ & $2\times10^3$ pairs\\
\ref{fig:mean}, \ref{fig:var} & $3$ & $1.25$ & $0.2$--$0.5$ & $0,\,\tfrac14,\,\tfrac12,\,\tfrac34,\,1$ & $5\times10^{-3}$, $10^{-2}$ & $10$ & $84$--$48$ & $\leq10^3$ pairs\\
 & & & $0.6$--$4$ & & $10^{-4}$--$10^{-2}$ & $20$ & $50$ & $\leq10^3$\\
 & & & $5$--$10$ & & $10^{-2}$, $2\times10^{-2}$ & $12$ & $48$ & $2\times10^3$ pairs\\
 & & $10$ & $0.2$--$0.5$ & & $5\times10^{-3}$, $10^{-2}$ & $10$ & $84$--$48$ & $\leq10^3$ pairs\\
 & & & $0.6$--$2$ & & $10^{-4}$--$10^{-2}$ & $20$ & $50$ & $\leq10^3$\\
 & & & $2.5$--$10$ & & $10^{-2}$, $2\times10^{-2}$ & $12$ & $48$ & $10^3$ pairs\\
\ref{fig:numsimiso}, \ref{fig:numsimisore} & $2/3$ & $1$ & $1$ & $0,\,\tfrac12,\,1$ & $10^{-1}$ & $20$ & $100/40$ & $2.5\times10^4$\\
\ref{fig:prob3d}(a),(b) & $2/3$ & $1$ & $1$ & $0,\,\tfrac12,\,1$ & $10^{-3}/10^{-2}$ & $20$ & $100/40$ & $2.5\times10^4$\\
\ref{fig:numsimqiso}, \ref{fig:numsimqisore} & $2/3$ & $1.04$ & $1$ & $0,\,\tfrac12,\,1$ & $10^{-1}$ & $20$ & $100/40$ & $2.5\times10^4$\\
\ref{fig:qisoscal} & $2/3$ & $1.0001$--$1.12$ & $1,\,2,\,4/1,\,2$ & $0,\,\tfrac12,\,1$ & $10^{-3}$--$10^{-1}$ & $30/20$ & $120/40$ & $10^3$\\
\ref{fig:qisostrong} & $2$ & $1.05$--$4.2$ & $2$ & $0,\,\tfrac12,\,1$ & $1$ & $24$ & $60$ & $4\times10^3$ pairs\\
 & & $1.05$--$3.2$ & $2.5$ & & & $28$ & $70$ & \\
 & & $1.05$--$2.6$ & $3$ & & & $30$ & $76$ & \\
 & & $1.05$--$2.4$ & $3.5$ & & & $34$ & $86$ & \\
 & & $1.05$--$2.1$ & $4$ & & & $36$ & $90$ & \\
\end{tabular}
\end{ruledtabular}
\end{table*}
\end{document}